\documentclass[a4paper,11pt]{article}
\pdfoutput=1 
\usepackage{jcappub} 
\usepackage[T1]{fontenc}
\usepackage[utf8]{inputenc}
\usepackage{float}
\usepackage{enumitem}
\usepackage{graphics}
\usepackage{mdframed}
\usepackage{mathtools}
\usepackage[dvipsnames]{xcolor}
\usepackage{soul}
\usepackage{amstext}
\usepackage{bm}	
\usepackage{booktabs}
\usepackage{cleveref}
\usepackage[section]{placeins}
\usepackage{pgfplots} 
\usepackage{array,collcell} 
\usepackage{siunitx} 
\usepackage{caption}
\usepackage{subcaption}
\usepackage{makecell}
\usepackage{ctable} 
\usepackage{adjustbox}
\usepackage{parskip}
\usepackage{graphicx}
\usepackage{tabularx}
\usepackage{multirow}
\usepackage{arydshln}
\usepackage{rotating}
\usepackage{pstricks}
\usepackage{color}
\usepackage{amsfonts}
\usepackage{mathrsfs}
\usepackage{epsfig}\usepackage{amsmath,amssymb,amsthm,graphicx,latexsym,braket}
\usepackage{dutchcal} 
\usepackage{ulem}
\usepackage{pifont}
\renewcommand{\arraystretch}{2}

\newcolumntype{H}{>{\centering\arraybackslash}X}

\newcommand{\be}{\begin{equation}}
\newcommand{\ee}{\end{equation}}
\newcommand{\bea}{\begin{eqnarray}}
\newcommand{\eea}{\end{eqnarray}}
\newcommand{\dd}{\mathrm{d}} 

\title{\textsc{\fontsize{30}{65}\selectfont \sffamily \bfseries Investigating the Constraints on Primordial Features with Future Cosmic Microwave Background and Galaxy Surveys}}
\author[a,1]{Debabrata Chandra\note{Electronic address: {deb.iitdelhi@gmail.com}}}
\author[a,b,2]{{ and Supratik Pal}\note{Electronic address: {supratik@isical.ac.in}}}

\affiliation[a]{\textit{Physics and Applied Mathematics Unit, Indian Statistical Institute,  203 B.T.Road, Kolkata 700108, India}}
\affiliation[b]{\textit{Technology Innovation Hub on Data Science, Big Data Analytics and Data Curation, Indian Statistical Institute, 203 B.T.Road, Kolkata 700108, India}}

\abstract{In this article, we do a thorough investigation of the competency of the forthcoming Cosmic Microwave Background (CMB) and Galaxy surveys in probing the features in the primordial power spectrum. Primordial features are specific model-dependent corrections on top of the standard power-law inflationary power spectrum; the functional form being given by different inflationary scenarios. Signature of any significant departure from the feature-less power spectrum  will enable us to decipher the intricacies of the inflationary Universe. Here, we delve into three major yet distinct features, namely, \textit{Bump feature}, \textit{Sharp feature signal}, and \textit{Resonance feature signal}. To analyse the features, we adopt a specific template for each feature model. We estimate the possible constraints on the feature parameters by employing Fisher matrix forecast analysis for the upcoming CMB missions such as \texttt{CMB-S4}, \texttt{CORE-M5}, \texttt{LiteBIRD}, \texttt{PICO} conjointly with \texttt{DESI}, and \texttt{EUCLID} galaxy surveys. To this end, we make use of four distinct observations to forecast on the bounds on the model parameters, namely, \textsf{CMB}, \textsf{Baryon Acoustic Oscillations~(BAO)}, \textsf{Galaxy Clustering} and \textsf{Gravitational Weak Lensing} or \textsf{Cosmic Shear} and their permissible synergy. For large scale structure (LSS) information, we consider different upper limits of scale for different redshifts for the purpose of circumventing the propagation of the errors stemming from the uncertainties on nonlinear scales into the constraints on the feature parameters. A comparative analysis of all three features has been done to estimate relative capabilities of these upcoming observations in shedding light on this crucial aspect of precision cosmology.} 

\DeclareUnicodeCharacter{2212}{-}
\begin{document}
\maketitle

\section{\textbf{Introduction}}\label{intro}
In the last few decades, we have gradually shifted to precision cosmology owing to the continuous improvement in the sensitivity of the experiments conducted to pin down the subtle mysteries of the Universe. The myriad of experiments previously conducted have already provided us with a great deal of information about different phases of the Universe, ranging from the primordial era to late time evolution of the Universe. The Planck mission~\cite{Planck:2018vyg,Planck:2018jri} has found with a high degree of confidence that  the Universe is described by  the following salient points:
\begin{itemize}[itemsep=-.3em]
\item[\ding{113}] The spatial curvature of the Universe is nearly zero, $ \Omega_K = 0.0007 \pm 0.0037 $ (95~\%~CL).  \item[\ding{113}] The statistical distribution of primordial perturbations is very much close to Gaussian in nature; no significant evidence of primordial non-Gaussianity has yet been found.
\item[\ding{113}] Primordial perturbations are adiabatic in nature.
\item[\ding{113}] Within the domain of comoving scales $0.005~\mathrm{Mpc}^{-1 } \lesssim k \lesssim 0.2~\mathrm{Mpc}^{-1}$, the primordial power spectrum exhibits a power-law behaviour, which is nearly scale-invariant with slight scale-dependence attributed to spectral index ($ n_s= 0.965 \pm 0.004$).
\end{itemize}
The above results collectively have paved the way to draw significant inferences about possible theoretical framework necessary to elucidate the physics of the primordial Universe by subjecting various inflationary models to experimental verification with high accuracy, distinguishing among them, as well as, comparing between the inflationary paradigm and alternative  theories.
These results derived from Planck have strongly advocated for the simplest inflationary models having single canonical scalar field rolling slowly over a smooth concave potential (potential having a negative second derivative) with Einstein gravity are the most favoured models among all the available inflationary candidates in addition to the other alternatives to inflationary hypothesis for providing a good fit to Planck's temperature and polarization data, thereby ruling out  several classes of models of  inflation or its alternatives~\cite{Planck:2018jri,Martin:2013tda,Domenech:2018bnf,Domenech:2020qay}.

Despite having such an exhaustive understanding of the physics of the primordial Universe, we are still left with several unresolved issues regarding the inflationary paradigm, which include the following aspects:
\begin{itemize}[itemsep=-.3em]
\item[\ding{113}]  What is the real nature of the initial vacuum of inflaton fluctuations, whether it is the standard Bunch-Davies vacuum or some non-trivial exotic one? 
\item[\ding{113}]  We are yet unsure about the exact energy scale of inflation; we are only provided with the upper limit of the energy scale, $V^{1/4}<1.7\times10^{16}~\mathrm{Gev}$~(95~\%~CL).
\item[\ding{113}] What is the exact form of interaction Hamiltonian of the field(s) involved in inflationary dynamics? 
\item[\ding{113}] How to explain low-$ \ell $ deficit at multipole $ \ell \sim$ 20 along with other interesting features present in the temperature power spectrum of CMB using inflationary theory?
\end{itemize}
Thus, we need to investigate each and every minute detail of the physics of the inflationary epoch in order to find the solution to these questions and this could be the primary target for the upcoming experiments concerning primordial perturbations and, for that matter, future experiments, in  a broader perspective, have one or more of the following targets at varied levels of importance:
\begin{itemize}[itemsep=-.3em]
\item[\ding{113}] A confirmed detection of primordial gravitational waves or the B-mode signals in the CMB polarization data would settle the uncertainties regarding the energy scale of inflation. 
\item[\ding{113}] A definite probe of primordial non-Gaussianity from CMB signals would solve the puzzle related to the structure of the inflationary interaction Hamiltonian.
\item[\ding{113}] The necessity to have a mathematically elegant quantum gravity theory, which will accommodate the long-standing problem of defining a vacuum state for quantum fields interacting with non-static background geometry. Consequently, more sensitive next-generation observations of the early universe are imperative to unveil new energy scale with profound certainty. This measure will provide an impetus to meticulously speculate about the physics of either the near Planck scale or the GUT scale. Furthermore, it would help us to a certain degree to make a conjecture about a more complete theory to find a self-consistent explanation of the inflationary paradigm from a quantum gravity theory. This would enable us to predict the observational results of future experiments and contribute towards scaling up the process of dealing with the quantum gravity problem. 
\item[\ding{113}] An exhaustive search for statistically significant deviations from the standard power-law power spectrum for simple inflationary models abiding by the slow-roll condition is essential to gauge any sort of non-triviality present in the physics of the primordial Universe, thereby, understanding the physics of the features present in the CMB temperature power spectrum at different multipoles~\cite{Braglia:2021sun,Braglia:2021rej}. However, it must be ensured from next-generation CMB experiments whether these features are of statistical origin or have some physical sources. 
\end{itemize}

So, a thorough investigation of different non-trivial features in the power spectrum originated from 
different background models of inflation in the light of next generation CMB missions like,  CMB-S4, CORE-M5, LiteBIRD and PICO in combination with galaxy surveys including DESI and EUCLID surveys, is the need of the hour. In this work we have addressed this important aspect methodically and stepwise, by taking into account three different features encompassing several inflationary models and by doing a forecast analysis on the above CMB missions, their combinations and combination with LSS surveys. Experimental specifications for the mentioned experiments are provided in Section~\ref{COS:EXPSPEC}. Although inflationary features are not too much statistically significant in the latest Planck data, it is quite likely that more precise data from upcoming missions in the coming decades will improve upon the existing results to a good extent and/or act as a complementary to them to improve our understanding of the primordial features to a considerable amount. Besides, next-generation galaxy measurements will also prove to be effective, as expected, in narrowing down the present uncertainties on features. Considering the feasibility of detecting primordial features in the near future, we have investigated the prospects of such experiments to identify the features. As already mentioned, the investigation would be undertaken by performing Fisher forecast analysis using mock likelihood, details of which are given in subsequent Sections. For the Fisher matrix analysis, we use the publicly available cosmological code, \texttt{MontePython}\footnote{https://github.com/brinckmann/montepython\_public}~\cite{Audren:2012wb,Brinckmann:2018cvx}, which is interfaced with the Boltzmann code \texttt{CLASS}\footnote{http://class-code.net}~\cite{Blas:2011rf}.
 
In the literature, a bulk of research exists along with sound theoretical motivation to opt for non-trivial mechanisms involved in inflationary physics, which results in departure from the featureless primordial power spectrum. However, it offers a good fit to the CMB temperature data despite not being selected statistically with strong confidence level by CMB data~\cite{Planck:2018jri}. Some of the works in this category can be listed as follows:
\begin{itemize}[itemsep=-.3em]
\item[\ding{113}]Inflationary models where there are steps or bumps in the potential, inflaton fields experience a shift from slow rolling for a small interval that results in producing features in the power spectrum~\cite{Starobinsky:1992ts,Adams:2001vc,Gong:2005jr,Chen:2006xjb,Hazra:2014goa,Hazra:2014jka,Hazra:2016fkm,GallegoCadavid:2015hld,GallegoCadavid:2017bzb,GallegoCadavid:2017pol,GallegoCadavid:2016wcz}. \item[\ding{113}] In the inflationary models involving multiple fields, the auxiliary scalar degrees of freedom often generates twists and turns in the trajectory of inflaton field as result of mutual interactions that bring about non-trivial change in the dynamics of inflaton field, thereby, allowing deviation from simple featureless power-law power spectrum~\cite{Achucarro:2010jv,Achucarro:2010da,Chen:2012ge,Pi:2012gf,Noumi:2013cfa,Burgess:2012dz,Battefeld:2013xka,Konieczka:2014zja,Mizuno:2014jja,Saito:2012pd,Gao:2013ota,Cespedes:2012hu,Shiu:2011qw,Gao:2012uq}. \item[\ding{113}] A sudden production of particles during an inflationary epoch for a short span of time, too, can introduce features in the power spectrum~\cite{Chung:1999ve,Elgaroy:2003hp,Mathews:2004vu,Romano:2008rr,Barnaby:2009mc,Barnaby:2009dd,Barnaby:2010ke,Naik:2022mxn,Furuuchi:2020klq}. \item[\ding{113}] Inflationary models, motivated from string theory producing short-term breaches of the slow-roll condition lead to deviations from feature-free power spectrum~\cite{Kobayashi:2012kc,Ashoorioon:2008qr,Cai:2015xla,Bean:2008na}. 
\end{itemize} 
We can broadly classify all these features into two different ways; firstly, in terms of the physical process generating the features and secondly, the extension of the features in the spectrum of the comoving scale. With regard to the former category, features are roughly divided into the following classes:
\begin{itemize}
\item[\ding{113}] Features generating due to momentary deviation from slow-roll condition during inflationary phase. 
\item[\ding{113}] Features due to the presence of oscillations in the inflationary dynamics itself.
\end{itemize}
In another way, we can crudely tag features into two broad categories: 
\begin{itemize} 
\item[\ding{113}] Local features, where departure from standard feature-free power spectrum occurs at a narrow range of scale.
\item[\ding{113}] Global features, here the signature of new physics is disseminated over an extended region of the mode in the spectrum.  
\end{itemize}
The features considered here will be categorized and briefly reviewed in section~\ref{Feature:Template}. Here, the main motive is to search for the sensitivity of future CMB experiments with diverse strengths and weaknesses in identifying the features. At the same time, we have studied how various permissible synergy of experiments affect the bounds on the feature parameters. The allowed combinations investigated here include CMB-CMB combinations and CMB-LSS combinations. In this direction, there are already many such forecasts in the literature; see for instance~\cite{Huang:2012mr,Chen:2016vvw,Chen:2016zuu,Ballardini:2016hpi,Xu:2016kwz,Palma:2017wxu}. However, CMB experiments have several innate limitations such as they are restricted by cosmic variance and it reflects the information of three dimensional primordial Universe from a two dimensional surface known as last scattering surface, consequently limiting the accessible mode numbers for performing statistical analysis and mitigating the subtle structure of the primordial power spectrum to some extent in projecting it on a two dimensional surface. These limitations of CMB experiments can not be evaded. However, LSS experiments can act as a complementary to CMB experiments since LSS experiments give a tomographic view of the Universe, along with that LSS surveys can act as separate probe for the primordial features since it probes the underlying matter power spectrum, which carries the information of primordial fluctuations and in this case the three dimensional aspect is staying intact~\cite{Beutler:2019ojk}. 

The plan of the paper is as follows: In Section~\ref{Feature:Template} we briefly review three different types of features and present the associated templates parameterizing the features in the primordial power spectrum. Section~\ref{COS:EXPSPEC} is dedicated to introduce the CMB and LSS experiments considered in this study, and discuss their respective experimental specifications. We present Fisher method and the analysis of the obtained results in Section~\ref{Analysis} and Section~\ref{Results}, respectively. In Section~\ref{Con} we draw the conclusions coming from this analysis. We summarize the results and the experimental specifications of the CMB experiments in tabular format in Appendix~\ref{Tables} and Appendix~\ref{CMB:expspec}, respectively. 
Finally, in Appendix~\ref{pwrspec:likhd} we review the construction of power spectra and likelihoods of the CMB and LSS experiments.

\section{Feature Templates}\label{Feature:Template}

This section provides a brief review of the features investigated in this article. For our analysis and comparison, we have taken under consideration three major features, namely, sharp features, resonance features and bump features. A template for each feature  is presented, which is actually a scale-dependent function that encodes all the detailed physics of inflationary models in the curvature power spectrum. These features are expressed as an extension over the standard feature-free power spectrum. An extensive review of these features belonging to different classes can be found in~\cite{Palma:2017wxu,Chen:2016vvw}. The definition of feature-free curvature power spectrum is as follows:
\be\label{pwrlw}
P_{0}(k)=A_{s}\left(\frac{k}{k_{*}}\right)^{\left(n_{s}-1 \right)}\ee
where, $ A_{s} $, $ n_{s} $ and $ k_{*} $ are the amplitude of scalar power spectrum, scalar spectral index and the pivot scale, respectively. 
The power spectrum with features can be described as follows:
\be\label{flpwr}
P(k)=P_{0}(k)\left[1+F(k)\right]\ee
where, $ F(k) $ takes into account any modification on the standard featureless power spectrum. Depending upon different inflationary models, the functional form of $ F(k) $ varies. Details about the form of $ F(k) $ for the features are shown in Table~\ref{tab:Tamplates} and elaborated in subsequent subsections.

\newcommand\AddLabel[1]{%
  \refstepcounter{equation}
  (\theequation)
  \label{#1}
}
\newcolumntype{M}{>{\hfil$\displaystyle}X<{$\hfil}} 
\newcolumntype{L}{>{\collectcell\AddLabel}r<{\endcollectcell}}

\newcommand\PV{P_{\text{pv,rated}}}
\begin{table}[h!]
\small
\centering
\captionsetup{font=footnotesize}
\parbox{15cm}{\captionof{table}{Templates of the feature models}\label{tab:Tamplates}}
 \begin{tabularx}\textwidth{@{}lML@{}}
 \toprule
 \textbf{Model} & \multicolumn{1}{l}{~~~~~~~~~~~~~~~~~~~~\textbf{Template}~[$F(k)=\dfrac{\Delta P}{P_{0}}(k)$]}
                              & \multicolumn{1}{l}{}\\ \midrule

  \textbf{Sharp Feature Signal}    &F(k)=S \sin \left( \frac{2 k}{k_s} + \phi_{s} \right)
                      & Template:Sharp \\
                        \textbf{Bump Feature}             &F(k)=B \left( \dfrac{\pi e}{3} \right)^{3/2} \left( \frac{k}{k_b}\right)^3 e^{-\frac{\pi}{2}(\frac{k}{k_b})^2}
                      & Template:Bump \\
  \textbf{Resonance Feature Signal}        &F(k)=R \sin \left[ k_r \log \left( 2k \right) + \phi_{r} \right]
                      & Template:Resonance \\
 
\bottomrule
\end{tabularx}
\end{table}

\subsection{Template I: Sharp Feature (Linear Sinusoidal Feature)}
\label{Sec:Sharp}
This is an oscillatory type feature, where, the deviation from the smooth power spectrum is extended over the entire momentum space. There is a wide range of inflationary models with different physical considerations that produce such oscillatory features in the spectrum. In all these models, the prime factor causing the genesis of such oscillatory patterns in the primordial power spectrum is the momentary violation of slow-roll condition due to the presence of sharp local features. Depending upon where such sharp local features are appearing we can have different inflationary scenarios, as in, the presence of sharp features in sound speed of inflaton fields can produce such sinusoidal behaviour in the spectrum~\cite{Bean:2008na,Miranda:2012rm}, presence of sharp features in the form of abrupt flex in the trajectory of inflaton field in multi-field space leads to oscillatory features in the power spectrum too~\cite{Gao:2012uq,Achucarro:2010da}, sharp features occurring in the inflationary potential generate oscillatory features in primordial power spectrum~\cite{Hazra:2014goa,Chen:2006xjb,Adams:2001vc,Starobinsky:1992ts,Domenech:2019cyh}. In this article,  we have taken under consideration the template~(\ref{Template:Sharp})~\cite{Chen:2011zf,Chen:2016vvw,Palma:2017wxu,Fergusson:2014hya,Fergusson:2014tza}; where, $S$, $1/k_s$ and $\phi_s$ represents amplitude, characteristic oscillation frequency and phase of the sharp feature or linear sinusoidal feature, respectively.

\subsection{Template II: Bump Feature}
\label{Sec:Bump}
The template~(\ref{Template:Bump}), which we have examined here belongs to the class of Bump-Like features; this particular form of the template has been well studied as a feature in~\cite{Barnaby:2009dd,Barnaby:2009mc,Barnaby:2010ke,Romano:2008rr,Chantavat:2010vt}. These kinds of features  appear as sudden bumps spread over a small region in the power spectrum of curvature perturbations. Bump-like features can be sourced from several models; each has their own physical motivations to generate it. For example, it can be generated from  production of particles during inflation and can be well expressed by the template~(\ref{Template:Bump})~\cite{Barnaby:2009mc,Palma:2017wxu,Chen:2016vvw}, where $ B $ denotes the amplitude of the feature and $ k_b $ denotes the scale where the bump shows up.

\subsection{Template III: Resonance Feature (Logarithmic Sinusoidal Feature)}\label{Sec:Resonance}

This is another broad class of features, where, the striking difference lies in the very nature of the source term that generates the output signature, in comparison to the features, as discussed in template I. The logarithmic features are the effect of periodic characteristics present in the inflationary picture, be it in the inflationary potential, or in the internal field space. For instance, if the sound speed or the slow-roll parameter shows periodic behaviour with time, then it can lead to such signals in the power spectrum of the curvature perturbations. Such features can be seen in inflationary scenarios like, brane inflation~\cite{Bean:2008na}, axion monodromy inflation~\cite{Flauger:2009ab} and natural inflation~\cite{Freese:1990rb}. To model resonance feature signal or sinusoidal logarithmic feature, we have considered the template~(\ref{Template:Resonance})~\cite{Chen:2008wn,Palma:2017wxu,Chen:2016vvw}. The model parameters $R$, $k_r$  and $\phi_{r}$ represent the amplitude, oscillation frequency and phase of the oscillation, respectively.

\begin{figure}[h!]
\begin{mdframed}
\captionsetup{font=footnotesize}
\center
\begin{minipage}[b1]{1.0\textwidth}
$\begin{array}{rl}
    \includegraphics[width=0.5\textwidth]{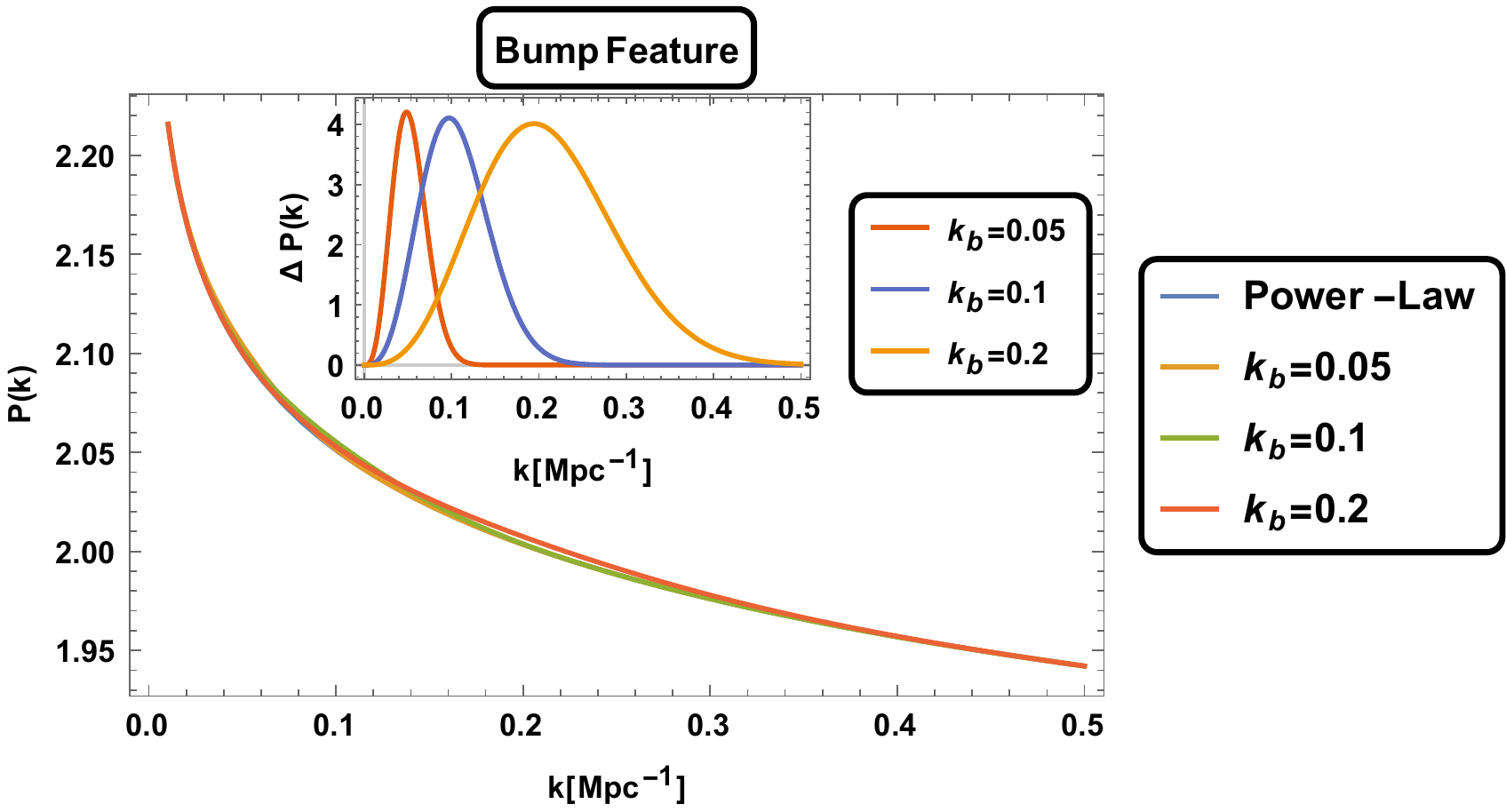} &
    \includegraphics[width=0.5\textwidth]{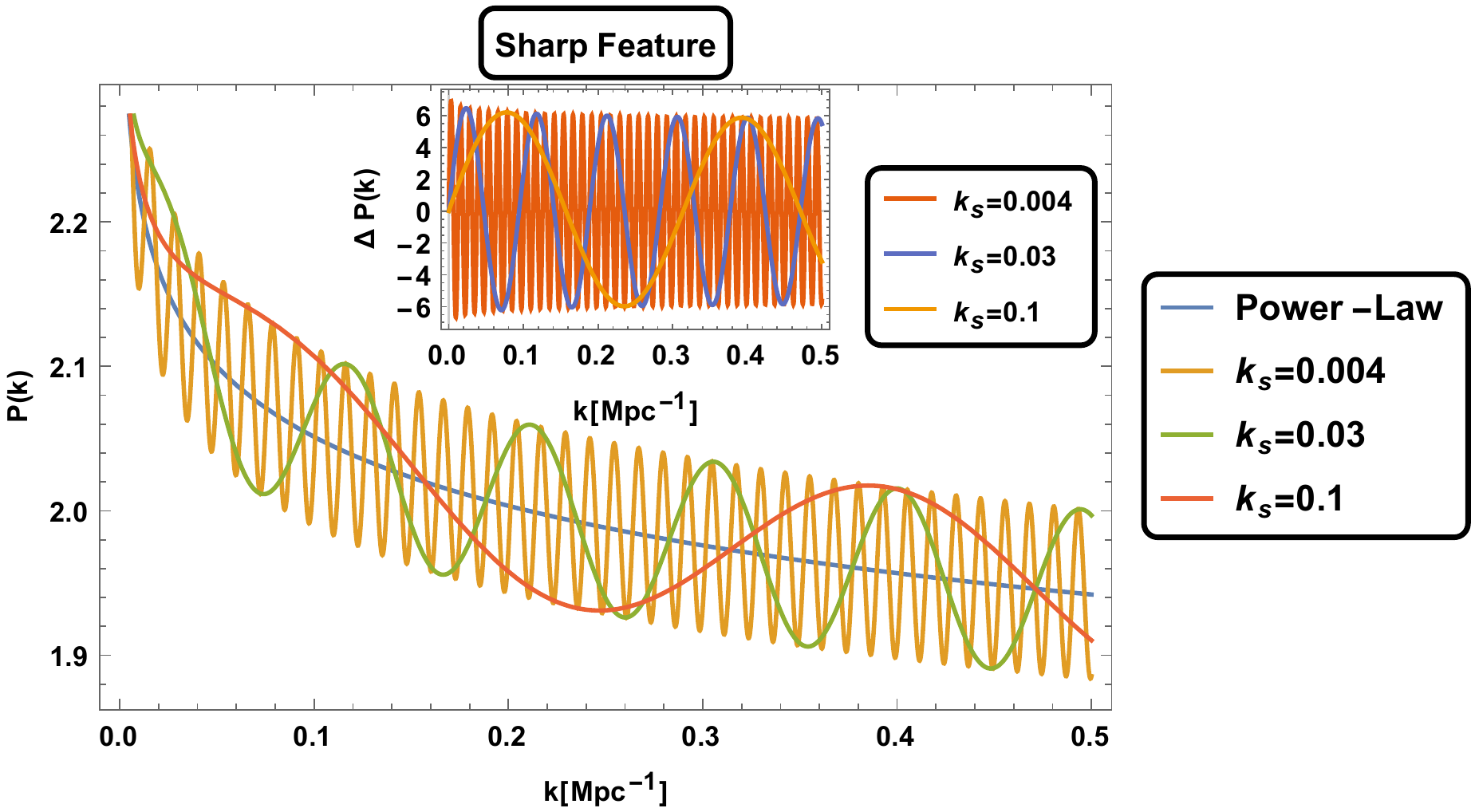}\\
    \multicolumn{2}{c}{\includegraphics[width=0.5\textwidth]{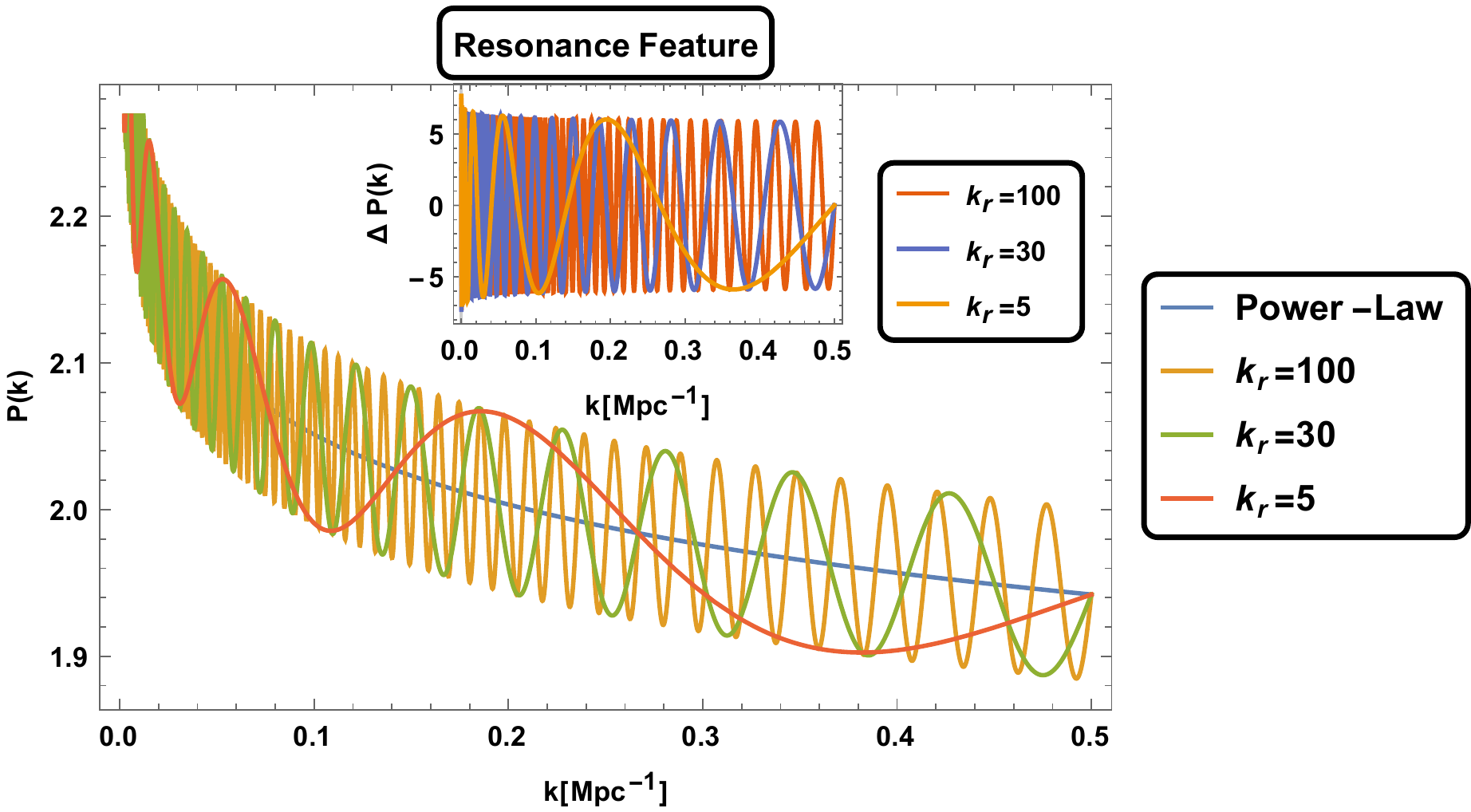}}
\end{array}$
\caption[]{\label{fig:features} The $P(k)$ and $\Delta P(k)$ for the templates of feature models have been plotted, where the used fiducial values are tabulated in Table~\ref{tab:BaseFeaFid}.}
\end{minipage}
\hfill
\end{mdframed}
\end{figure}

In this work, the possible bounds on the feature model parameters are estimated using Fisher method. To perform Fisher forecast analysis for these model parameters, the  fiducial values are separately given in Table~\ref{tab:BaseFeaFid} of Section~\ref{Analysis}. We have plotted the $P(k) $ and $ \Delta P(k)$ for the templates shown in Table~\ref{tab:Tamplates} using equations~from~(\ref{pwrlw}) to~(\ref{Template:Resonance}), which is shown in Figure~\ref{fig:features}.

\section{Cosmological Missions and Instrumental Specifications}\label{COS:EXPSPEC}

In this section, we describe the cosmological missions considered for our analysis and their respective instrumental specifications. To perform Fisher forecast analysis for the aforementioned features, four distinct types of experiments are taken into consideration: CMB anisotropies, baryon acoustics oscillations, galaxy clustering, and weak lensing. The individual missions under our consideration are as follows: 

\begin{itemize}[itemsep=-.3em]
\item[\ding{113}] \texttt{Cosmic Microwave Background Observations~(CMB)}: \textbf{Planck}, \textbf{CMB-S4}, \textbf{LiteBIRD}, \textbf{CORE-M5}, \textbf{PICO}
\item[\ding{113}] \texttt{Baryon Acoustic Oscillation Observations~(BAO)}: \textbf{DESI}
\item[\ding{113}] \texttt{Galaxy Clustering Observations~(GC)}: \textbf{EUCLID}
\item[\ding{113}] \texttt{Cosmic Shear Observations~(CS)}: \textbf{EUCLID}
\end{itemize}
Below we give a very brief description of each individual mission, that will help us do a comparative analysis subsequently.

\subsection{Cosmic Microwave Background Missions~(CMB)}\label{CMB:EXSPEC}
Here, we briefly introduce the CMB experiments  and elaborate on the scheme considered to combine two separate CMB experiments for our analysis. The instrumental specifications for the CMB experiments are adopted from Ref.~\cite{Brinckmann:2018owf}. The detailed specifications of the CMB experiments are  tabulated in Appendix~\ref{CMB:expspec}.
\begin{itemize}[itemsep=-.3em]
\item[\ding{113}] $\textsf{\textbf{LiteBIRD:}}$ This is a satellite-based project~\cite{Matsumura:2013aja,Suzuki:2018cuy}, designed to perform optimally in CMB B-mode measurment. It has a moderate resolving capacity but better sensitivity.  
\item[\ding{113}] $\textsf{\textbf{CORE-M5:}}$ CORE-M5 is also a possible future satellite proposal to measure CMB sky. It has nearly equal sensitivity but much better resolution compared to LiteBIRD.
\item[\ding{113}] $\textsf{\textbf{CMB-S4:}}$ This is a Stage-4 ground-based CMB experiment~\cite{CMB-S4:2016ple,CMB-S4:2017uhf} having a smaller sky coverage compared to satellites, whereas the beam resolution and the sensitivity is significantly better than LiteBIRD and CORE-M5.     
\item[\ding{113}] $\textsf{\textbf{PICO:}}$ This is another satellite project~\cite{Sutin:2018onu,Young:2018aby}. For different channel it has different resolution and sensitivity in comparison to other future projects. It has better beam resolution than LiteBIRD for all channels, but compared to CORE-M5, resolutions for few channels are slightly better, and for few slightly poorer. However, the CMB-S4 has better resolution than PICO. About sensitivity for few channels, it is comparable with CMB-S4 but better than CORE-M5 and LiteBIRD.     
\end{itemize}
For our analysis we take into account these individual CMB missions (except LiteBIRD, which we consider for combined case only) as well as possible combinations of them.  
As is well-known, CMB-S4 is supposed to perform well at small angular scales, whereas, LiteBIRD and CORE-M5 are proposed to perform well for large angular scales. Consequently, a quantitative analysis is highly significant to understand how their combinations will work in constraining the feature parameters. In this direction, apart from these future CMB missions, we take into consideration the existing Planck mission in combination with the upcoming CMB-S4 mission, as well, to gauge how combining a future mission with existing one would affect the bounds on the feature parameters. For our analysis, the combinations of our consideration are as follows  Ref.~\cite{Brinckmann:2018owf}:
\begin{itemize}[itemsep=-.3em]
\item[\ding{113}] $\textsf{\textbf{Planck + CMB-S4:}}$ In combining Planck and CMB-S4, we have considered Planck data for $ \ell\leq50 $ and CMB-S4 data for $ \ell>50 $ for their entire sky coverage plus high-$ \ell $ data of Planck for its $17\%$ sky coverage. Here, in this work we have considered the mock Planck likelihood instead of the full Planck mission~\cite{Planck:2018vyg}.
\item[\ding{113}] $\textsf{\textbf{LiteBIRD + CMB-S4 /CORE-M5 + CMB-S4:}}$ To combine LiteBIRD/CORE-M5 and CMB-S4, we have done similarly as above. We have considered LiteBIRD/CORE-M5 data for low-$ \ell $ ($ \ell\leq50 $) and CMB-S4 data for high-$ \ell $ ($ \ell>50 $) for their entire sky coverage along with that high-$ \ell $ data from LiteBIRD/CORE-M5 for its $30\%$ sky coverage.    
\end{itemize}

\setlength{\tabcolsep}{2.5pt} 
\renewcommand{\arraystretch}{1.2} 
\newcolumntype{C}[1]{>{\Centering}m{#1}}
\renewcommand\tabularxcolumn[1]{C{#1}}
\begin{minipage}{\linewidth}
\small
\centering
\captionsetup{font=footnotesize}
\begin{tabular}{|cccccccc|}
\hline
\multicolumn{8}{|c|}{\textbf{Galaxy Clustering (GC)}}                                                                                                                                                                                                \\ \hline
\multicolumn{1}{|c|}{$z_{\text {min}}$} & \multicolumn{1}{c|}{$z_{\text {max}}$} & \multicolumn{1}{c|}{$f_{\text {sky}}$} & \multicolumn{1}{c|}{$ \Delta z$}   & \multicolumn{1}{c|}{Radial Error~[$\sigma_{\shortparallel}$]}           & \multicolumn{1}{c|}{Angular Error~[$\sigma_{\perp}$]}            & \multicolumn{1}{c|}{Flux Limit~[$\text{F}_\mathrm{H\alpha}$]} & Galaxy Bias~[$b(z)$]      \\ \hline
\multicolumn{1}{|c|}{0.7}  & \multicolumn{1}{c|}{2.0}  & \multicolumn{1}{c|}{0.3636}    & \multicolumn{1}{c|}{0.1} & \multicolumn{1}{c|}{$0.001[1+z]c/H$} & \multicolumn{1}{c|}{0}              & \multicolumn{1}{c|}{$3\times10^{-16}$~$[\text{erg/s-}\rm{cm^2}] $}       & ${\mathscr N}_1[1+z]^{\frac{{\mathscr N}_2}{2}}$ \\ \hline
\multicolumn{8}{|c|}{\textbf{Cosmic Shear (CS)}}                                                                                                                                                                                                     \\ \hline
\multicolumn{1}{|c|}{$ z_{\text {min}}$} & \multicolumn{1}{c|}{$ z_{\text {max}}$} & \multicolumn{1}{c|}{$f_{\text {sky}}$}   & \multicolumn{2}{c|}{Galaxy~No.~Density~[${\cal n}_{\text{gal}}$]}                           & \multicolumn{1}{c|}{Redshift~error~[$\sigma_{\text{z}}$]} & \multicolumn{2}{c|}{{Median Redshift}~[$z_{\text {median}}$=$1.412z_{0}$]}        \\ \hline
\multicolumn{1}{|c|}{0}    & \multicolumn{1}{c|}{3.5}  & \multicolumn{1}{c|}{0.3636} & \multicolumn{2}{c|}{30~[$\text {per arcmin}^2$]}                                          & \multicolumn{1}{c|}{$0.05[1 + z]$}            & \multicolumn{2}{c|}{0.9}                    \\ \hline
\end{tabular}\par
\bigskip
\parbox{17cm}{\captionof{table}{Experimental specifications for the EUCLID galaxy clustering and cosmic shear experiments.}}\label{tab:GCCSExpspec} 
\end{minipage}

\subsection{Large Scale Structure Survey~(LSS)}\label{LSS:EXSPEC}
\begin{itemize}[itemsep=-.3em]
\item[\ding{113}] $\textsf{\textbf{EUCLID:}}$ Euclid  will probe nearly $10^7$ galaxies in a spectroscopic survey approximately within the redshift range $0.7<z<2.0$~\cite{EuclidTheoryWorkingGroup:2012gxx,Amendola:2016saw}. Along with spectroscopic survey, it will have photometric survey as well for measuring the cosmic shear; in this regard it will measure huge number of galaxies, approximately 30 galaxies per $\text {arcmin}^2$ spanning the redshift range of $0<z<3.5$ over 15,000 $\text {deg}^2$ of the sky. Both galaxy clustering and the weak lensing measurement are considered in this present work. A concise review on the galaxy power spectrum and weak lensing angular power spectrum is provided in Appendix~\ref{pwrspec:likhd}. The experimental specifications for both are mentioned below. 
\begin{itemize}[itemsep=-.3em]
\item[•] $\textsf{\textbf{Galaxy Clustering~(GC):}}$
The specifications for the galaxy clustering experiment of EUCLID are illustrated in Table~\ref{tab:GCCSExpspec}, where, $z$, $ \Delta z$, $f_{\text {sky}}$ are redshift, redshift bin width, and sky fraction, respectively. The fiducial value of $ \sigma_{\text{nl}} $ in eq.~(\ref{eqn:Pgal}) associated with the FoG effect is taken to be 7 Mpc in this work. Two nuisance parameters~(${\mathscr N}_1, {\mathscr N}_2$) with the fiducial value of 1 are considered in the galaxy bias in order to accommodate possible uncertainties in the form of the bias as per Ref.~\cite{Sprenger:2018tdb}. The galaxy number count distributions~($\dd N(z)/\dd z$) per $\text{deg}^2$ are summarized in Table~\ref{tab:GalDist}. In a given redshift bin, the total number of galaxies can be computed by
\begin{equation}
N(\bar{z}) = \frac{360^2}{\pi} f_{\text{sky}}\cdot\int_{\bar{z}-\frac{\Delta z}{2}}^{\bar{z}+\frac{\Delta z}{2}}\dfrac{\dd N(z)}{\dd z}\dd z \ .
\end{equation}
\begingroup
\setlength{\tabcolsep}{4.0pt} 
\renewcommand{\arraystretch}{1.0} 
\begin{table}[h]
\captionsetup{font=footnotesize}
\caption{Galaxy Number Count Distribution} 
\small
\centering 
\begin{tabular}{|c|rrrrrrrrrrrrr|} 
\hline 
\text{Mean Redshift}~($\bar{z}$) & 0.75 & 0.85 & 0.95 & 1.05 & 1.15 & 1.25 & 1.35 & 1.45 & 1.55 & 1.65 & 1.75 & 1.85 & 1.95\\ 
\hline
$\dd N(z)/\dd z$ & 4825 & 4112 & 3449 & 2861 & 2357 & 1933 & 1515 & 1140 & 861 & 654 & 499 & 382 & 295\\
\hline 
\end{tabular}
\label{tab:GalDist}
\footnotesize{This data has been taken from Table 3 of article~\cite{Pozzetti:2016cch}} for model 1 and flux limit $3 \times 10^{-16}~\text{erg}\cdot \text{s}^{-1}\cdot \text{cm}^{-2}$. \\
\end{table}
\endgroup
\item[•] $\textsf{\textbf{Cosmic Shear~(CS):}}$
The specifications for the cosmic shear experiment of EUCLID are summarized in Table~\ref{tab:GCCSExpspec}. The entire redshift range is divided into 10 bins with equal number of galaxies and number of galaxies per bin per steradian is designated by $ n_i $. The assumed fiducial value of $\sigma_{\text{lensing}}$ in eq.~(\ref{Shear:Noise}) is 0.3.
The galaxy number density distribution~($ \dd {\cal n}_\text{gal}/\dd z $) and associated Gaussian error function $ {\cal E}(z,z_m) $ are given below, where, $z$ and $z_m$ are respectively the true and measured redshifts.
\end{itemize}
\be
\label{Galaxy:Number}
{\cal n}_i = \frac{{\cal n}_{\text{gal}}}{10} \times 3600\left(\frac{180}{\pi}\right)^2
\ee
\noindent\begin{minipage}{.5\linewidth}
\begin{equation}
\label{Number:Density}
\frac{\dd {\cal n}_{\text{gal}}}{\dd z} \propto z^{2} e^{-\left(\frac{z}{z_0}\right)^{1.5}}
\end{equation}
\end{minipage}%
\begin{minipage}{.5\linewidth}
\begin{equation}
\label{Error:Func}
{\cal E}(z,z_m) = \frac{1}{\sqrt{2\pi}\sigma_{\text{z}}} e^{-\left(\frac{z-z_{m}}{\sqrt{2}\sigma_{\text{z}}}\right)^{2}}
\end{equation}
\end{minipage}
\item[\ding{113}] $\textsf{\textbf{DESI:}}$ DESI is designed to study the expansion history of the Universe by measuring the signature of the Baryon Acoustic Oscillations embedded on the cosmic structures and the growth rate of large scale structures through redshift-space distortion measurements caused by the peculiar velocities~\cite{DESI:2013agm,DESI:2016fyo}. It is a Stage IV ground-based experiment primarily probes for four different tracers or extragalactic objects over 14000 $\text{deg}^2$ of sky. It is proposed to measure around 10 million bright galaxies (BGS) in the redshift range $0.05<z<0.4$, 4 million luminous red galaxies (LRG) in redshift range $0.4<z<1.0$, 17 million emission line galaxies (ELG) in redshift range $0.6<z<1.6$ and roughly 2 million quasars (QSO) in redshift range $0.9<z<2.1$. Furthermore, it will also observe higher redshift quasars ($2.1<z$) to probe the absorption lines of the Lyman-$\alpha$  forest. In this analysis we have adopted the same experimental configuration as considered in the article~\cite{Archidiacono:2016lnv} for $ \Lambda \text{CDM} $ scenario.
\end{itemize}

\setlength{\tabcolsep}{1.5pt} 
\renewcommand{\arraystretch}{1.2} 
\newcolumntype{C}[1]{>{\Centering}m{#1}}
\renewcommand\tabularxcolumn[1]{C{#1}}
\begin{minipage}{\linewidth} 
\small
\centering
\captionsetup{font=footnotesize}
\begin{tabular}{|cllcccccc|}
\hline
\multicolumn{9}{|c|}{\textbf{Baseline Model Parameters}}                                                                                                                                                                 \\ \hline
\multicolumn{3}{|c|}{\text{Baseline Model}}  & \multicolumn{1}{c|}{$\omega_{\rm cdm}=\Omega_{\rm cdm}h^2$}       & \multicolumn{1}{c|}{$\omega_{\rm b}=\Omega_{\rm b}h^2$} & \multicolumn{1}{c|}{$ H_0$}    & \multicolumn{1}{c|}{$ \tau_\mathrm{reio} $}    & \multicolumn{1}{c|}{$n_{s}$} & $ A_{s}$      \\ \hline
\multicolumn{3}{|c|}{\text{Fiducial Value}} & \multicolumn{1}{c|}{0.12011}     & \multicolumn{1}{c|}{0.022383}  & \multicolumn{1}{c|}{67.32~$[\text{km/s-Mpc}] $} & \multicolumn{1}{c|}{0.0543} & \multicolumn{1}{c|}{0.96605}     & $2.100 \times 10^{-9}$ \\ \hline
\multicolumn{9}{|c|}{\textbf{Feature Model Parameters}}                                                                                                                                                                  \\ \hline
\multicolumn{3}{|c|}{\text{Feature Model}}  & \multicolumn{1}{c|}{\text{Bump Feature}} & \multicolumn{2}{c|}{\text{Sharp Feature Signal}}                  & \multicolumn{3}{c|}{\text{Resonance Feature Signal}}                           \\ \hline
\multicolumn{3}{|c|}{\text{Fiducial Value}} & \multicolumn{1}{c|}{B=0.002; $k_b$=0.05,~0.1,~0.2}           & \multicolumn{2}{c|}{S=0.03; $k_s$=0.004,~0.03,~0.1; $\phi_{s}$=0}                                  & \multicolumn{3}{c|}{R=0.03; $k_r$=5,~30,~100; $\phi_{r}$=0}                                                \\ \hline
\end{tabular}\par
\bigskip
\parbox{14cm}{\captionof{table}{Fiducial values of the baseline model parameters and feature model parameters.}}\label{tab:BaseFeaFid}
\end{minipage}
As it will be found out, in our analysis, we will take different combinations of \textbf{CMB + LSS} missions, and compare them vis-a-vis \textbf{CMB}-only / 
\textbf{CMB + CMB} missions.

\section{\textbf{Fisher Matrix Forecast Analysis}}\label{Analysis}
In this section, Fisher Matrix Forecast Analysis and the fiducial values of the parameters  are discussed briefly. In the subsequent sections, the results obtained from this analysis are presented. In this study, we have considered $ \Lambda $CDM as our baseline cosmology. The fiducial values for the baseline and the feature model parameters are given in Table~\ref{tab:BaseFeaFid}, which are adopted from the Planck results\footnote{This fiducial values are taken from Table 1. Column 2 of article~\cite{Planck:2018vyg}.}~\cite{Planck:2018vyg} and Ref~\cite{Palma:2017wxu}, respectively. With these fiducial values, we here adopted the Fisher forecast technique~\cite{Tegmark:1996bz} to provide possible constraints on the parameters for aforesaid experiments. The Fisher matrix is by definition the second derivatives of the log-likelihood function or the effective chi-square function with respect to the cosmological model parameters around their best-fit values. This can be expressed as follows:
\be
\label{Fisher Matrix}
\mathbf{F}_{ij}= -\left\langle \dfrac{\partial^{2} \ln {\mathscr{L}}}{\partial \alpha_{i}\partial \alpha_{j}} \right\rangle = - \dfrac{\partial^{2} \ln {\mathscr{L}}}{\partial \alpha_{i}\partial \alpha_{j}} \Bigg|_{\alpha_{0}} 
\ee
\noindent\begin{minipage}{.5\linewidth}
\begin{equation}
\label{Cov:Matrix}
\mathbf{Cov}~(\alpha_i,\alpha_j)\geq [\mathbf{F}^{-1}]_{ij}
\end{equation}
\end{minipage}%
\begin{minipage}{.5\linewidth}
\begin{equation}
\label{Para:Error}
\sigma(\alpha_i) = \sqrt{[\mathbf{F}^{-1}]_{ii}}  ~.
\end{equation}
\end{minipage}

In the Fisher matrix method the log-likelihood function is expanded in Taylor series around its peak and truncating the higher order terms (higher than second order) enables us to approximate the log-likelihood function as a multi-variate Gaussian function of cosmological parameters and the coefficients of the second order terms produce the elements of the Fisher matrix written in eq.~(\ref{Fisher Matrix}).
This equation encapsulates the information about possible constraints on the parameters and their mutual correlations. The inverse of the Fisher matrix  is the covariance matrix, and the square root of the diagonal elements of the covariance matrix gives the bounds on the corresponding parameters, which have been shown in eq.~(\ref{Cov:Matrix}) and~(\ref{Para:Error}), respectively. 
For our computations, we have used publicly available code, \texttt{MontePython}\footnote{https://github.com/brinckmann/montepython\_public}~\cite{Audren:2012wb,Brinckmann:2018cvx} \texttt{v3.4}, interfaced with the Boltzmann solver \texttt{CLASS}\footnote{http://class-code.net}~\cite{Blas:2011rf} \texttt{v2.9.4}. We made the necessary changes in the Boltzmann solver \texttt{CLASS} to incorporate the feature models for our analysis and used \texttt{MontePython} to compute the Fisher matrices. \texttt{MontePython} computes Fisher matrix directly from likelihood by generating fiducial data to mimic observed data and using eq.~(\ref{Fisher Matrix}); for further details we refer the reader to the Ref.~\cite{Brinckmann:2018cvx}. The likelihoods in eq.~(\ref{Fisher Matrix}) for different observables are illustrated in Appendix~\ref{pwrspec:likhd}.

In galaxy clustering observation~(\textbf{Euclid GC}), we have introduced upper and lower bounds on the scales. On the lower limit of scales we have imposed a cut-off at $k_{\text{min}}=0.02~\mathrm{Mpc}^{-1}$ to preserve the small angle approximation or to filter out the scales which are larger than the assumed bin width. On the upper limit of scales, we have adopted the conservative scheme by imposing a non-linear cut-off at $k_{\text{NL}}(z) = k_{\text{NL}}(0)\cdot(1+z)^{2/(2+n_s)}$, where, the theoretical uncertainty for \textbf{Euclid GC}, $k_{\text{NL}}(0)=0.2 $\,$h/\text{Mpc}$. In a similar manner for cosmic shear observation~(\textbf{Euclid CS}) we have considered multipole range from $\ell_{\text{min}}=5$ to $\ell_{\text{max}}^i(z)$. The redshift dependence of $\ell_{\text{max}}^i(z)$ is given as $\ell_{\text{max}}^i = k_{\text{NL}}(z) \cdot \bar{r}^{i}_{\text{peak}}$, where, $k_{\text{NL}}(0)=0.5 $\,$h/\text{Mpc}$ and $\bar{r}^{i}_{\text{peak}}$ is given by
\begin{equation}
\bar{r}^{i}_{\text{peak}}\equiv  \frac{1}{(N-i)}\sum_{j>i} \frac{\displaystyle \int_0^{\infty}\frac{\dd r \cdot r}{r^2} {\cal K}_i(r) {\cal K}_j(r)}{\displaystyle \int_0^{\infty}\dfrac{\dd r}{r^2} {\cal K}_i(r) {\cal K}_j(r)} \ ~,
\end{equation}
where, $ i, j $ are pair of redshift bins and $N$ is the number of bins.

The details of the mock likelihoods for the experiments under consideration can be found in Appendix~\ref{pwrspec:likhd}.

\section{\textbf{Results and Discussions}}\label{Results} 
Let us now investigate for the performance of different CMB surveys as well as possible combined surveys based on the above Fisher analysis and fiducial values.  In Appendix~\ref{Tables}, all results are elaborated in tabular format.  These results represent the marginalized 1-$\sigma$ constraints on the parameters~(cosmological + feature model). The interested reader can refer to them for any particular numerical value of his/her interest. Below we discuss the results for different features based on the plots obtained from those tables.

Here, in the following Figures~(\ref{fig:Sharp-1}-\ref{fig:Resonance-3}), we have summarised the obtained results in a graphical way to perform a comparative analysis of the 1-$\sigma$ errors on the feature parameters coming from different experiments. For each feature, we have considered three separate cases for three representative fiducial values of the parameters under consideration, For all these plots~(\ref{fig:Sharp-1}-\ref{fig:Resonance-3}),  different CMB experiments are listed on the right side. Number of experiments under consideration have been labeled along \textit{X-axes} and the corresponding 1-$\sigma$ uncertainties for relevant parameters for individual features are represented along \textit{Y-axes}. The numbers \textbf{1}, \textbf{2}, \textbf{3}, \textbf{4}, \textbf{5} and \textbf{6} along \textit{X-axes}  stand for \textbf{CMB}-Only, \textbf{CMB + DESI}, \textbf{CMB + EUCLID-CS}, \textbf{CMB + DESI + EUCLID-CS}, \textbf{CMB + EUCLID-GC}, \textbf{CMB + EUCLID-(GC + CS)}, respectively. In other words,   the variation of the 1-$\sigma$ errors with each LSS experiment is shown for a given CMB experiment horizontally, whereas  the change in uncertainties on the model parameters is highlighted for a given LSS experiment with different CMB experiments along vertical direction.

\begin{figure}[h!]
\begin{mdframed}
\captionsetup{font=footnotesize}
\center
\begin{minipage}[b1]{1.0\textwidth}
$\begin{array}{rl}
\includegraphics[width=0.5\textwidth]{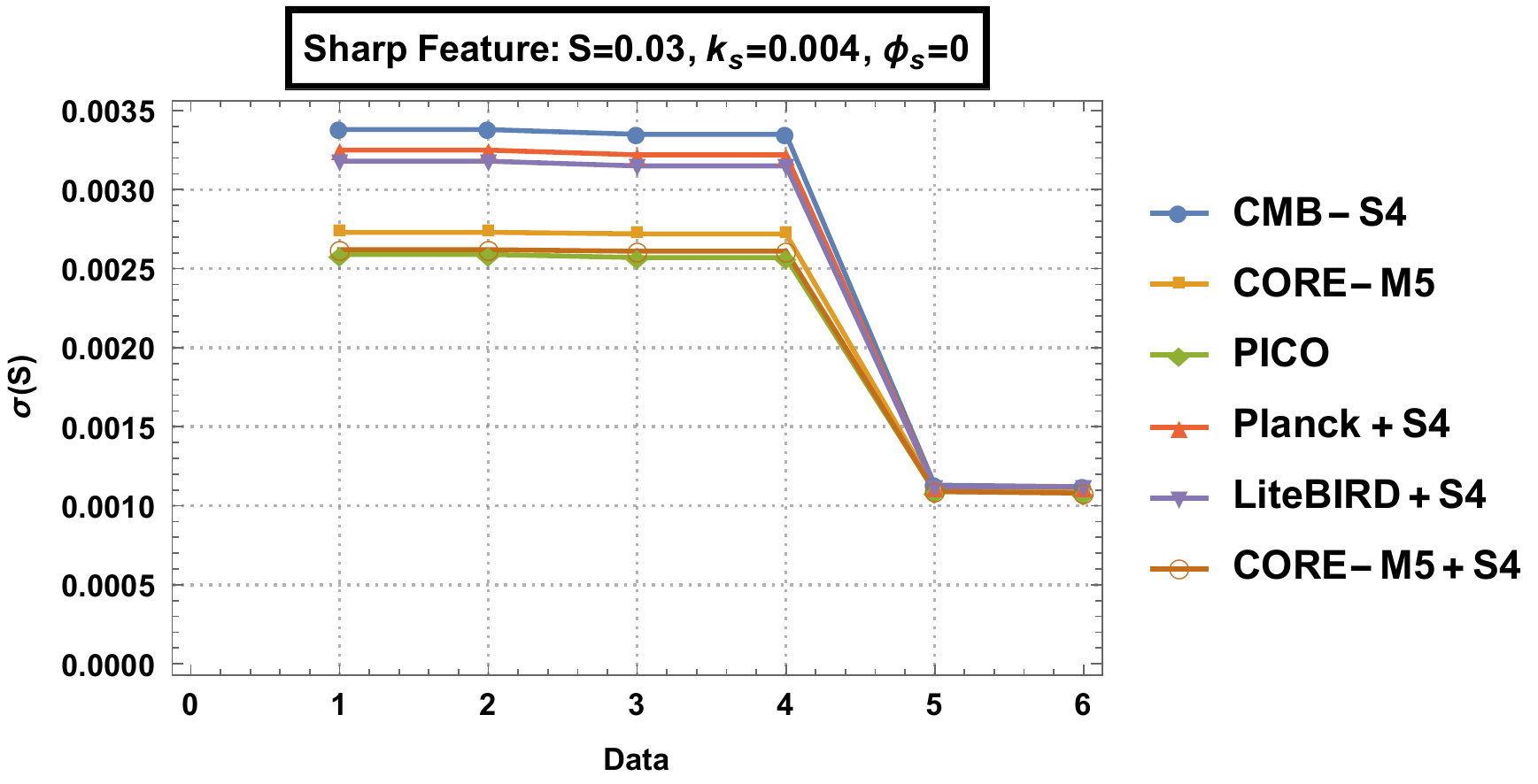} &
\includegraphics[width=0.5\textwidth]{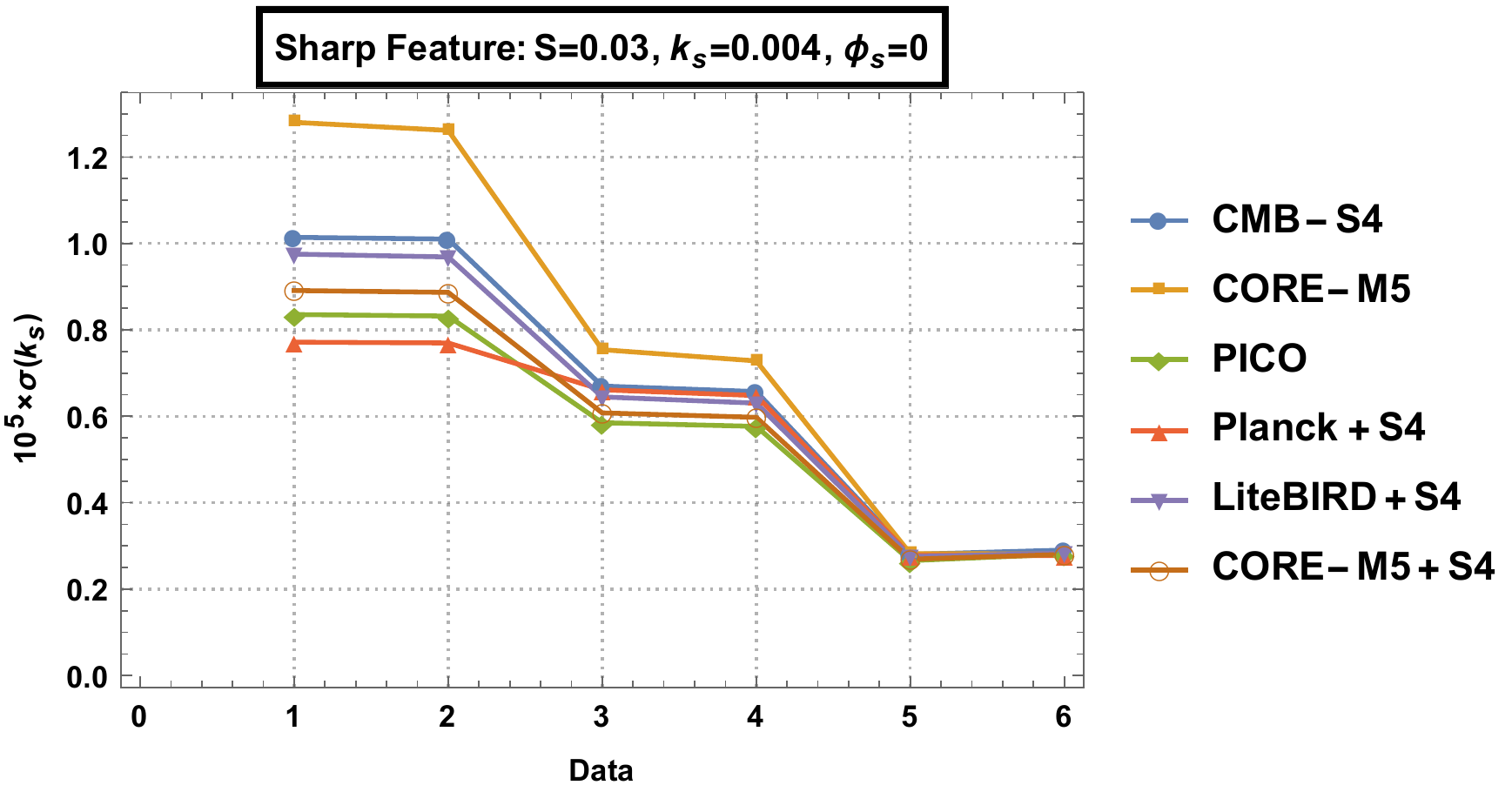}\\
\multicolumn{2}{c}{\includegraphics[width=0.5\textwidth]{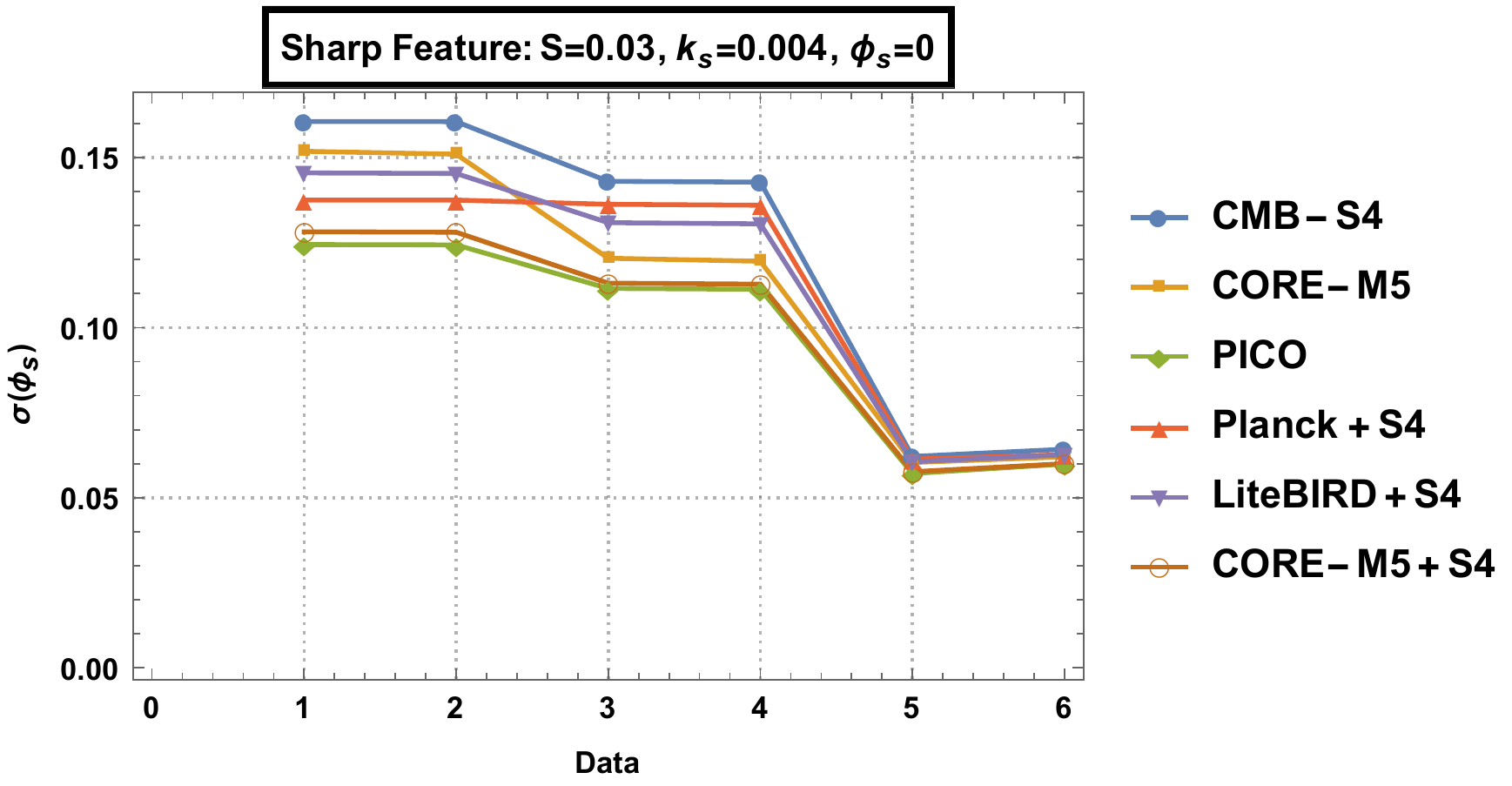}}
\end{array}$
\caption[]{\label{fig:Sharp-1}Experimental sensitivity of each CMB experiment for CMB-only~(1) and combination with DESI~(2), EUCLID-CS~(3), DESI+EUCLID-CS~(4), EUCLID-GC~(5) and EUCLID-(CS+GC)~(6) on the parameters of \textbf{Sharp Feature} template~(\ref{Template:Sharp}) has been depicted here. The 1-$\sigma$ uncertainties on the amplitude~(S), oscillation frequency~($k_s$) and phase~($\phi_s$) of the linear sinusoidal signal are shown in the \textit{Left} plot, \textit{Right} plot and \textit{Middle} plot, respectively. The plots are for these given fiducial values of the model parameters: S=0.03; $k_s$=0.004; $\phi_{s}$=0.}
\end{minipage}
\hfill
\end{mdframed}
\end{figure}

\subsection{Sharp Feature}
Let us first discuss the results for sharp feature signal. As mentioned earlier, we will consider three different cases for three different sets of fiducial values of the parameters associated with sharp feature signal. 
\subsubsection{Case I}
Fig.~\ref{fig:Sharp-1} represents the first case with the fiducial values S=0.03; $k_s$=0.004; $\phi_{s}$=0. From Fig.~\ref{fig:Sharp-1} we can summarise our results as follows:
\begin{itemize}[itemsep=-.3em]
\item[•] For CMB-only experiments, PICO and CORE-M5 + CMB-S4 are giving the best constraints; both are providing similar bounds on the amplitude~(S) of the linear oscillatory signal if CMB-only experiments are considered.
\item[•] DESI, EUCLID-CS and their combination are not improving the results in comparison to CMB-only bounds; this is the same picture for all the CMB experiments.
\item[•] For LSS experiments in combination with CMB experiments, EUCLID-GC and EUCLID-(GC+CS) are giving similar bounds; even  combination of GC and CS is not working in improving the results. Yet, in comparison to CMB-only results, the best improvement is coming for CMB-S4 data, it is improving the bounds by more than factor of 3, and for PICO it is more than factor of 2. All CMB experiments in combination with EUCLID-GC and EUCLID-(GC+CS) give rise to similar constraints on the amplitude~(S).
\item[•] For the characteristic frequency ($1/k_s$), the best bound on $k_s$ is coming from Planck + CMB-S4 and the highest value from CORE-M5 for CMB-only scenario.
\item[•] Here as well, the DESI BAO experiment is not playing any role in improving the bounds for any given combination with  CMB experiments. However, EUCLID-CS is somewhat sensitive here; it slightly improves the bounds, unlike the oscillation amplitude.
\item[•] On considering LSS experiments, the best constraints are from the EUCLID-GC and EUCLID-(GC+CS); for all CMB experiments, both experiments are giving similar bounds.
\item[•] For CORE-M5, the combination with EUCLID-GC and EUCLID-(GC+CS) improves the results by a factor of 4. 
\item[•] For phase angle~($ \phi_s $), the bounds are between CMB-S4 and PICO. PICO is providing the best constraint. Phase is also not sensitive to DESI BAO experiment but shows slight improvements for EUCLID-CS except Planck + CMB-4; the relatively best improvement is occurring for CORE-M5 when combined with EUCLID-CS.
\item[•] For all CMB experiments, the best constraints are coming when it is combined with EUCLID-GC and EUCLID-(GC+CS), though they are giving almost the same bounds but taking together EUCLID-CS with EUCLID-GS in combination with CMB is slightly deteriorating the results. Combination of GC data with CMB is shrinking the bounds by a factor between 2-2.5 depending upon the CMB experiments.  
\end{itemize}
\begin{figure}[h!]
\begin{mdframed}
\captionsetup{font=footnotesize}
\center
\begin{minipage}[b1]{1.0\textwidth}
$\begin{array}{rl}
\includegraphics[width=0.5\textwidth]{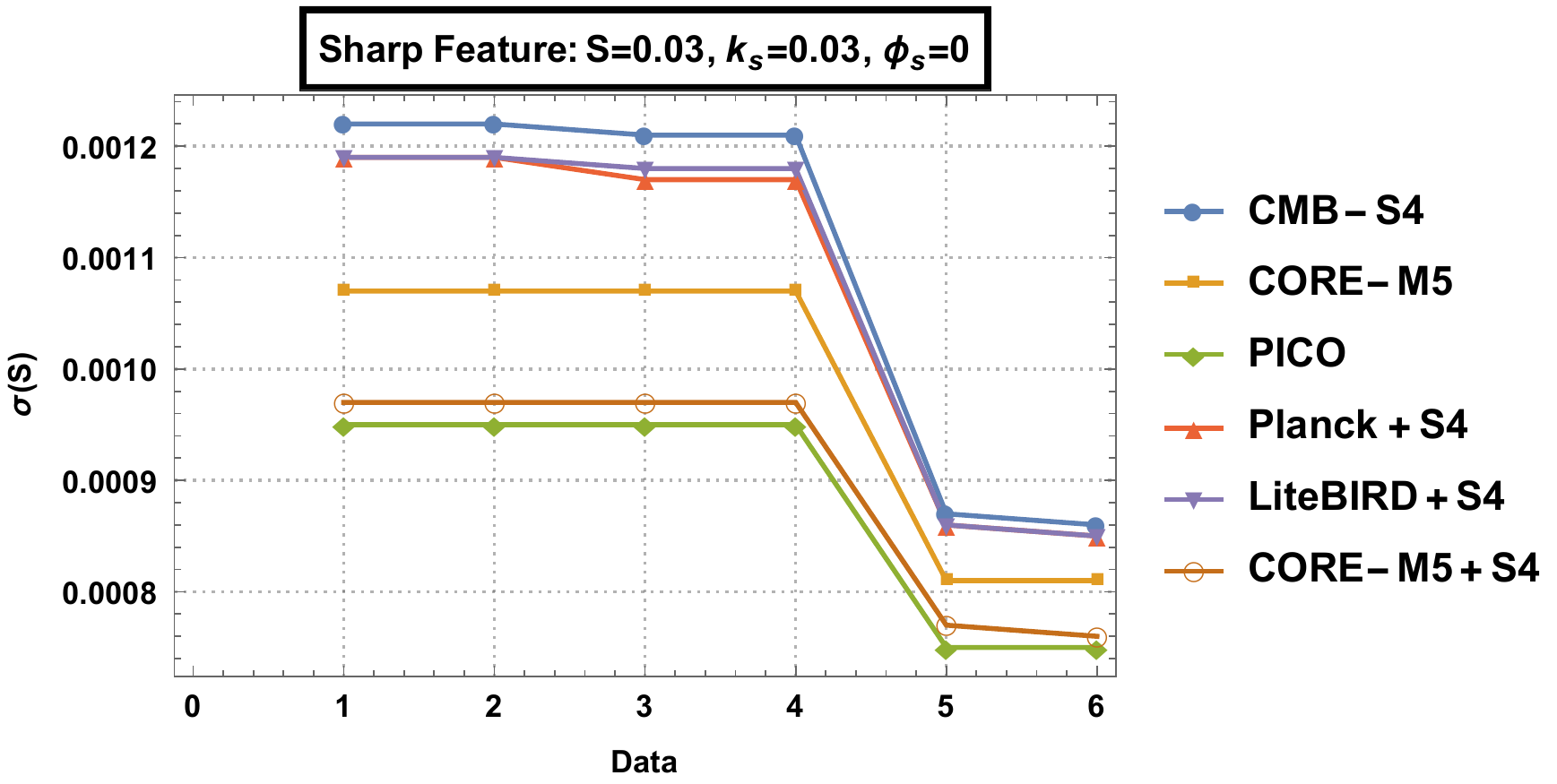} &
\includegraphics[width=0.5\textwidth]{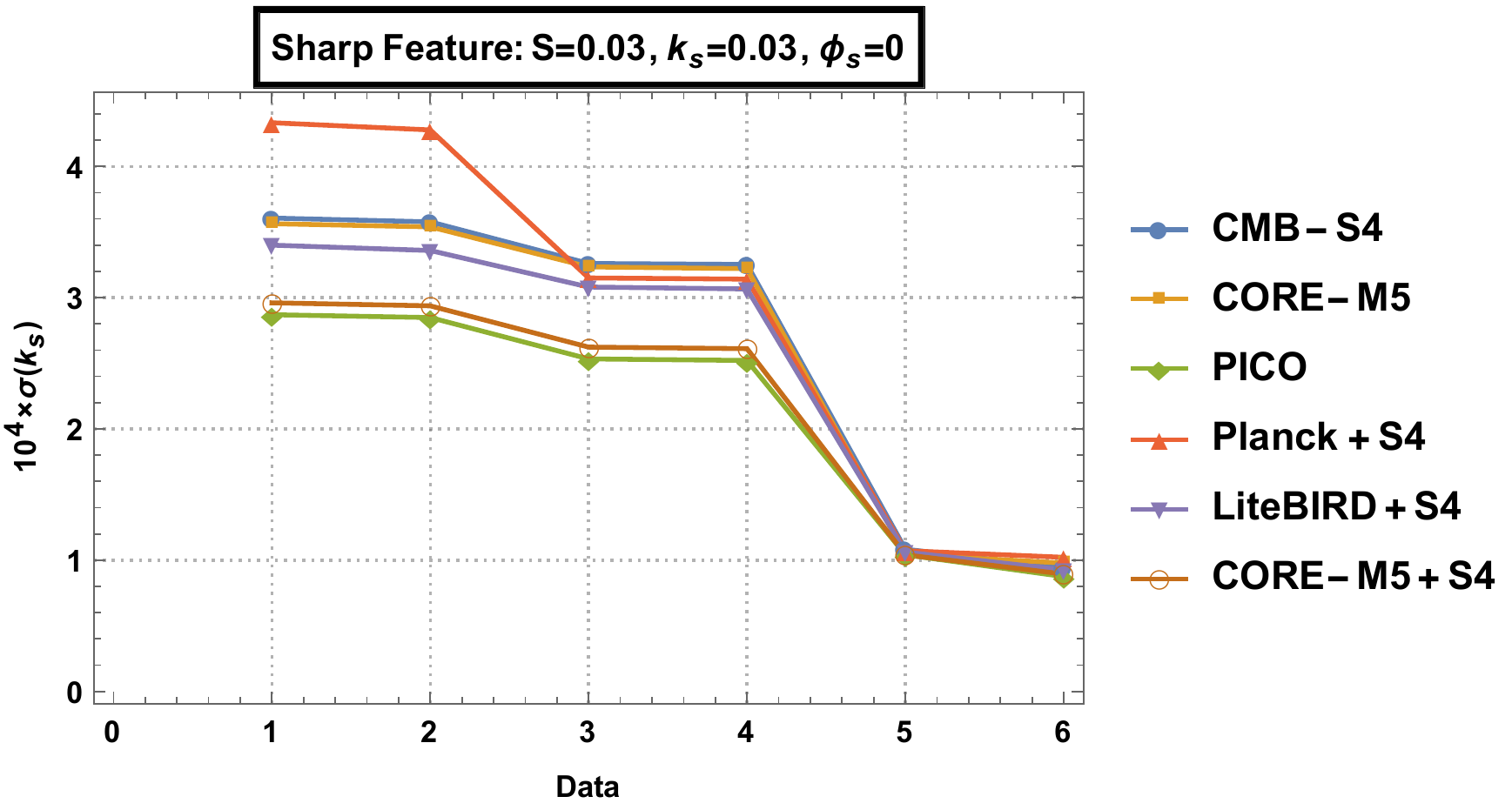}\\
\multicolumn{2}{c}{\includegraphics[width=0.5\textwidth]{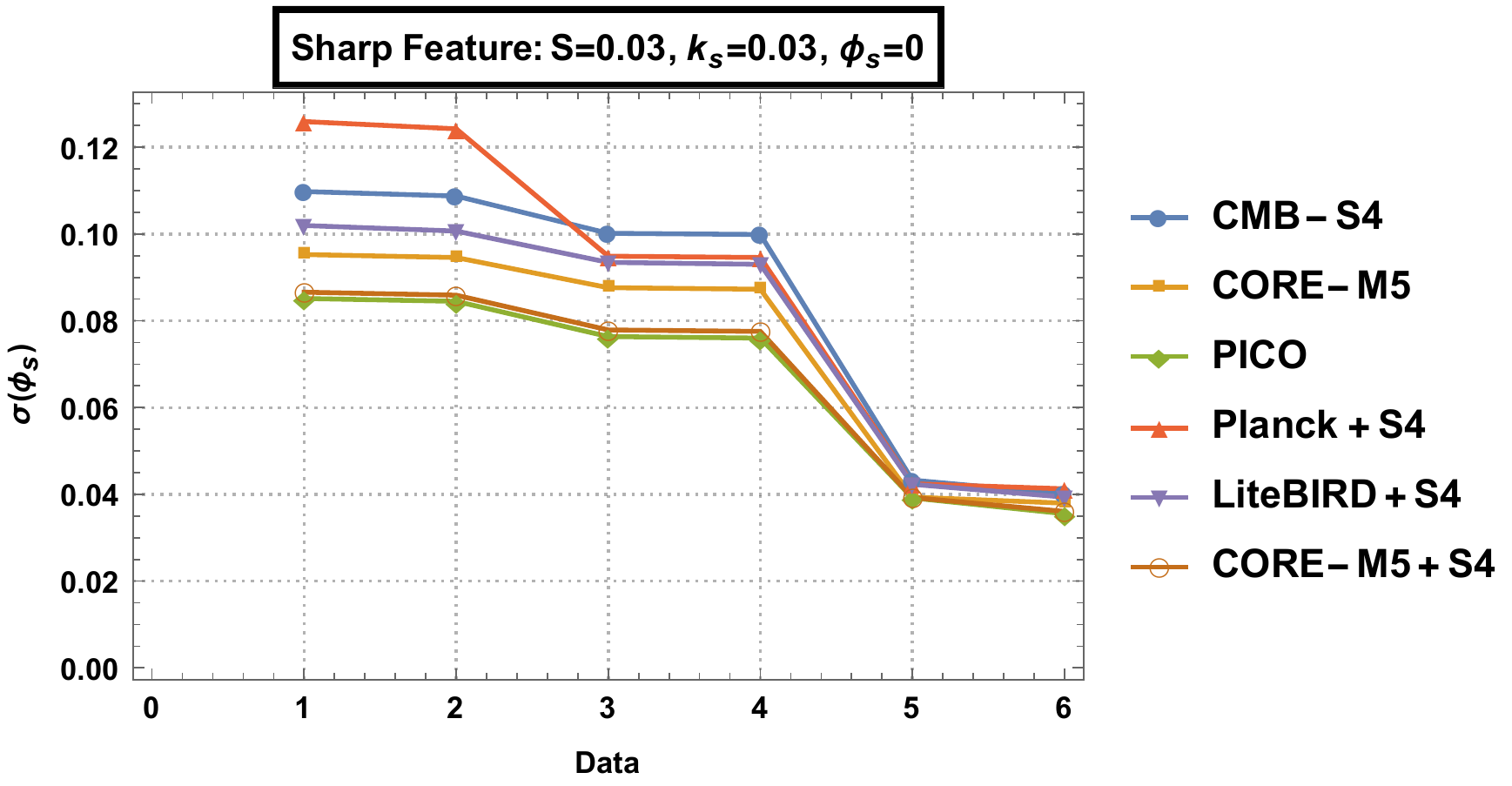}}
\end{array}$
\caption[]{\label{fig:Sharp-2}Experimental sensitivity of each CMB experiment for CMB-only~(1) and combination with DESI~(2), EUCLID-CS~(3), DESI+EUCLID-CS~(4), EUCLID-GC~(5) and EUCLID-(CS+GC)~(6) on the parameters of \textbf{Sharp Feature} template~(\ref{Template:Sharp}) has been depicted here. The 1-$\sigma$ uncertainties on the amplitude~(S), oscillation frequency~($k_s$) and phase~($\phi_s$) of the linear sinusoidal signal are shown in the \textit{Left} plot, \textit{Right} plot and \textit{Middle} plot, respectively. The plots are for these given fiducial values of the model parameters: S=0.03; $k_s$=0.03; $\phi_{s}$=0.}
\end{minipage}
\hfill
\end{mdframed}
\end{figure}
\subsubsection{Case II}
The next case of sharp feature, which is for the fiducial values  
S=0.03; $k_s$=0.03; $\phi_{s}$=0 has been depicted in Fig.~\ref{fig:Sharp-2}. From Fig.~\ref{fig:Sharp-2} the following inferences can be drawn:
\begin{itemize}[itemsep=-.3em]
\item[•] In this scenario for characteristic scale $k_s$=0.03, the weakest constraint on the oscillation amplitude~(S) is coming from CMB-S4, and PICO is improving the result the most and giving rise to the lowest error.   
\item[•] Oscillation amplitude~(S) is insensitive to BAO and CS data. For CMB-S4, Planck + CMB-S4 and LiteBIRD + CMB-S4, inclusion of CS data is showing slight variation in the results in comparison to CMB-only results.
\item[•] The GC data and GC+CS data further improve the bounds on combining with CMB data.
\item[•] LiteBIRD + CMB-S4 and Planck + CMB-S4 follow each other, where CMB-S4 alone catches them when combined with GC and GC+CS data. PICO and CORE-M5 + CMB-S4 have similar constraining ability with regard to oscillation amplitude~(S).
\item[•] In the context of characteristic scale~($k_s$) bounds are between Planck + CMB-S4 and PICO, where the lowest error is coming from PICO. 
\item[•] CMB-S4 and CORE-M5 are showing similar sensitivity for characteristic scale~($k_s$) at the fiducial value $k_s$=0.03, where their combination meets the sensitivity of PICO. 
\item[•] Bounds on the characteristic scale~($k_s$) are not affected by BAO data but show slight improvement on using CS data. For Planck + CMB-S4, the improvement is maximum.
\item[•] Here as well, the introduction of GC data is improving the constraints, and further inclusion of CS data along with GC data is slightly improving the bounds coming from the CMB + GC combinations alone. The combination of CMB + EUCLID(GC+CS) reduces the errors almost by a factor of 3-4 depending upon different CMB experiments.
\item[•] For the phase angle~($ \phi_s $), the bounds are between Planck + CMB-S4 and PICO. PICO is giving the best bound. A negligible sensitivity is shown to DESI BAO experiment by phase factor but exhibits little improvement for EUCLID-CS; the best improvement is arising from Planck + CMB-S4 + EUCLID-CS case.
\item[•] For all CMB experiments, the best constraints are coming when it is combined with EUCLID-GC and EUCLID-(GC+CS), though they are providing almost the same bounds but the addition of EUCLID-CS with EUCLID-GS is improving the results by a tiny amount. Combination of GC data with CMB shrinks the bounds by a factor between 2-4 depending upon the CMB experiments.
\item[•] When CMB-S4 is added to CORE-M5, it reaches the sensitivity of PICO in constraining the phase angle.
\end{itemize}
\begin{figure}[h!]
\begin{mdframed}
\captionsetup{font=footnotesize}
\center
\begin{minipage}[b1]{1.0\textwidth}
$\begin{array}{rl}
\includegraphics[width=0.5\textwidth]{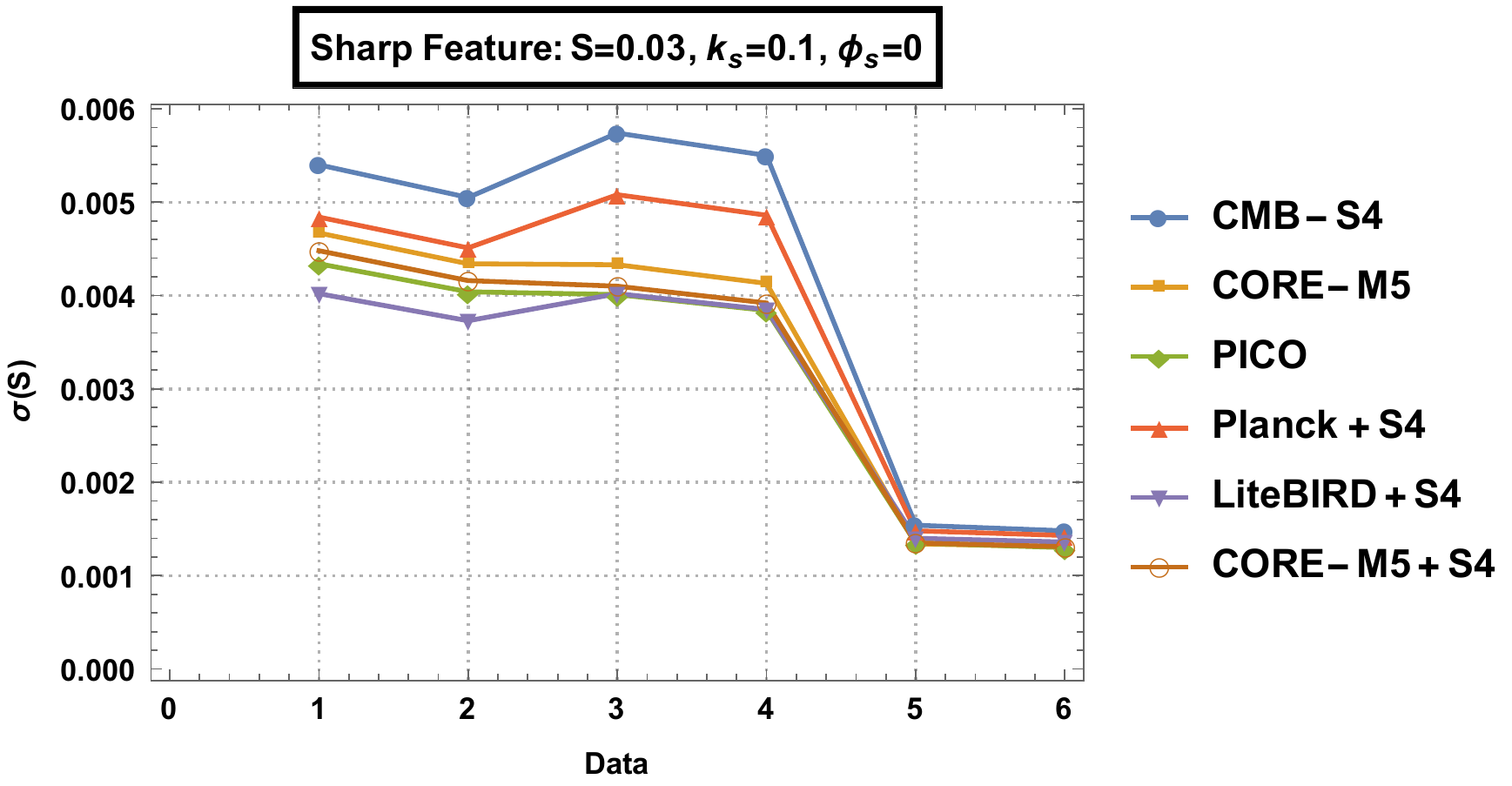} &
\includegraphics[width=0.5\textwidth]{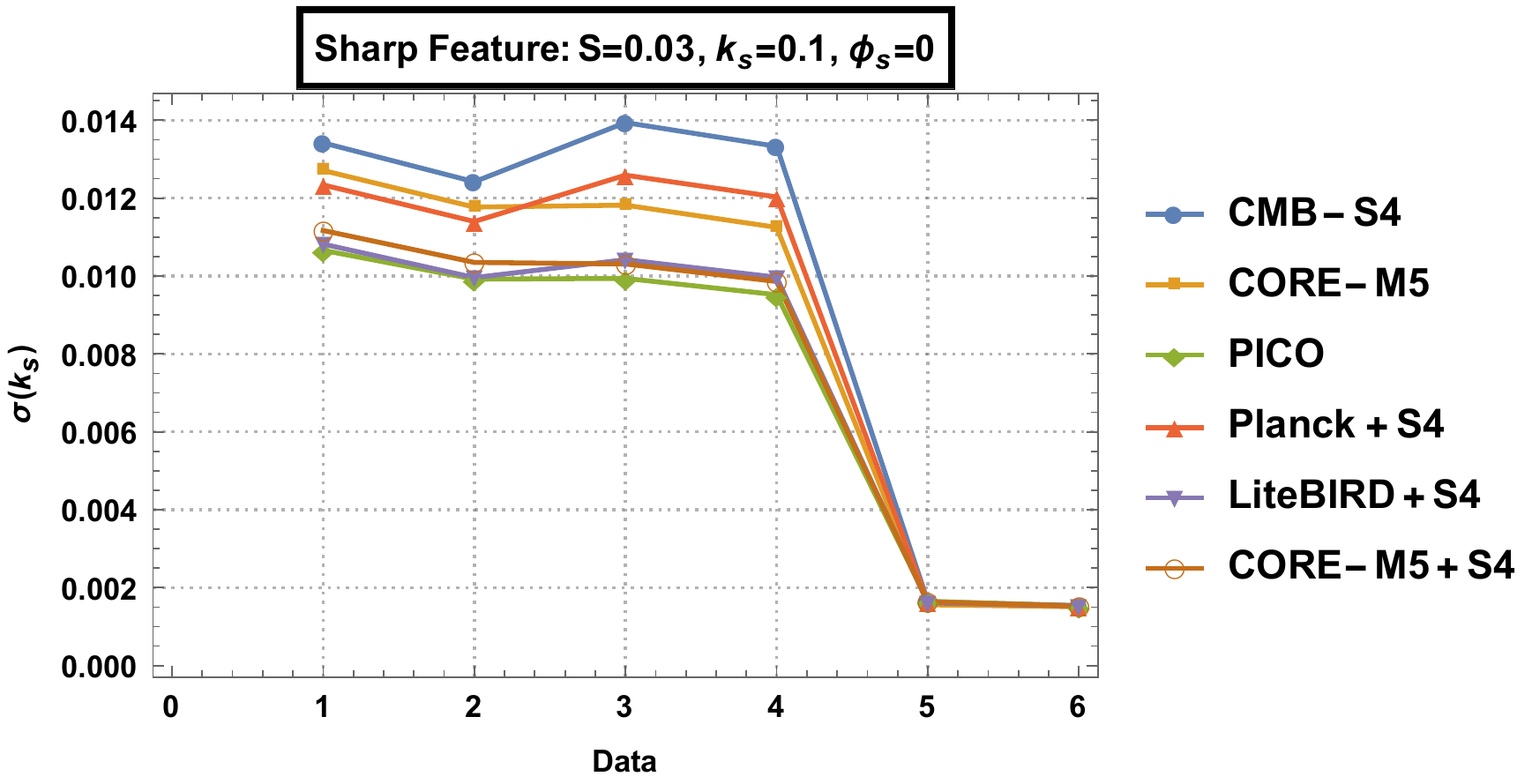}\\
\multicolumn{2}{c}{\includegraphics[width=0.5\textwidth]{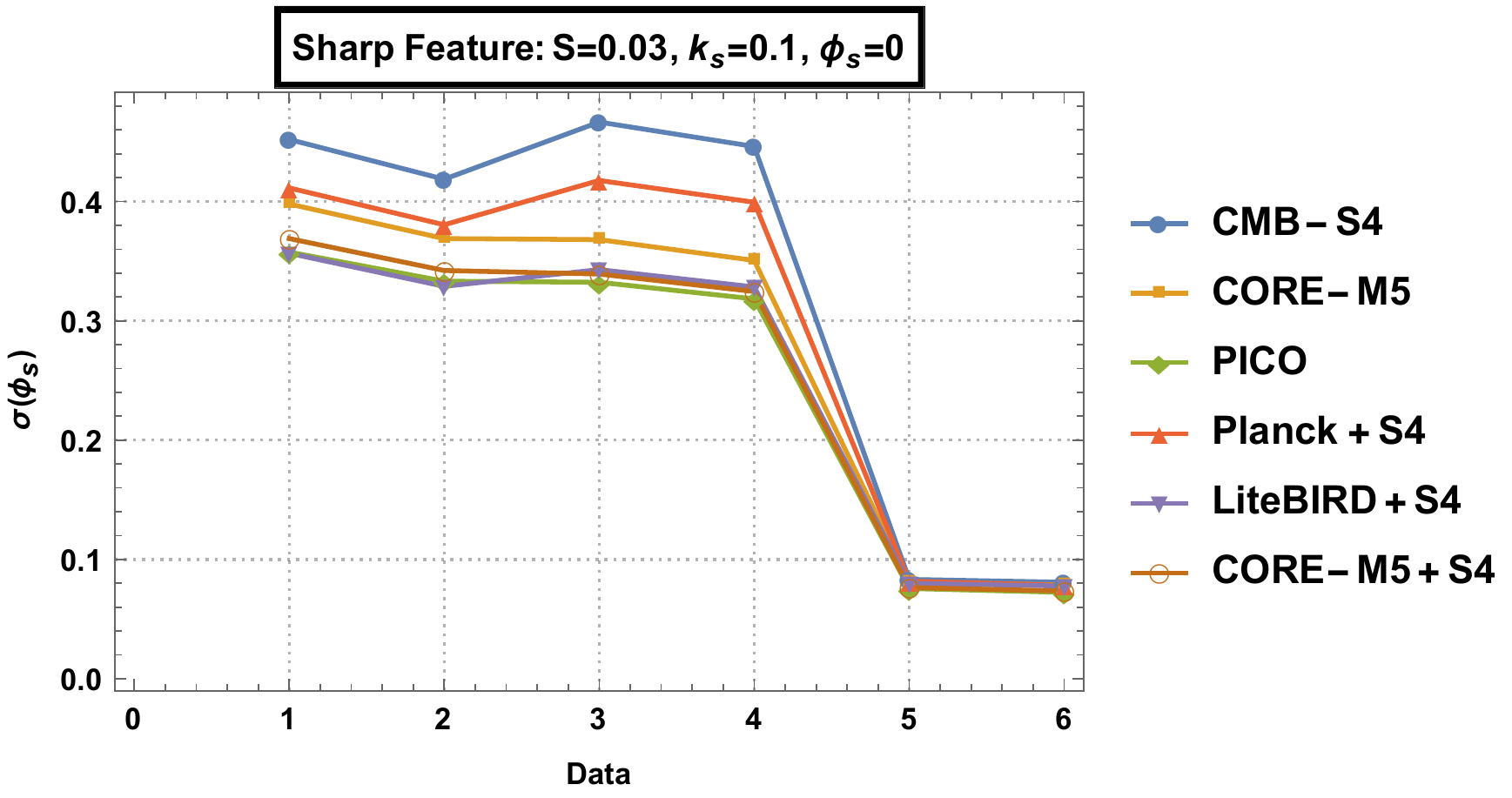}}
\end{array}$
\caption[]{\label{fig:Sharp-3}Experimental sensitivity of each CMB experiment for CMB-only~(1) and combination with DESI~(2), EUCLID-CS~(3), DESI+EUCLID-CS~(4), EUCLID-GC~(5) and EUCLID-(CS+GC)~(6) on the parameters of \textbf{Sharp Feature} template~(\ref{Template:Sharp}) has been depicted here. The 1-$\sigma$ uncertainties on the amplitude~(S), oscillation frequency~($k_s$) and phase~($\phi_s$) of the linear sinusoidal signal are shown in the \textit{Left} plot, \textit{Right} plot and \textit{Middle} plot, respectively. The plots are for these given fiducial values of the model parameters: S=0.03; $k_s$=0.1; $\phi_{s}$=0.}
\end{minipage}
\hfill
\end{mdframed}
\end{figure}
\subsubsection{Case III}
The third case of the sharp feature is for the fiducial values,  
S=0.03; $k_s$=0.1; $\phi_{s}$=0, which has been shown in Fig.~\ref{fig:Sharp-3}. From Fig.~\ref{fig:Sharp-3} we can infer the following results:
\begin{itemize}[itemsep=-.3em]
\item[•] If we consider the CMB experiments only, then for characteristic scale $k_s$=0.1, the best constraint on oscillation amplitude~(S) is coming from LiteBIRD + CMB-S4 and the weakest from CMB-S4.   
\item[•] Here, oscillation amplitude~(S) is showing sensitivity to BAO and CS data. For CMB-S4, Planck + CMB-S4 and LiteBIRD + CMB-S4, inclusion of CS data is worsening the results in comparison to CMB + DESI results. \item[•] Combination of GC and GC+CS data with CMB improves the bounds to a great extent, and provides almost similar bounds for all CMB experiments; however, CMB + (GC+CS) combination improves the results compared to CMB + GC combination.  
\item[•] PICO and CORE-M5 + CMB-S4 have similar constraining capacity for oscillation amplitude~(S), when CS data is added then constraining capability of LiteBIRD + CMB-S4 also matches with PICO and CORE-M5 + CMB-S4.
\item[•] In the context of characteristic scale~($k_s$) bounds are between CMB-S4 and PICO, where the strongest bound is coming from PICO. 
\item[•] CORE-M5+CMB-S4, PICO and LiteBIRD+CMB-S4 are showing nearly similar sensitivity for the characteristic scale~($k_s$) at the fiducial value $k_s$=0.1. 
\item[•] Bounds on characteristic scale~($k_s$) are affected by BAO data, showing little improvement in the results in compared to CMB-alone scenario. CMB + CS and CMB + DESI are behaving closely, except CMB-S4 and Planck + CMB-S4.
\item[•] Like previous cases, here also, the introduction of GC data is improving the constraints greatly, and the addition of CS data along with GC data is further improving the bounds, although by a marginal amount. The combination of CMB + EUCLID(GC+CS) reduces the errors almost by a factor of 7-8 depending upon different CMB experiments.
\item[•] For the phase angle~($ \phi_s $) the bounds are almost equal for CORE-M5 + CMB-S4, LiteBIRD + CMB-S4 and PICO.
\item[•] For all CMB experiments, the best constraints are coming when it is combined with EUCLID-GC and EUCLID-(GC+CS); though they are giving almost equal bounds; still, the addition of EUCLID-CS with EUCLID-GS is improving the results slightly. Combination of GC data with CMB improves the bounds by a factor of 5 approximately.
\end{itemize}
\begin{figure}[h!]
\begin{mdframed}
\captionsetup{font=footnotesize}
\center
\begin{minipage}[b1]{1.0\textwidth}
$\begin{array}{rl}
\includegraphics[width=0.5\textwidth]{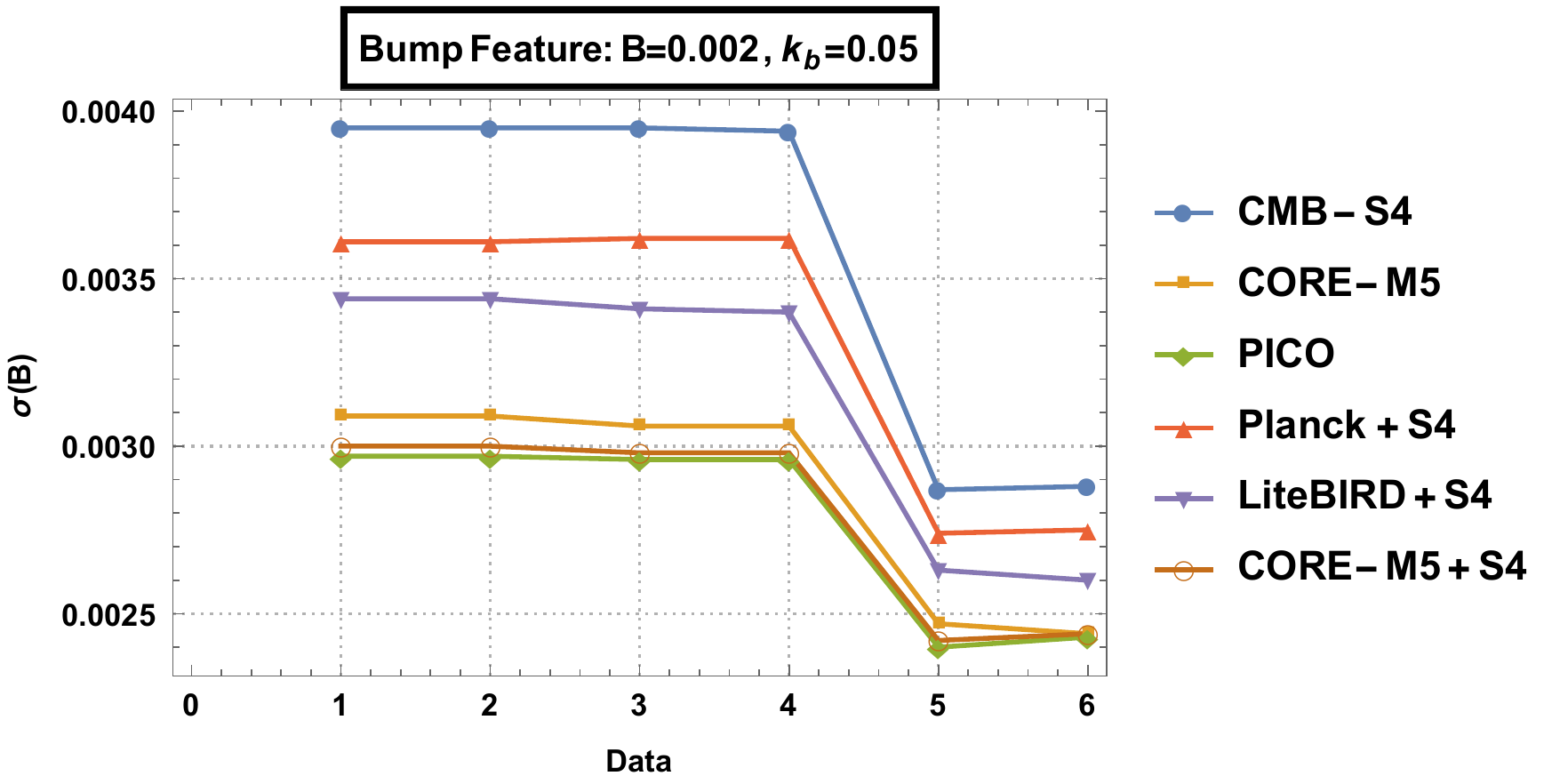} &
\includegraphics[width=0.5\textwidth]{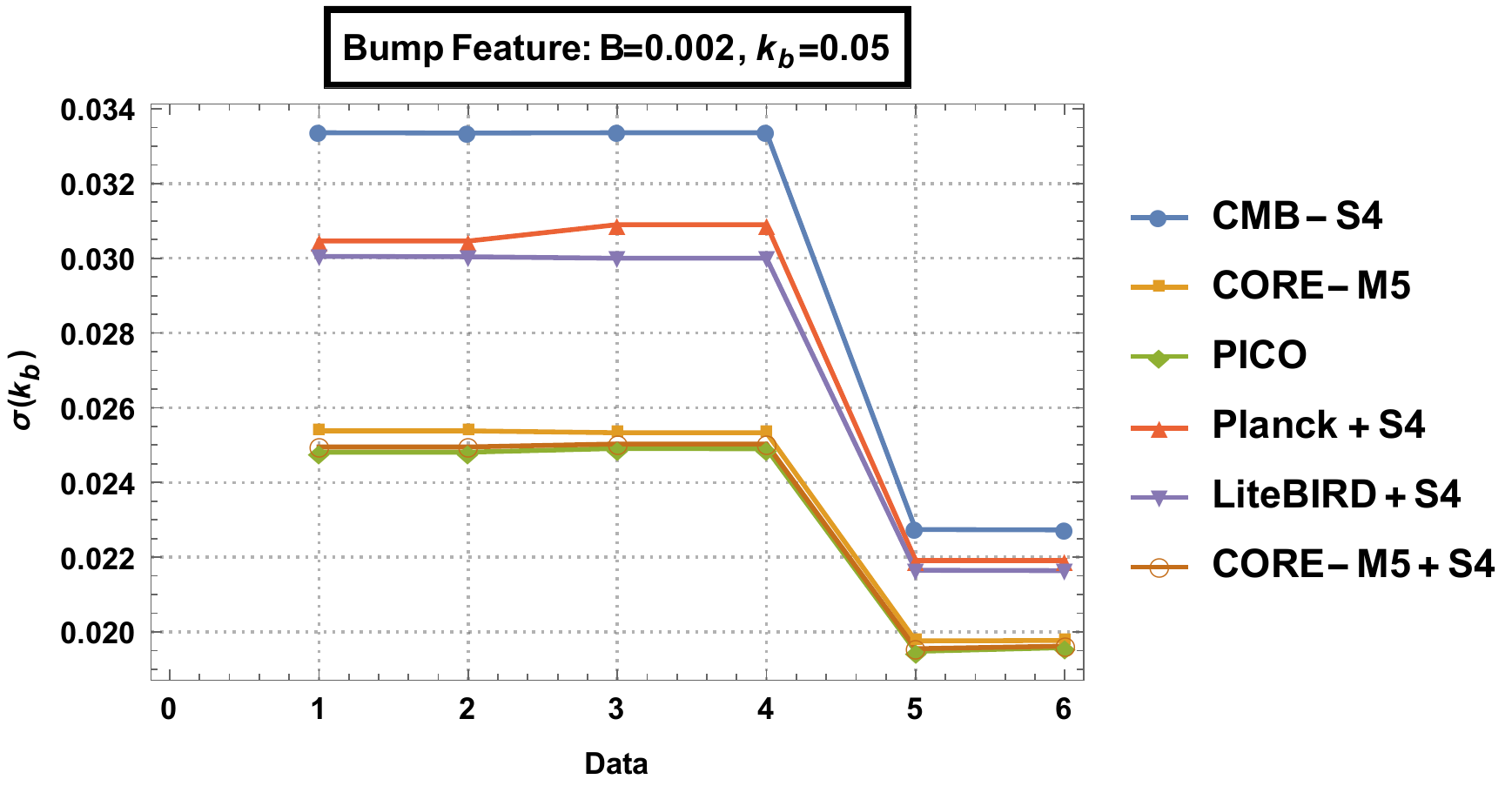}\\
\end{array}$
\caption[]{\label{fig:Bump-1}Experimental sensitivity of each CMB experiment for CMB-only~(1) and combination with DESI~(2), EUCLID-CS~(3), DESI+EUCLID-CS~(4), EUCLID-GC~(5) and EUCLID-(CS+GC)~(6) on the parameters of \textbf{Bump Feature} template~(\ref{Template:Bump}) has been depicted here. The 1-$\sigma$ uncertainties on the amplitude~(B) and the position of the Bump in k-space~($k_b$) of the Bump feature are shown in the \textit{Left} plot and \textit{Right} plot, respectively. The plots are for these given fiducial values of the model parameters: B=0.002; $k_b$=0.05.}
\end{minipage}
\hfill
\end{mdframed}
\end{figure}

\subsection{Bump Feature}
Let us now proceed to analyzing the results obtained for bump features. We will have three separate scenarios for three different sets of fiducial values of the bump feature model parameters, just like in the prior case. 
\subsubsection{Case I}
First, let us consider the  fiducial values  
B=0.002; $k_b$=0.05 that is shown in Fig.~\ref{fig:Bump-1}. From Fig.~\ref{fig:Bump-1} we can draw the following conclusions:
\begin{itemize}[itemsep=-.3em]
\item[•] For the characteristic scale $k_b$=0.05, the strongest bound on the amplitude~(B) of bump feature is coming from PICO, and the weakest bound from CMB-S4, for the CMB-only case.   
\item[•] DESI, EUCLID-CS and their combinations are not significantly improving the results in comparison to bounds for CMB-only surveys; this is the same for all the CMB experiments. 
\item[•] Combination of GC and GC+CS data with CMB data improves the bounds on the bump amplitude~(B).  
\item[•] PICO and CORE-M5 + CMB-S4 have similar constraining capacity for bump amplitude~(B), when GC+CS data is added then constraining capability of CORE-M5 also matches with PICO and CORE-M5 + CMB-S4 data.
\item[•] In the context of characteristic scale~($k_b$) bounds are between CMB-S4 and PICO, if only CMB experiments are considered, where the strongest bound is coming from PICO. 
\item[•] CORE-M5+CMB-S4, PICO and CORE-M5 are showing almost similar sensitivity for the characteristic scale~($k_b$) at the fiducial value $k_b$=0.05. 
\item[•] Bounds on the characteristic scale~($k_b$) are not influenced by BAO and CS data; but for Planck + CMB-S4 data, the addition of CS data for both the cases Planck + CMB-S4 + CS and Planck + CMB-S4 + BAO + CS, is making the results worse compared to the scenarios without CS. 
\item[•] For all CMB experiments the best constraints are coming when it is combined with EUCLID-GC and EUCLID-(GC+CS). However, addition of CS data with CMB + GC combination is not improving the bounds further.
\end{itemize}
\begin{figure}[h!]
\begin{mdframed}
\captionsetup{font=footnotesize}
\center
\begin{minipage}[b1]{1.0\textwidth}
$\begin{array}{rl}
\includegraphics[width=0.5\textwidth]{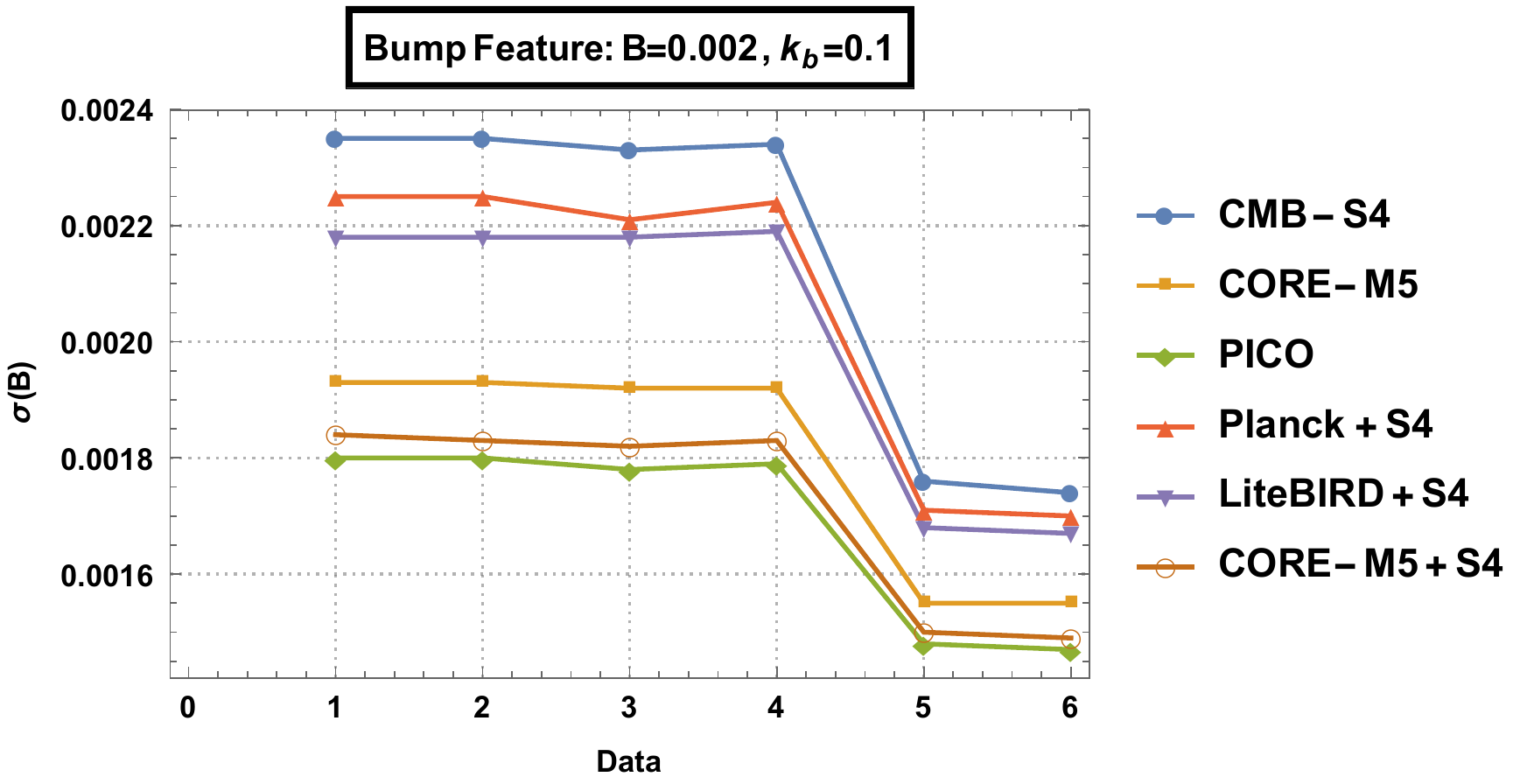} &
\includegraphics[width=0.5\textwidth]{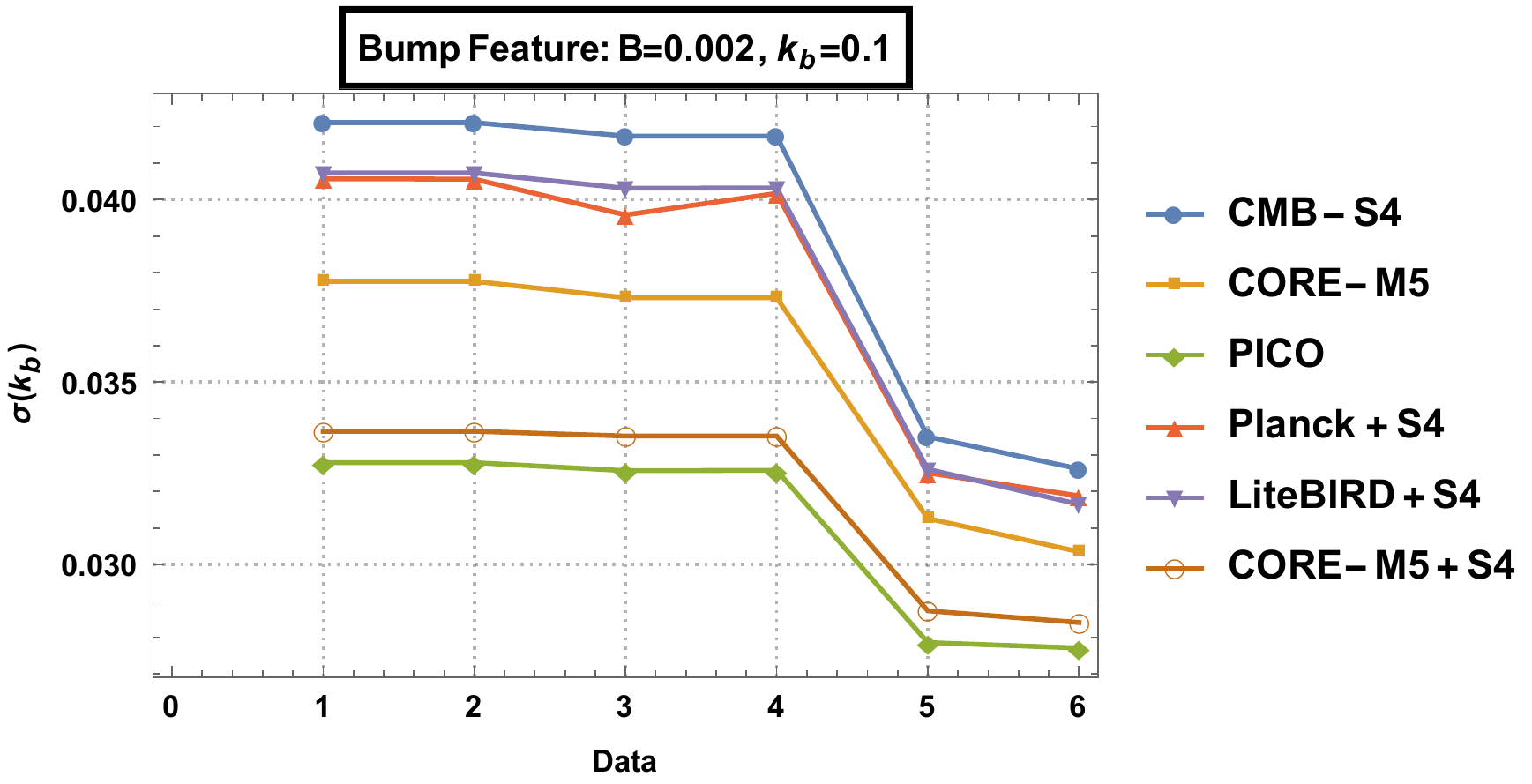}\\
\end{array}$
\caption[]{\label{fig:Bump-2}Experimental sensitivity of each CMB experiment for CMB-only~(1) and combination with DESI~(2), EUCLID-CS~(3), DESI+EUCLID-CS~(4), EUCLID-GC~(5) and EUCLID-(CS+GC)~(6) on the parameters of \textbf{Bump Feature} template~(\ref{Template:Bump}) has been depicted here. The 1-$\sigma$ uncertainties on the amplitude~(B) and the position of the Bump in k-space~($k_b$) of the Bump feature are shown in the \textit{Left} plot and \textit{Right} plot, respectively. The plots are for these given fiducial values of the model parameters: B=0.002; $k_b$=0.1.}
\end{minipage}
\hfill
\end{mdframed}
\end{figure}
\subsubsection{Case II}
The second scenario of bump feature model arises for the fiducial values,  
B=0.002; $k_b$=0.1. The results for this characteristic scale $k_b$=0.1, are summarized in Fig.~\ref{fig:Bump-2}. Here is the analysis of the results from Fig.~\ref{fig:Bump-2}:
\begin{itemize}[itemsep=-.3em]
\item[•] The best bound on the amplitude~(B) of bump feature for these fiducial values is coming from PICO, and the weakest bound from CMB-S4, for the CMB-only data.   
\item[•] DESI, EUCLID-CS and their combination are not altering the bounds in comparison to CMB-only constraints; this is the same for all the CMB experiments. Slight improvement can be seen for Planck+CMB-S4 data when CS data is added.  
\item[•] Combination of GC and GC+CS data with CMB data improves the bounds on the bump amplitude~(B). 
\item[•] In the context of characteristic scale~($k_b$) bounds are between CMB-S4 and PICO, if only CMB experiments are considered, where the strongest bound is coming from PICO. 
\item[•] Bounds on the characteristic scale~($k_b$) are not significantly sensitive towards BAO and CS data; but for Planck + CMB-S4 data, the addition of CS data is improving the result a bit compared to CMB-only scenario. 
\item[•] Except Planck + CMB-S4 + CS case, Planck + CMB-S4 and LiteBIRD + CMB-S4 are showing similar constraining capacity towards characteristic scale~($k_b$).
\item[•] For all CMB experiments, the best constraints are coming when it is combined with EUCLID-(GC+CS) and the strongest bound on $k_b$ is coming from PICO + EUCLID-(GC+CS) data sets.  
\end{itemize}
\begin{figure}[h!]
\begin{mdframed}
\captionsetup{font=footnotesize}
\center
\begin{minipage}[b1]{1.0\textwidth}
$\begin{array}{rl}
\includegraphics[width=0.5\textwidth]{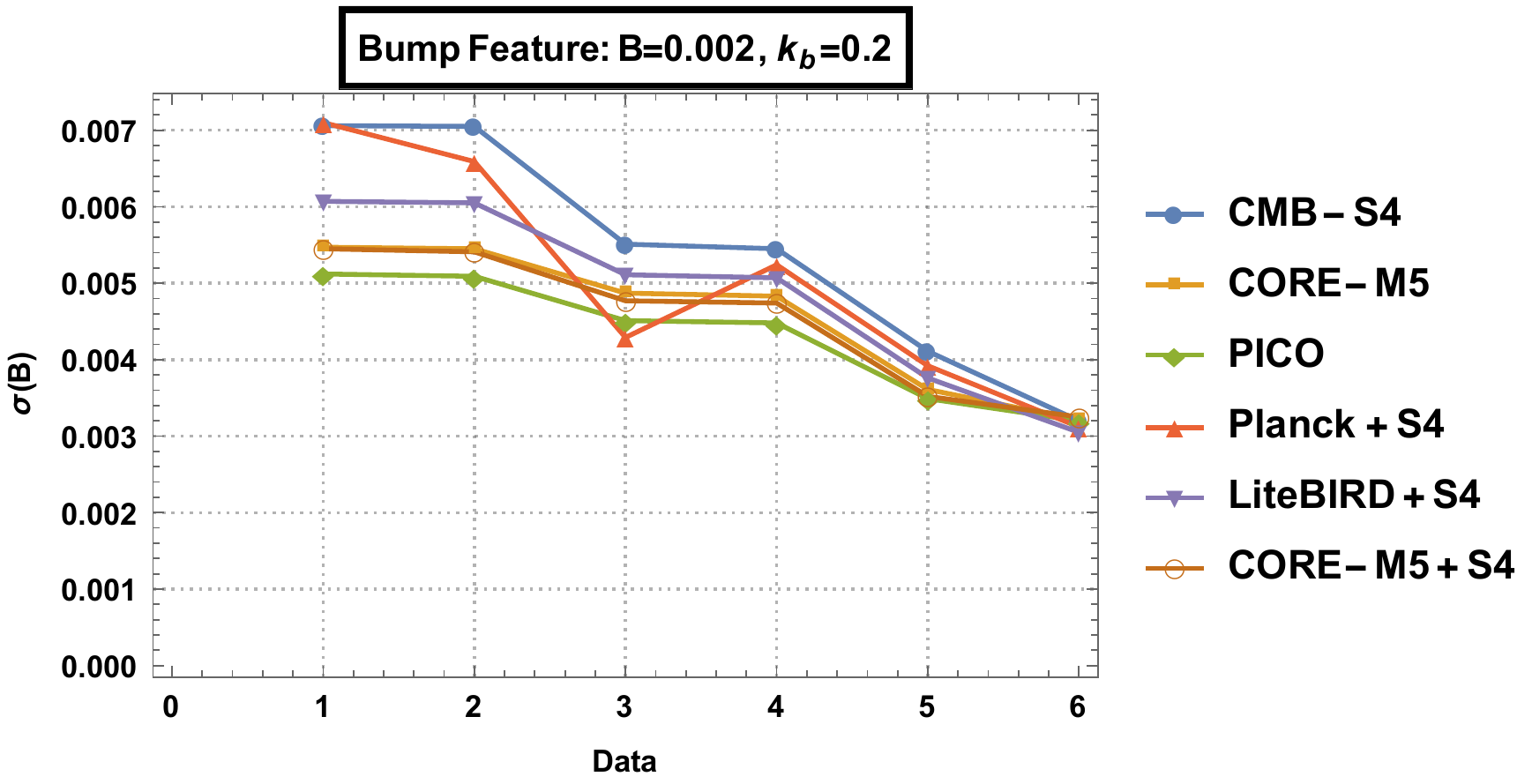} &
\includegraphics[width=0.5\textwidth]{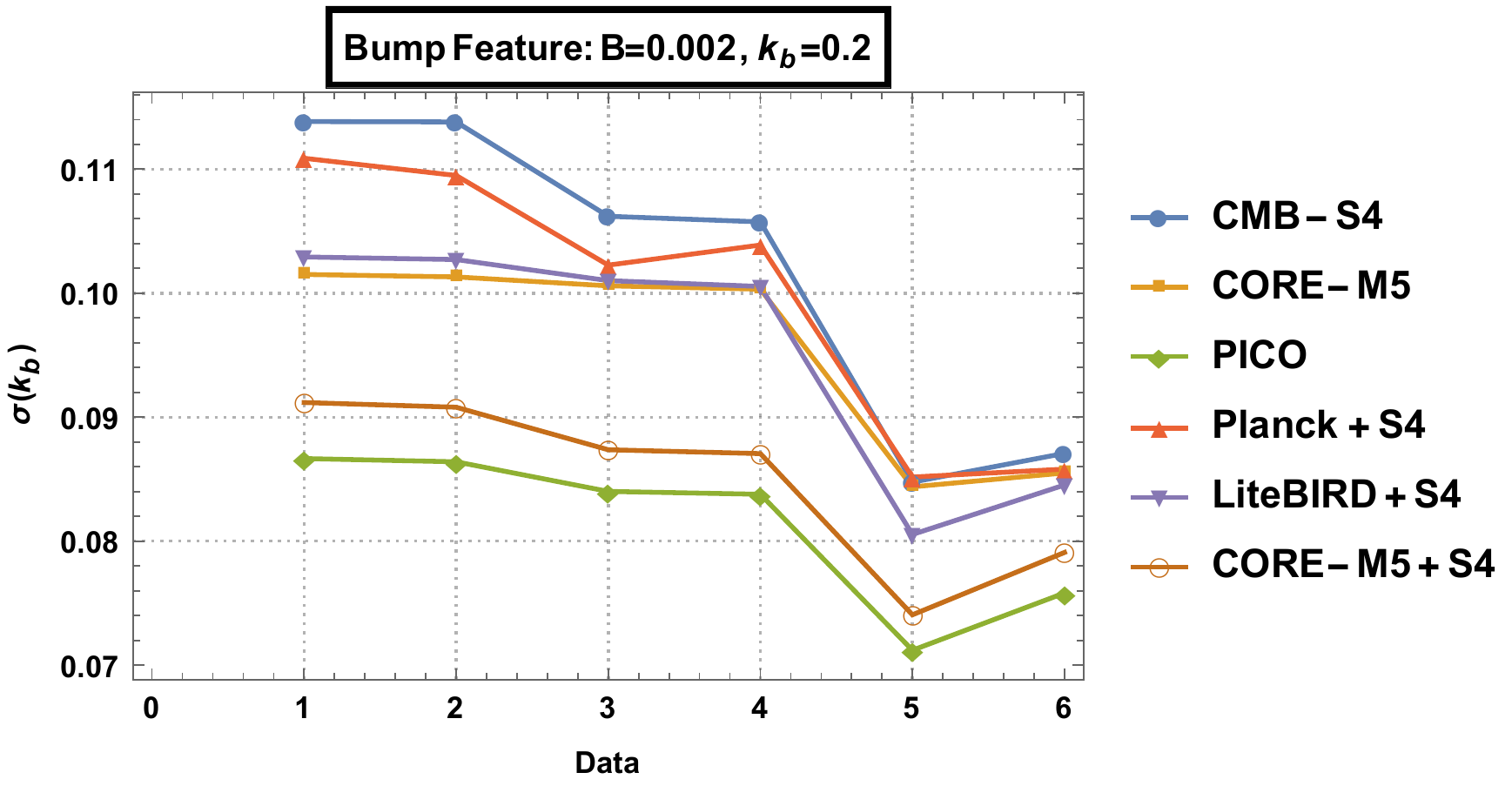}\\
\end{array}$
\caption[]{\label{fig:Bump-3}Experimental sensitivity of each CMB experiment for CMB-only~(1) and combination with DESI~(2), EUCLID-CS~(3), DESI+EUCLID-CS~(4), EUCLID-GC~(5) and EUCLID-(CS+GC)~(6) on the parameters of \textbf{Bump Feature} template~(\ref{Template:Bump}) has been depicted here. The 1-$\sigma$ uncertainties on the amplitude~(B) and the position of the Bump in k-space~($k_b$) of the Bump feature are shown in the \textit{Left} plot and \textit{Right} plot, respectively. The plots are for these given fiducial values of the model parameters: B=0.002; $k_b$=0.2.}
\end{minipage}
\hfill
\end{mdframed}
\end{figure}
\subsubsection{Case III}
The third scenario of bump feature model arises for the fiducial values,  
B=0.002; $k_b$=0.2. The results for this characteristic scale, $k_b$=0.2, are summarized in Fig.~\ref{fig:Bump-3}. Here is the analysis of the results from Fig.~\ref{fig:Bump-3}:
\begin{itemize}[itemsep=-.3em]
\item[•] The best bound on the amplitude~(B) of bump feature for fiducial values, B=0.002; $k_b$=0.2 is coming from PICO, and the weakest bound from CMB-S4 and Planck + CMB-S4, for the CMB-only case.   
\item[•] DESI is unable to improve the bounds, except for Planck + CMB-S4 case. EUCLID-CS is improving the constraints in comparison to CMB-only results; maximum improvement is coming for Planck + CMB-S4 data sets. 
\item[•] Combination of GC+CS data with CMB data improves the bounds on the bump amplitude~(B) nearly by a factor of 2. 
\item[•] In the context of characteristic scale~($k_b$) bounds are between CMB-S4 and PICO, if only CMB experiments are considered, where the strongest bound is coming from PICO. 
\item[•] Bounds on the characteristic scale~($k_b$) are not sensitive towards BAO data, except Planck + CMB-S4 case.
\item[•] Bounds on the characteristic scale~($k_b$) are showing varied sensitivity towards the CMB + CS combination. 
\item[•] For all CMB experiments, the best constraints are coming when it is combined with EUCLID-GC and the strongest bound on $k_b$ is coming from PICO + EUCLID-GC data sets. However, inclusion of CS data is weakening the bounds compared to CMB+GC results.   
\end{itemize}
\begin{figure}[h!]
\begin{mdframed}
\captionsetup{font=footnotesize}
\center
\begin{minipage}[b1]{1.0\textwidth}
$\begin{array}{rl}
\includegraphics[width=0.5\textwidth]{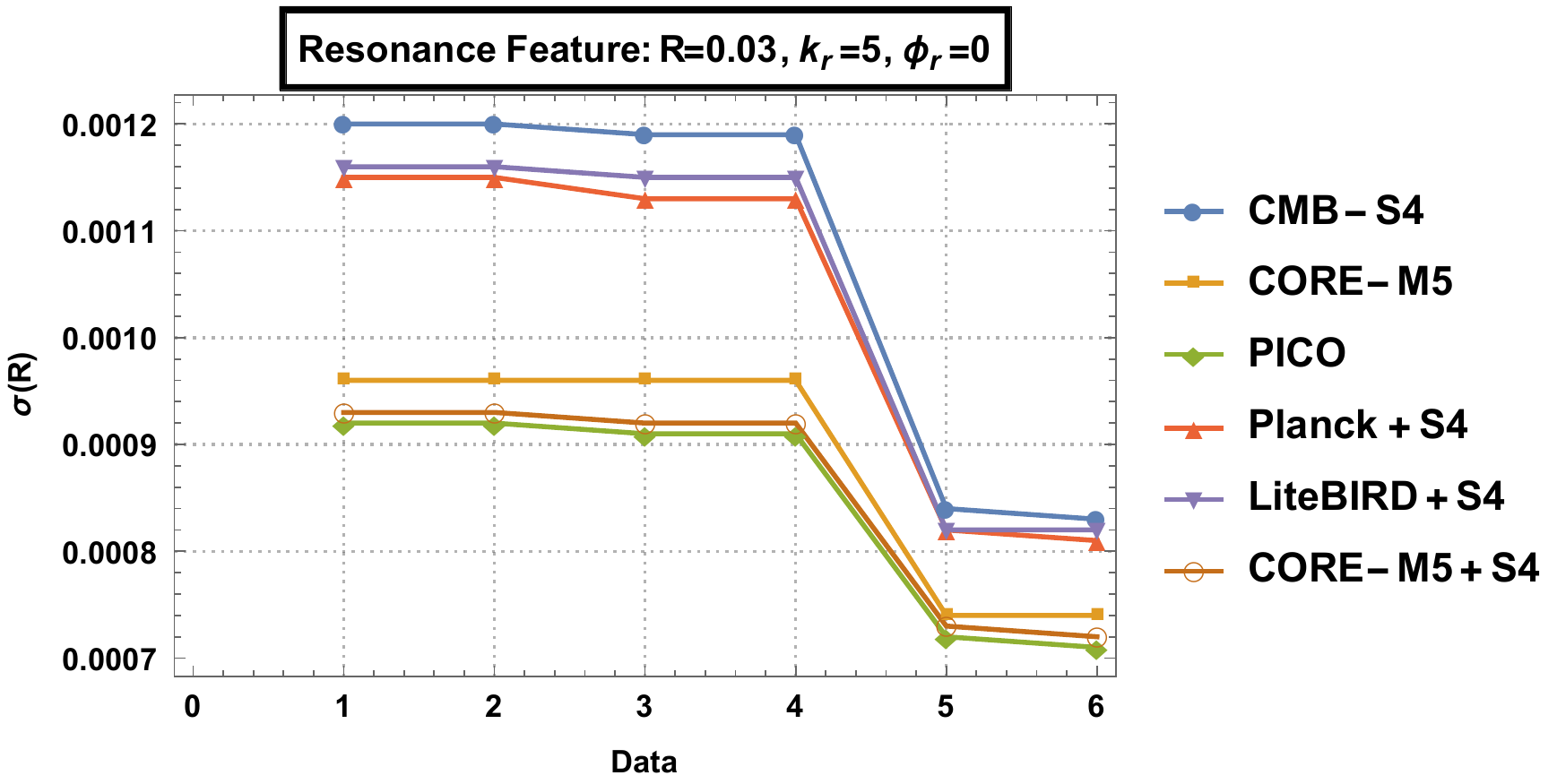} &
\includegraphics[width=0.5\textwidth]{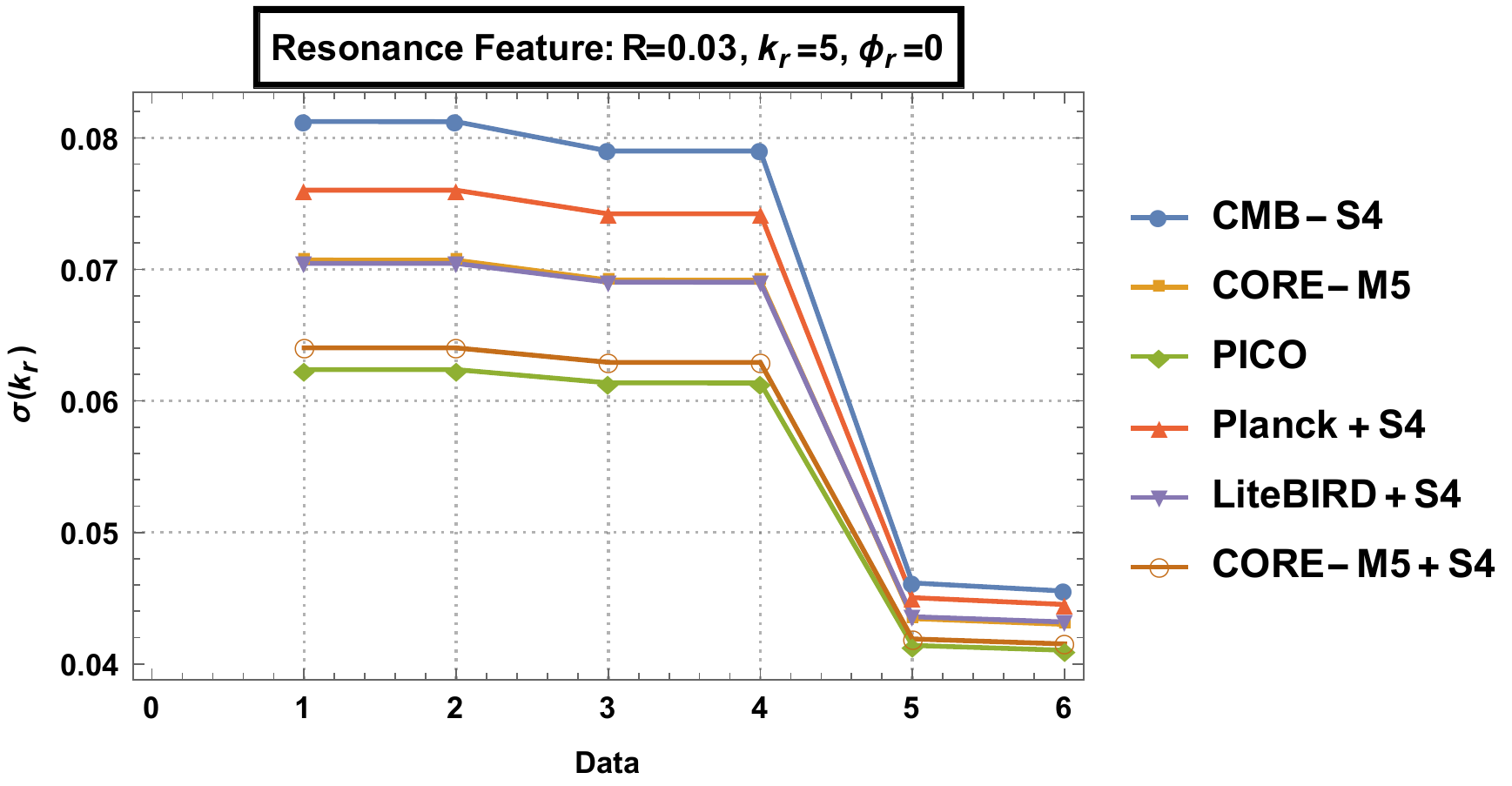}\\
\multicolumn{2}{c}{\includegraphics[width=0.5\textwidth]{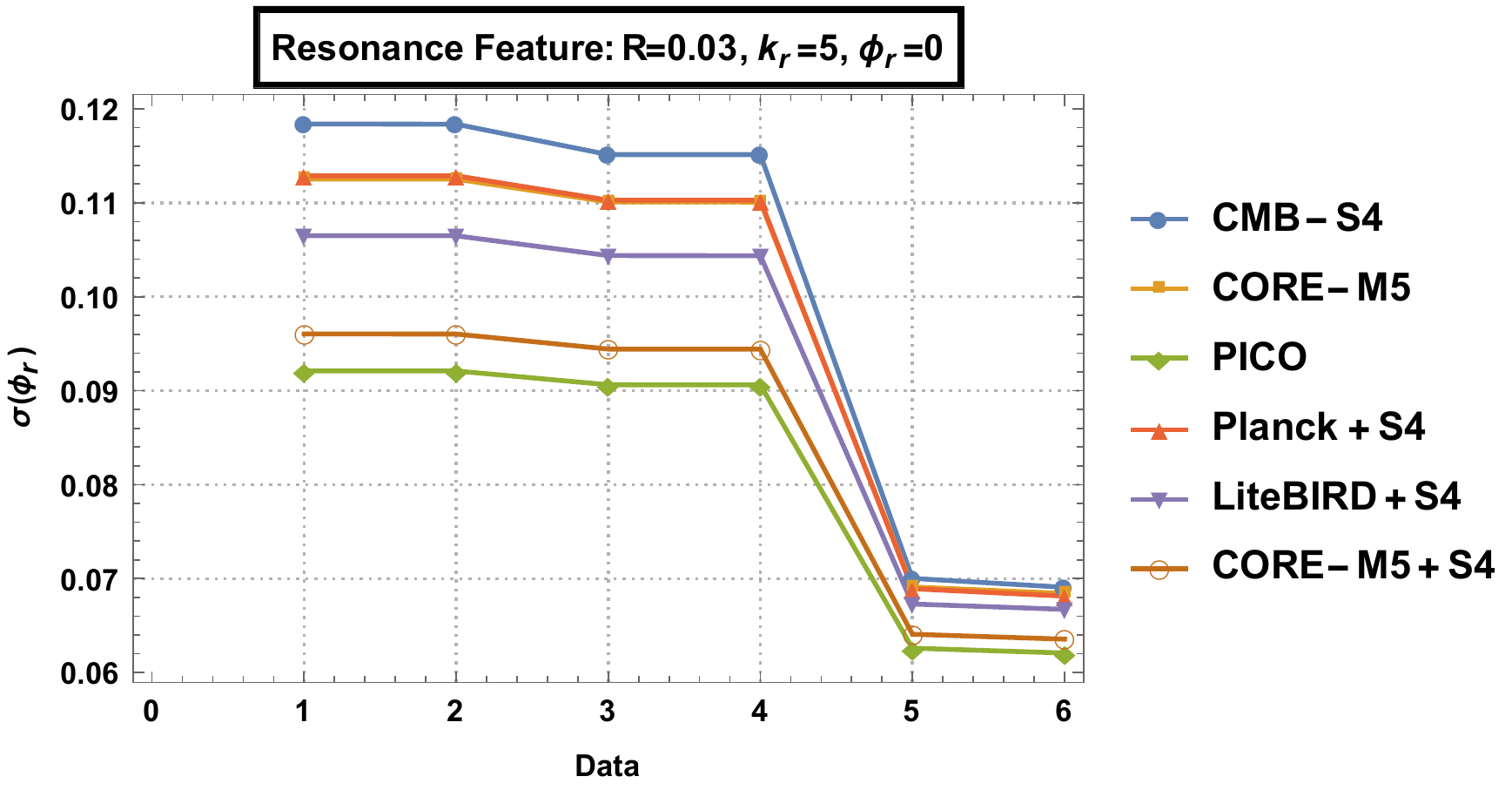}}
\end{array}$
\caption[]{\label{fig:Resonance-1}Experimental sensitivity of each CMB experiment for CMB-only~(1) and combination with DESI~(2), EUCLID-CS~(3), DESI+EUCLID-CS~(4), EUCLID-GC~(5) and EUCLID-(CS+GC)~(6) on the parameters of \textbf{Resonance Feature} template~(\ref{Template:Resonance}) has been depicted here. The 1-$\sigma$ uncertainties on the amplitude~(R), oscillation frequency~($k_r$) and phase~($\phi_r$) of the logarithmic sinusoidal signal are shown in the \textit{Left} plot, \textit{Right} plot and \textit{Middle} plot, respectively. The plots are for these given fiducial values of the model parameters: R=0.03; $k_r$=5; $\phi_{r}$=0.}
\end{minipage}
\hfill
\end{mdframed}
\end{figure}

\subsection{Resonance Feature}
Let us now engage ourselves in analyzing the    results and discussion of the resonance feature model step-by-step for each individual case of fiducial values.

\subsubsection{Case I}
The first case of the logarithmic sinusoidal feature is for the fiducial values R=0.03; $k_r$=5; $\phi_{r}$=0. The results for this characteristic scale $k_r$=5 are graphically illustrated in Fig.~\ref{fig:Resonance-1}, from which the following inferences can be drawn:
\begin{itemize}[itemsep=-.3em]
\item[•] In this scenario for the characteristic scale $k_r$=5, the weakest constraint on the oscillation amplitude~(R) is coming from CMB-S4, and PICO is improving the result the most, which gives rise to the lowest error.   
\item[•] DESI, EUCLID-CS and their combination are not altering the bounds in comparison to CMB-only constraints; this is same for all the CMB experiments except Planck+CMB-S4; a very little improvement can be seen for Planck+CMB-S4 data when CS data is added.
\item[•] Combination of GC data and GC+CS data with CMB data improves the bounds for all CMB experiments; the best bound is coming from the PICO+GC+CS data combination. Both of these LSS data are giving almost the same bounds for a given CMB data.
\item[•] PICO and CORE-M5 + CMB-S4 have almost similar constraining ability for oscillation amplitude~(R).
\item[•] In the context of characteristic scale~($k_r$) bounds are between CMB-S4 and PICO, where the lowest error is coming from PICO. 
\item[•] LiteBIRD+CMB-S4 and CORE-M5 are showing nearly equal sensitivity for characteristic scale~($k_r$) at the fiducial value $k_r$=5. 
\item[•] Bounds on the characteristic scale~($k_r$) are not affected by BAO data but show slight improvement on using CS data.
\item[•] Here as well, the introduction of GC data is improving the constraints, and further inclusion of CS data along with GC data is slightly improving the bounds coming from the CMB + GC combinations alone.
\item[•] The bounds are between CMB-S4 and PICO for the phase angle~($ \phi_r $). PICO is giving the best constraint. No sensitivity is shown to the DESI BAO experiment by phase factor but shows little improvements for EUCLID-CS.
\item[•] For all CMB experiments, the best constraints on phase angle~($ \phi_r $) are coming when it is combined with EUCLID-GC and EUCLID-(GC+CS), though they are giving almost the same bounds but the addition of EUCLID-CS with EUCLID-GS is improving the results by a marginal amount only.
\item[•] For the phase angle~($ \phi_r $), when CMB-S4 is added with Planck, it reaches the sensitivity of CORE-M5 in constraining the phase angle.
\end{itemize}
\begin{figure}[h!]
\begin{mdframed}
\captionsetup{font=footnotesize}
\center
\begin{minipage}[b1]{1.0\textwidth}
$\begin{array}{rl}
\includegraphics[width=0.5\textwidth]{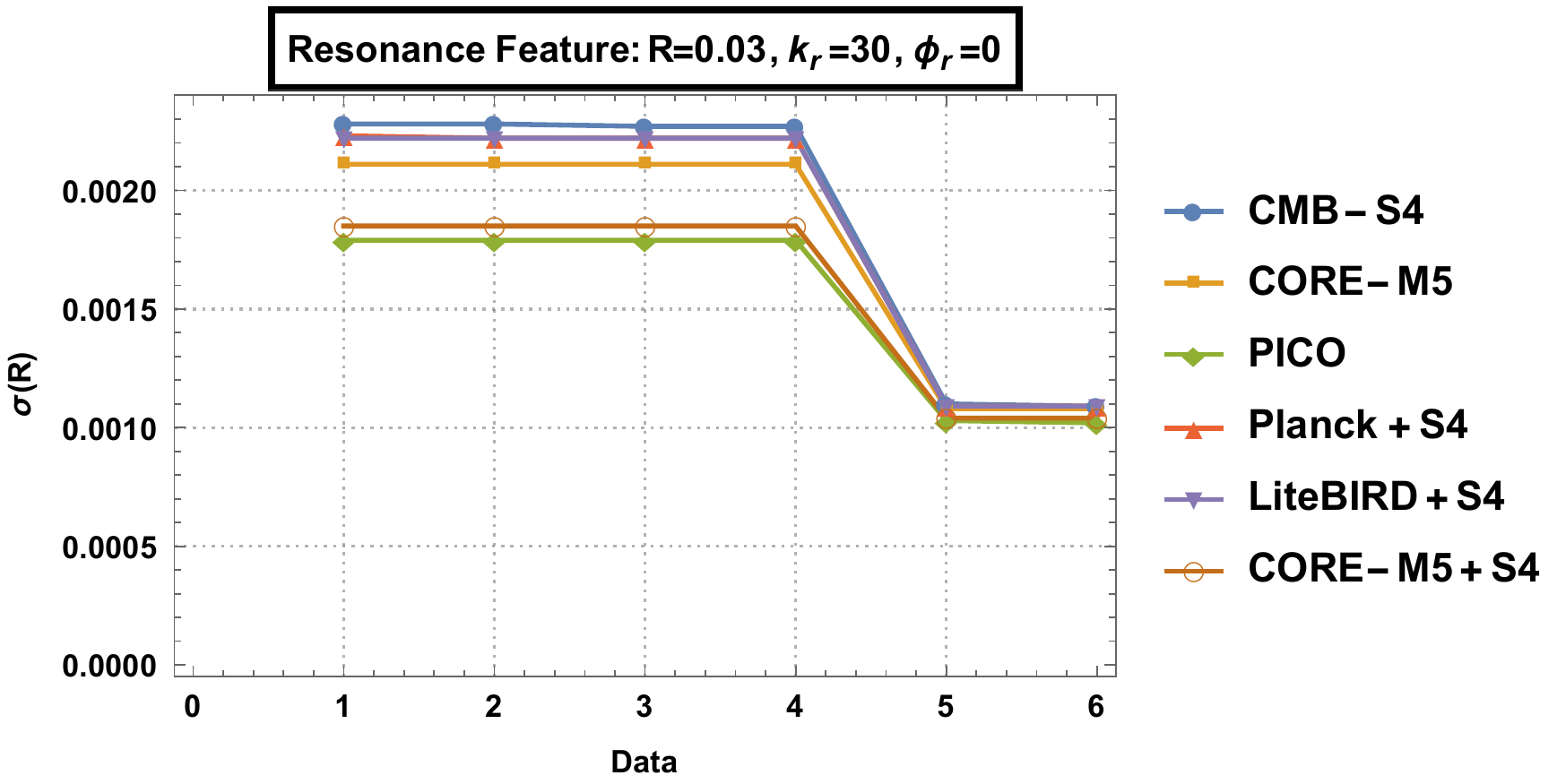} &
\includegraphics[width=0.5\textwidth]{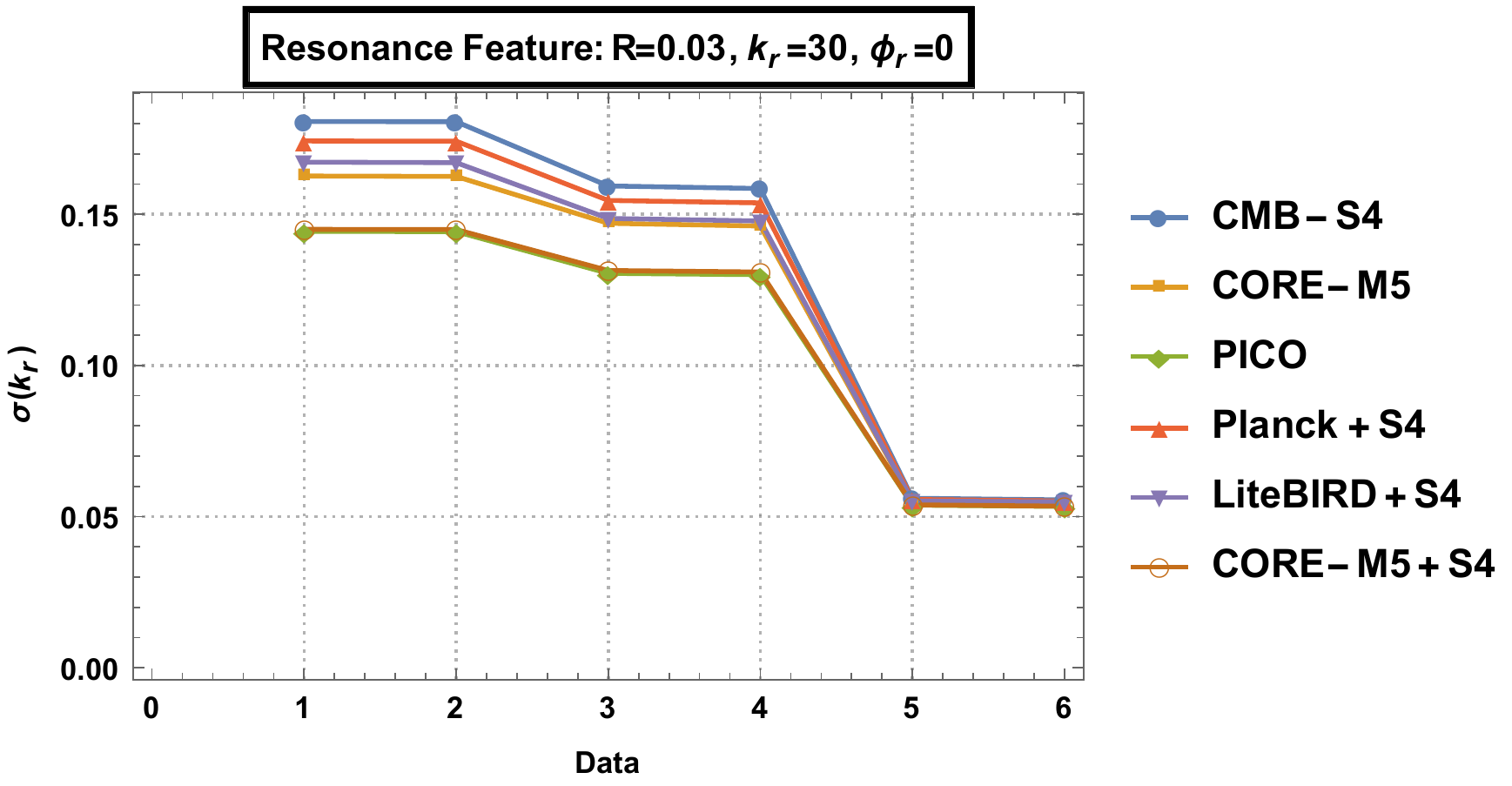}\\
\multicolumn{2}{c}{\includegraphics[width=0.5\textwidth]{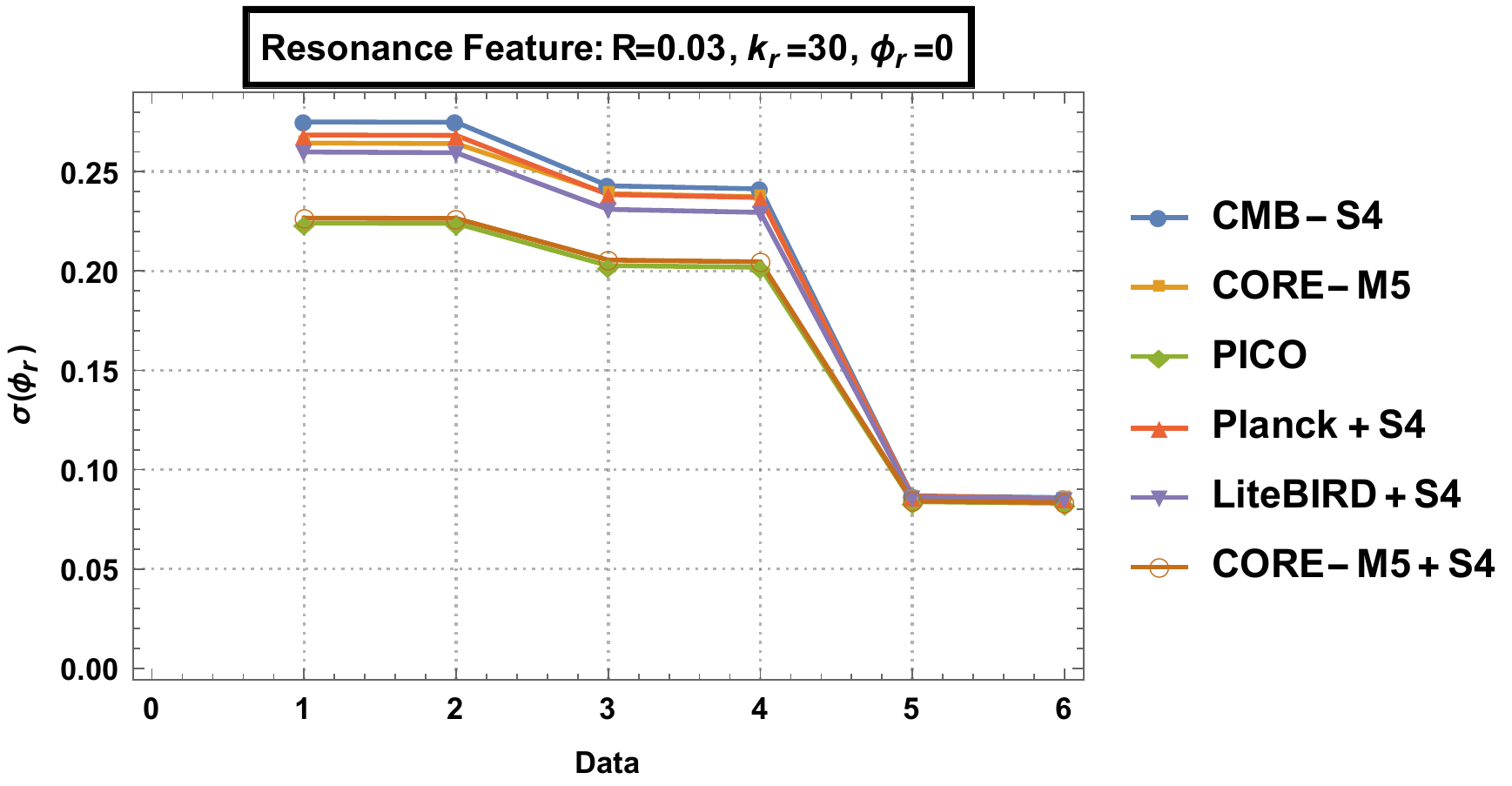}}
\end{array}$
\caption[]{\label{fig:Resonance-2}Experimental sensitivity of each CMB experiment for CMB-only~(1) and combination with DESI~(2), EUCLID-CS~(3), DESI+EUCLID-CS~(4), EUCLID-GC~(5) and EUCLID-(CS+GC)~(6) on the parameters of \textbf{Resonance Feature} template~(\ref{Template:Resonance}) has been depicted here. The 1-$\sigma$ uncertainties on the amplitude~(R), oscillation frequency~($k_r$) and phase~($\phi_r$) of the logarithmic sinusoidal signal are shown in the \textit{Left} plot, \textit{Right} plot and \textit{Middle} plot, respectively. The plots are for these given fiducial values of the model parameters: R=0.03; $k_r$=30; $\phi_{r}$=0.}
\end{minipage}
\hfill
\end{mdframed}
\end{figure}
\subsubsection{Case II}
The second case of the resonance feature signal is for the fiducial values R=0.03; $k_r$=30; $\phi_{r}$=0. The results for this characteristic scale $k_r$=30, are graphically expressed in Fig.~\ref{fig:Resonance-2}, from which the following inferences can be drawn:
\begin{itemize}[itemsep=-.3em]
\item[•] For this given fiducial values R=0.03; $k_r$=30; $\phi_{r}$=0, if CMB-only case is considered, then the weakest constraint on the oscillation amplitude~(R) comes from CMB-S4 and the best bound is given by PICO.    
\item[•] DESI, EUCLID-CS and their combination are not too sensitive towards the oscillation amplitude~(R); they do not improve the bounds in comparison to CMB-only constraints; this is same for all the CMB experiments.
\item[•] Combination of GC data and GC+CS data with CMB data improves the bounds for all CMB experiments compared to CMB-only results and they give almost the same bounds for all CMB experiments.
\item[•] PICO and CORE-M5 + CMB-S4 have almost similar constraining ability for oscillation amplitude~(R). On the other hand, CMB-S4, LiteBIRD + CMB-S4 and Planck + CMB-S4 have nearly equal sensitivity to oscillation amplitude~(R).
\item[•] In the context of characteristic scale~($k_r$) bounds are between CMB-S4 and PICO, where the lowest error is coming from PICO. 
\item[•] PICO and CORE-M5 + CMB-S4 are showing almost equal sensitivity for characteristic scale~($k_r$), at the fiducial value $k_r$=30. 
\item[•] Bounds on the characteristic scale~($k_r$) are not affected by BAO data but are showing slight improvement on using CS data.
\item[•] Here also, the introduction of GC data is improving the constraints, and further inclusion of CS data along with GC data is slightly improving the bounds coming from the CMB + GC combinations alone.
\item[•] For phase angle~($ \phi_r $), the bounds are between CMB-S4 and PICO; PICO is giving the best bounds. No sensitivity has been shown to the DESI BAO experiment by phase factor but showing little improvement for EUCLID-CS data.
\item[•] For all CMB experiments, the best constraints on phase angle~($ \phi_r $) are coming when it is combined with EUCLID-GC and EUCLID-(GC+CS), though they are giving almost the same bounds but the addition of EUCLID-CS with EUCLID-GS is improving the results by a small amount.
\item[•] PICO and CORE-M5 + CMB-S4 have almost similar constraining ability for phase angle~($ \phi_r $).
\end{itemize}
\begin{figure}[h!]
\begin{mdframed}
\captionsetup{font=footnotesize}
\center
\begin{minipage}[b1]{1.0\textwidth}
$\begin{array}{rl}
\includegraphics[width=0.5\textwidth]{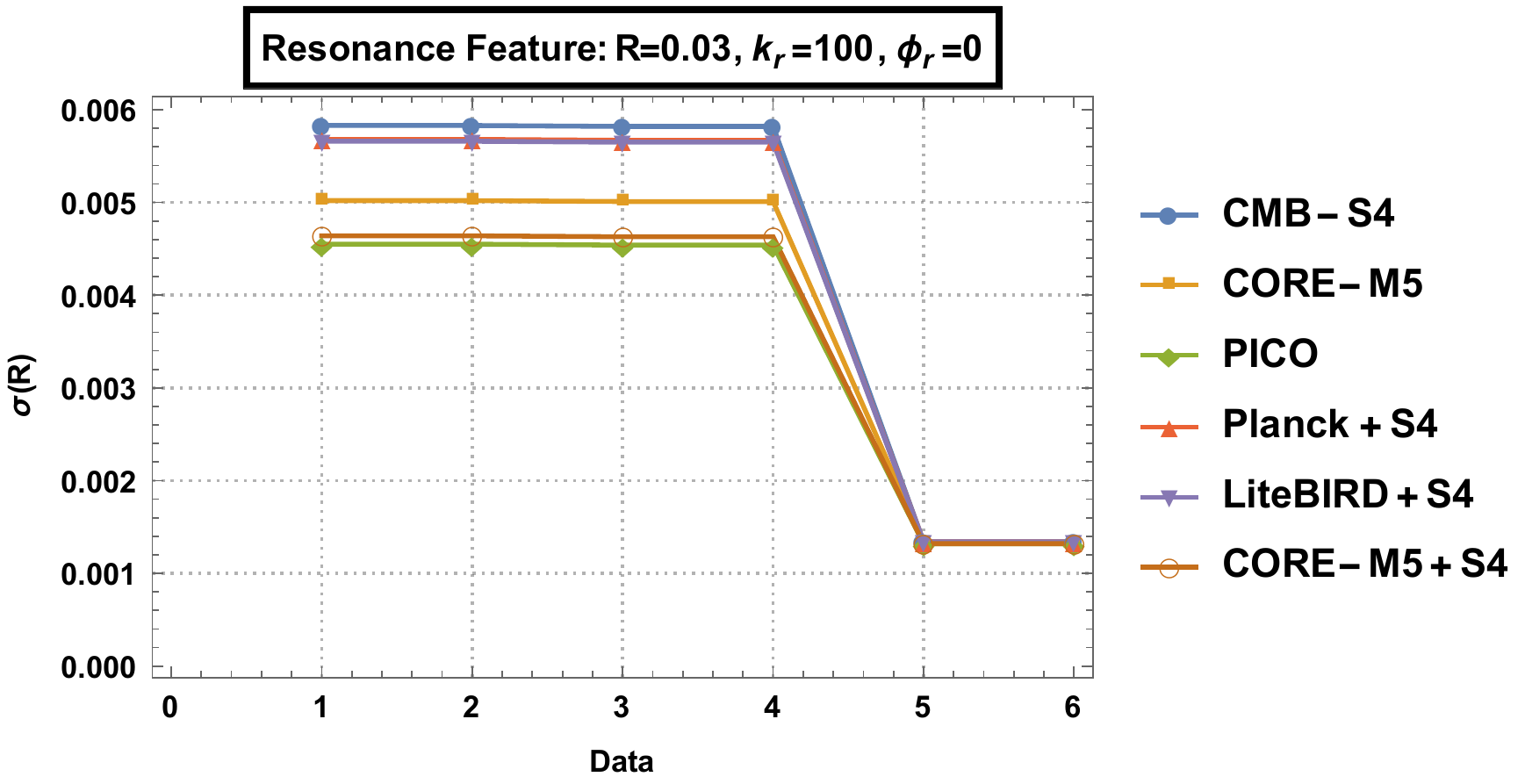} &
\includegraphics[width=0.5\textwidth]{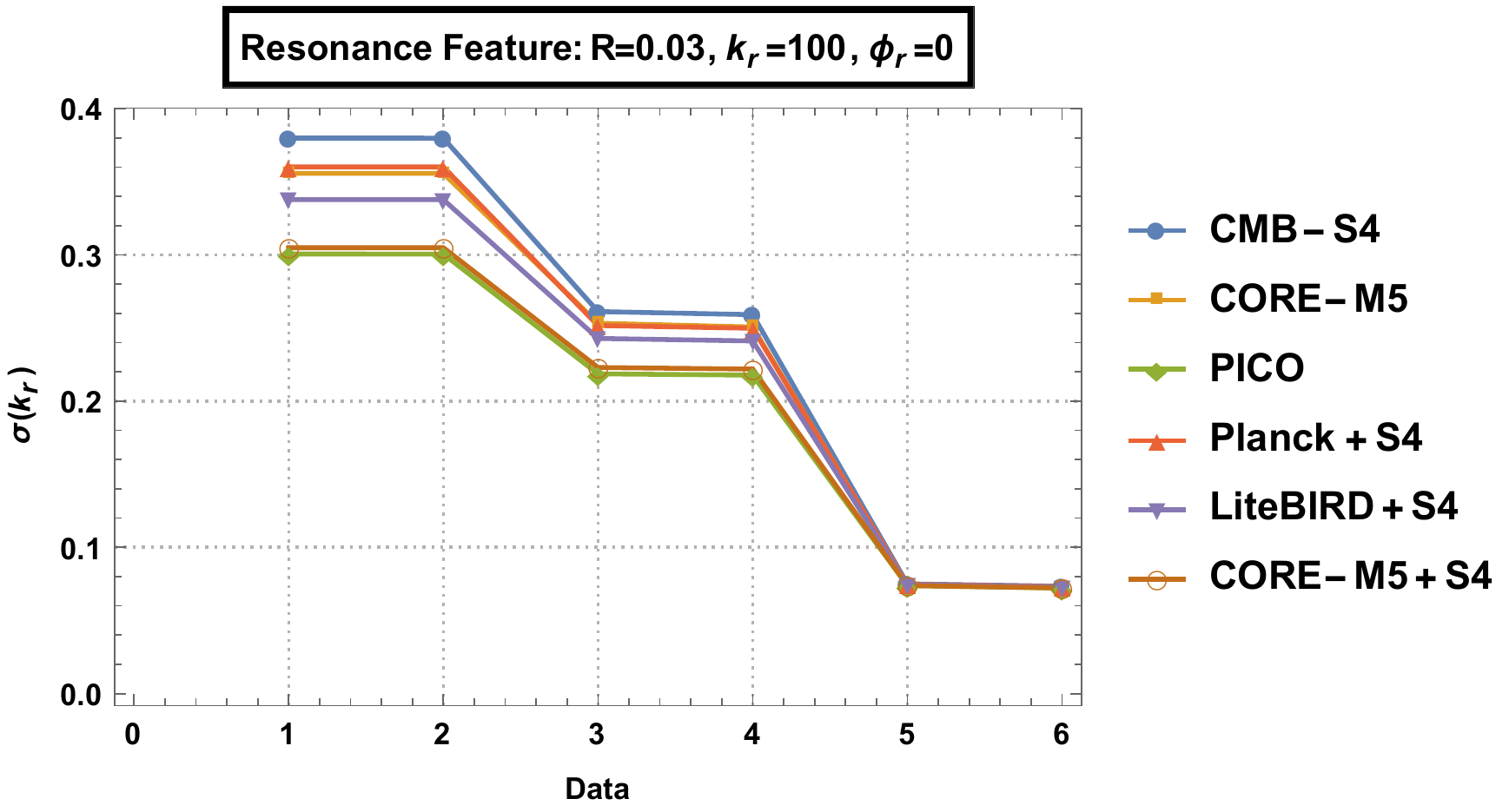}\\
\multicolumn{2}{c}{\includegraphics[width=0.5\textwidth]{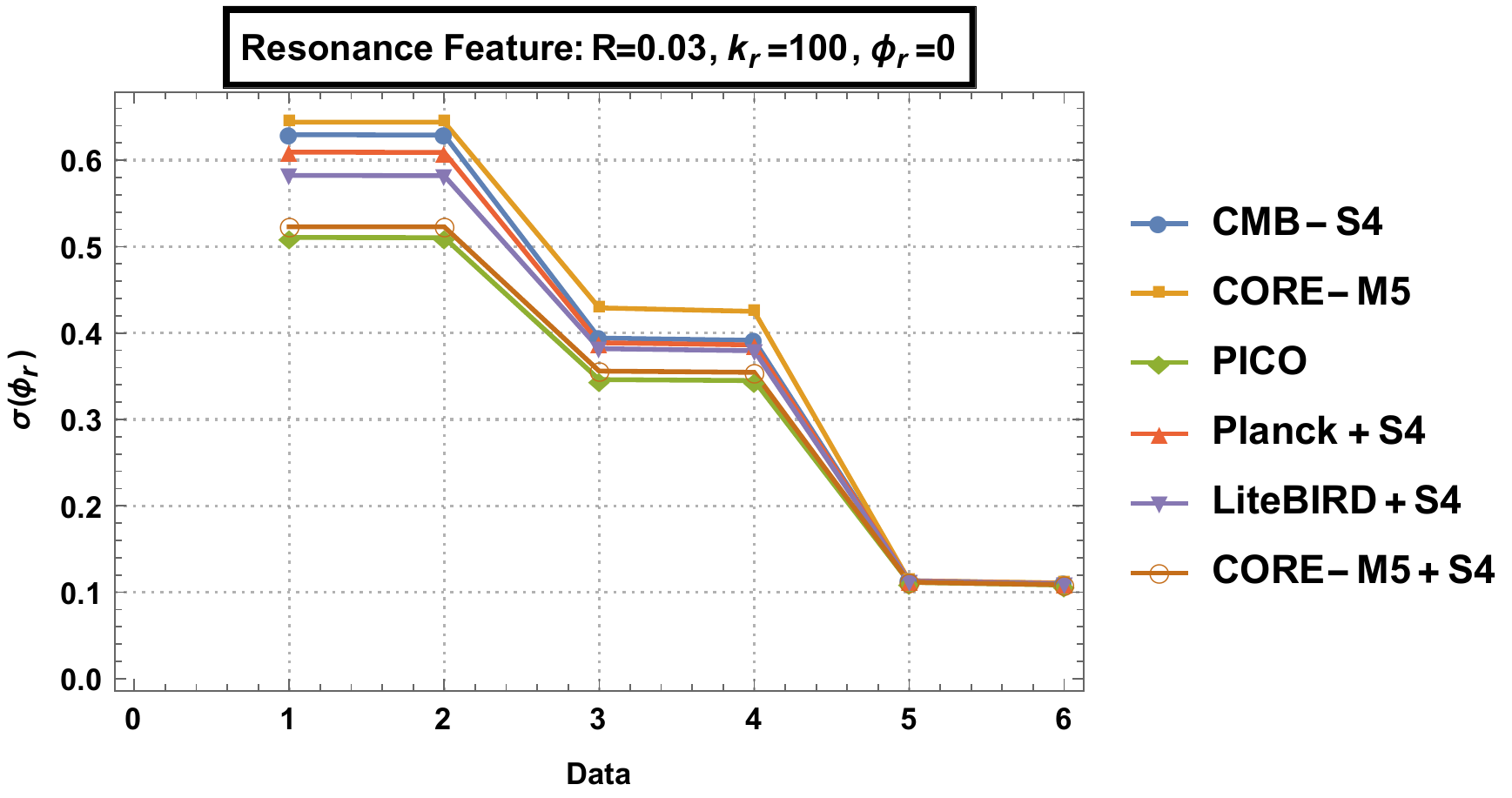}}
\end{array}$
\caption[]{\label{fig:Resonance-3}Experimental sensitivity of each CMB experiment for CMB-only~(1) and combination with DESI~(2), EUCLID-CS~(3), DESI+EUCLID-CS~(4), EUCLID-GC~(5) and EUCLID-(CS+GC)~(6) on the parameters of \textbf{Resonance Feature} template~(\ref{Template:Resonance}) has been depicted here. The 1-$\sigma$ uncertainties on the amplitude~(R), oscillation frequency~($k_r$) and phase~($\phi_r$) of the logarithmic sinusoidal signal are shown in the \textit{Left} plot, \textit{Right} plot and \textit{Middle} plot, respectively. The plots are for these given fiducial values of the model parameters: R=0.03; $k_r$=100; $\phi_{r}$=0.}
\end{minipage}
\hfill
\end{mdframed}
\end{figure}
\subsubsection{Case III}
The third case of the resonance feature signal is for the fiducial values R=0.03; $k_r$=100; $\phi_{r}$=0. The results for this characteristic scale, $k_r$=100 are graphically represented in Fig.~\ref{fig:Resonance-3}, from which the following inferences can be drawn:
\begin{itemize}[itemsep=-.3em]
\item[•] For fiducial values R=0.03; $k_r$=100; $\phi_{r}$=0, if we consider CMB-only case, then the weakest constraint on the oscillation amplitude~(R) comes from CMB-S4, and the best bound from PICO.   
\item[•] DESI, EUCLID-CS and their combination are insensitive towards the oscillation amplitude~(R); they do not alter the bounds in comparison to CMB-only constraints; this is the same for all the CMB experiments.
\item[•] Combination of GC data and GC+CS data with CMB data improves the bounds for all CMB experiments compared to CMB-only results and they give almost the same bounds for all CMB experiments.
\item[•] PICO and CORE-M5 + CMB-S4 have almost similar constraining ability for oscillation amplitude~(R). On the other hand, CMB-S4, LiteBIRD + CMB-S4 and Planck + CMB-S4 have nearly equal sensitivity to oscillation amplitude~(R).
\item[•] For characteristic scale~($k_r$) bounds are between CMB-S4 and PICO, where the lowest error is given by PICO. 
\item[•] PICO and CORE-M5 + CMB-S4 are showing similar sensitivity for characteristic scale~($k_r$), at the fiducial value $k_r$=100. 
\item[•] Bounds on the characteristic scale~($k_r$) are not affected by BAO data but show improvement on using CS data.
\item[•] The introduction of GC data is improving the constraints on $k_r$ as well, and further inclusion of CS data along with GC data is slightly improving the bounds compared to the case for CMB + GC combinations.
\item[•] For phase angle~($ \phi_r $), the bounds are between CORE-M5 and PICO; PICO gives the best bounds. No sensitivity has been shown to the DESI BAO experiment by phase factor but shows improvement for EUCLID-CS data.
\item[•] For all CMB experiments, the best constraints on phase angle~($ \phi_r $) are coming when it is combined with EUCLID-GC and EUCLID-(GC+CS); they are giving almost the same bounds, still, incorporation of CS data in CMB+GC data sets improves the results with a very small amount.
\item[•] PICO and CORE-M5 + CMB-S4 have almost similar constraining ability for phase angle~($ \phi_r $) too.
\end{itemize}

\subsection{Generic Impressions}
It is quite evident from this analysis that the constraining capacity of PICO is the highest and CMB-S4 is the lowest for most of the cases~(except a few instances, for both the experiments). It is mostly because PICO has couple of advantages over CMB-S4, such as, sky coverage of PICO is more than CMB-S4 and has more number of channels. If synergy of CMB missions is considered instead of individual CMB missions, then the best constraints arise from the CORE-M5+CMB-S4 combination~(except for a few instances) among all three CMB combinations~(Planck+CMB-S4, LiteBIRD+CMB-S4 and CORE-M5+CMB-S4) taken into consideration in this analysis; this is due to the fact that the CORE-M5 has better resolution, better sensitivity and possess more number of frequency channels compared to Planck and LiteBIRD. In fact, when CORE-M5 is supported by CMB-S4, it almost reaches the sensitivity of PICO by compensating for the mutual limitations. Another noteworthy point is that among all three LSS surveys~(DESI-BAO, EUCLID-CS and EUCLID-GC) taken into consideration, the inclusion of EUCLID-GC data significantly improves the results for all the CMB experiments if a single LSS survey is considered. Such relative improvements are maximum for the oscillatory signals compared to bump signals since bump-like features are unable to exploit the 3D-information coming from LSS surveys because of its non-oscillatory nature. However, if LSS survey combinations are considered, then the EUCLID-GC+EUCLID-CS combination yields the best results when combined with CMB experiments. Hence, in general, the best improvement in the overall sensitivity arises from the PICO+EUCLID-GC+EUCLID-CS combination (except for very few instances). Further, it is found that the constraining ability of different surveys depends on the frequency of the feature models, as in CMB experiments show less sensitivity for higher frequency feature models, whereas galaxy clustering surveys can perform well in a broad frequency range. Among these three types of features, the possibility of future detection of oscillatory features is the highest and that for bump features is negligible. To compare the improvements on the constraints of the feature model parameters with Planck bounds, one can refer to Ref.~\cite{Chen:2016vvw,Ballardini:2016hpi,Palma:2017wxu}.

\section{\textbf{Summary and Outlook}}\label{Con}
In this work, we have extensively investigated the ability of future CMB experiments in combination with upcoming LSS experiments in order to constrain the features on the primordial power spectrum for inflationary models.  Here we have studied the complementarity between ground-based CMB experiment with satellite CMB experiments as well as the complementarity between different LSS surveys~(BAO, CS and GC) and measurements of CMB anisotropies to identify the primordial features. 
Primordial features being primarily a scale-dependent correction on the primordial power spectrum over the standard scenario, its confirmed detection can help us have a better understanding of the  primordial physics, such as, it can unfold subtle behaviour of the inflationary potential, can discover new particles, can break the degeneracy among inflationary models as well as between inflation, and its alternative prescriptions, it can also help us have a clearer view of the inflationary dynamics as well.
Currently, we have the latest results coming from Planck mission. The latest data, namely, Planck18 has already improved our understanding of the possible cosmological model of the Universe, and are marginally giving hints about the presence of features in the primordial power spectrum. However, as of now, we do not have a very clear idea about whether these signals in the CMB power spectra originate from statistical reasons or they have some fundamental physical origin. Upcoming CMB experiments promise to improve our current understanding of the Universe to varied orders of magnitude, by providing more sensitive data compared to what we have from Planck18. Consequently, that they will be able to shed more light on primordial features and hence on the primordial universe as a whole, is a natural conclusion therefrom. 

With this motivation in mind, we have done a thorough investigation of the constraining capacity of major next generation CMB missions, such as \textbf{CMB-S4}, \textbf{CORE-M5}, \textbf{LiteBIRD} and \textbf{PICO} and their relevant combinations. We have further estimated the improvement, if any, in the sensitivity of each individual CMB experiment as well as their different combinations, when combined further with \textbf{DESI} and \textbf{EUCLID} galaxy surveys. Apart from these future CMB experiments, we have also investigated the capacity of the existing Planck + upcoming CMB-S4 combination, and its synergies with LSS experiments. To materialize this in concrete steps, we have taken into account three different types of features representing distinct classes of models, namely, \textbf{Bump feature}, \textbf{Sharp feature signal} and \textbf{Resonance feature signal}; and have done Fisher matrix forecast analysis for different CMB experiments and combinations as mentioned above. In this direction there are earlier studies in the literature as mentioned in the Introduction. However, in comparison to the previous studies, our analysis differs in many ways. Firstly, as mentioned above, in our analysis we have considered six different CMB scenarios, and for each CMB scenario we have considered five different LSS scenarios. Secondly, to combine two different CMB experiments, we have added one ground-based CMB mission~(CMB-S4) with a space-based mission, and to do that we have considered three different space-based missions~(Planck, LiteBIRD, CORE-M5) with completely different experimental specifications ranging from multipole range, sky coverage, beam resolution and sensitivity to the number of available frequency channels. Apart from considering two different types of CMB surveys~(one ground-based another satellite-based), we further considered three different types of LSS surveys~(BAO, CS and GC) and their permissible synergies with CMB experiments. On top of that, we have adopted up-to-date fiducial values for the cosmological parameters, which are consistent with the latest Planck 2018 release. For the EUCLID galaxy clustering survey also, we haved used updated information about the redshift bins and the number density of galaxies. Besides, in constructing the galaxy clustering power spectrum for EUCLID survey, we have assumed two nuisance parameters for modelling the galaxy bias which takes into account the uncertainties in modelling the systematics and finally marginalized over them to obtain the constraints on the model parameters. Furthermore, the scheme to define the maximum and minimum scale per bin is also different in this work compared to earlier works. Finally, the prescription to compute the Fisher matrices is also different in our work from the earlier works. Here, the MontePython package has been used to compute the Fisher matrices directly from the likelihood by generating the fiducial data, which mimics observed data. However, in earlier works, the scheme was to express the Fisher matrix in terms of the derivative of the observable quantities with respect to the free model parameters. There are other small differences as well. Thus, we can sum up the differences by stating that our work differs in several ways, ranging from assumed future surveys and their synergies, prescription of computation, assumed experimental specifications and instrumental sensitivity, number of addressed systematic errors to up-to-date fiducial values.

The results obtained in this analysis have been presented in details and discussed stepwise. The analysis reveals that in most of the cases, except for a few instances, for individual CMB missions, the PICO provides the best bounds, and the weakest bounds come from the CMB-S4. For all the CMB experiments, when combined with EUCLID-GC data, it is improving the results significantly, since in comparison to CMB experiments, LSS experiments have access to more information.
In particular, the EUCLID-GC experiment shows strong sensitivity towards power spectrum amplitude, as EUCLID has better redshift resolution, which allows it to provide a better measurement of the amplitude that also helps in improving the overall sensitivity. Except for a few cases, the CORE-M5+CMB-S4 combination is almost reaching the sensitivity of the PICO experiment. However, BAO data are unable to show any significant improvement in the results for any of the feature model parameters, be it amplitude, characteristic scale or the phase angle, except in very few instances. Among these three different features, the possibility of detecting oscillatory features is very high in the coming decades, as can be concluded from present forecasts. When ground-based experiments are accompanied by space-based missions of similar sensitivity with a wide range of channels and full-sky coverage, significant improvements are seen. Another point worth mentioning here is that, except for the characteristic scales and the phase factors of the feature models, the constraints on the model parameters do not significantly vary with the fiducial value of the amplitudes of the feature models. In fact, the bounds on the amplitudes remain almost invariant with different choices of the fiducial values of the amplitude as long as it is small enough.    

In a nutshell, in this article, we have presented a comparative analysis of different CMB experiments having different characteristics, such as, sky coverage, sensitivity, resolution, multipole range and number of channels; besides, we have shown how these CMB experiments with different strengths and weaknesses, behave when combined with different LSS missions. Considering several combinations of major experiments makes our analysis and conclusions robust.

\appendix

\section{\textbf{Tables}}\label{Tables}

\setlength{\tabcolsep}{4.5pt} 
\renewcommand{\arraystretch}{1.5} 
\newcolumntype{C}[1]{>{\Centering}m{#1}}
\renewcommand\tabularxcolumn[1]{C{#1}}
\begin{minipage}{\linewidth}
\centering
\captionsetup{font=footnotesize}
\begin{tabular}{|c|c|c|c|c|c|c|c|}
\hline
\textbf{Models}            & \textbf{Parameters}                                                       & \textbf{CMB}                                                              & \textbf{+ DESI}                                                           & \textbf{\begin{tabular}[c]{@{}c@{}}+ Euclid\\ (CS)\end{tabular}}            & \textbf{\begin{tabular}[c]{@{}c@{}}+ DESI +\\  Euclid \\ (CS)\end{tabular}} & \textbf{\begin{tabular}[c]{@{}c@{}}+ Euclid \\ (GC)\end{tabular}}           & \textbf{\begin{tabular}[c]{@{}c@{}}+ Euclid\\ (GC+CS)\end{tabular}}       \\ \hline
\textbf{\begin{tabular}[c]{@{}c@{}}Bump Feature\\ $k_b=0.05$\end{tabular}}           & \begin{tabular}[c]{@{}c@{}}$10^{5}\times\sigma \left(\omega_{\mathrm{b}}\right)$\\ $\sigma \left(\omega_{\mathrm{cdm}}\right)$\\ $\sigma \left( H_0 \right)$\\ $10^{12}\times\sigma \left( A_s \right)$\\ $\sigma \left( n_s \right)$\\ $\sigma \left( \tau_{\mathrm{reio}} \right)$\\ $\sigma \left(B\right)$\\ $\sigma \left(k_b\right)$\end{tabular}     & \begin{tabular}[c]{@{}c@{}}3.36941
\\ 0.00024\\ 0.09752
\\ 9.08845
\\ 0.00208
\\ 0.00218
\\ 0.00395
\\ 0.03336
\end{tabular}     & \begin{tabular}[c]{@{}c@{}}3.36551
\\ 0.00021
\\ 0.08774
\\ 8.69809\\ 0.00208
\\ 0.00207
\\ 0.00395
\\ 0.03335
\end{tabular}     & \begin{tabular}[c]{@{}c@{}}3.65849
\\ 0.00017
\\ 0.069
\\ 8.72239
\\ 0.00213
\\ 0.00186
\\ 0.00395\\ 0.03336
\end{tabular}     & \begin{tabular}[c]{@{}c@{}}3.65743
\\ 0.00016
\\ 0.0656
\\ 8.53596\\ 0.00213
\\ 0.00181
\\ 0.00394
\\ 0.03336
\end{tabular}     & \begin{tabular}[c]{@{}c@{}}3.20286
\\ 0.00020
\\ 0.07644
\\ 8.20796
\\ 0.00191
\\ 0.002
\\ 0.00287
\\ 0.02274
\end{tabular}     & \begin{tabular}[c]{@{}c@{}}3.30056
\\ 0.00015
\\ 0.05902
\\ 7.62035
\\ 0.00188
\\ 0.00163
\\ 0.00288
\\ 0.02273
\end{tabular}     \\ \hline
\textbf{\begin{tabular}[c]{@{}c@{}}Bump Feature\\ $k_b=0.1$\end{tabular}}           & \begin{tabular}[c]{@{}c@{}}$10^{5}\times\sigma \left(\omega_{\mathrm{b}}\right)$\\ $\sigma \left(\omega_{\mathrm{cdm}}\right)$\\ $\sigma \left( H_0 \right)$\\ $10^{12}\times\sigma \left( A_s \right)$\\ $\sigma \left( n_s \right)$\\ $\sigma \left( \tau_{\mathrm{reio}} \right)$\\ $\sigma \left(B\right)$\\ $\sigma \left(k_b\right)$\end{tabular}     & \begin{tabular}[c]{@{}c@{}}3.67828
\\ 0.00024
\\ 0.09719
\\ 8.39791
\\ 0.00184
\\ 0.00222
\\ 0.00235
\\ 0.04211
\end{tabular}     & \begin{tabular}[c]{@{}c@{}}3.67525
\\ 0.00021
\\ 0.08768
\\ 8.03081
\\ 0.00183
\\ 0.00211
\\ 0.00235
\\ 0.04211
\end{tabular}     & \begin{tabular}[c]{@{}c@{}}3.8931
\\ 0.00016
\\ 0.06634\\ 7.61143
\\ 0.00186\\ 0.00185
\\ 0.00233
\\ 0.04174
\end{tabular}     & \begin{tabular}[c]{@{}c@{}}4.00807
\\ 0.00016
\\ 0.06543
\\ 7.56123
\\ 0.00185
\\ 0.00185
\\ 0.00234
\\ 0.04174
\end{tabular}     & \begin{tabular}[c]{@{}c@{}}3.35341
\\ 0.00019
\\ 0.07241
\\ 7.44685
\\ 0.00172
\\ 0.00199
\\ 0.00176\\ 0.0335
\end{tabular}     & \begin{tabular}[c]{@{}c@{}}3.41939
\\ 0.00015
\\ 0.05805
\\ 6.67245
\\ 0.00163
\\ 0.00163
\\ 0.00174
\\ 0.03262
\end{tabular}     \\ \hline
\textbf{\begin{tabular}[c]{@{}c@{}}Bump Feature\\ $k_b=0.2$\end{tabular}}           & \begin{tabular}[c]{@{}c@{}}$10^{5}\times\sigma \left(\omega_{\mathrm{b}}\right)$\\ $\sigma \left(\omega_{\mathrm{cdm}}\right)$\\ $\sigma \left( H_0 \right)$\\ $10^{12}\times\sigma \left( A_s \right)$\\ $\sigma \left( n_s \right)$\\ $\sigma \left( \tau_{\mathrm{reio}} \right)$\\ $\sigma \left(B\right)$\\ $\sigma \left(k_b\right)$\end{tabular}     & \begin{tabular}[c]{@{}c@{}}4.26951
\\ 0.00025
\\ 0.10356
\\ 10.55538
\\ 0.00383
\\ 0.00256
\\ 0.00706\\ 0.11384
\end{tabular}     & \begin{tabular}[c]{@{}c@{}}4.26721
\\ 0.00022
\\ 0.09248
\\ 10.22007\\ 0.00383\\ 0.00244\\ 0.00705
\\ 0.11381
\end{tabular}     & \begin{tabular}[c]{@{}c@{}}4.13517
\\ 0.00018
\\ 0.06987
\\ 7.92601
\\ 0.00297
\\ 0.00189
\\ 0.00551
\\ 0.10620
\end{tabular}     & \begin{tabular}[c]{@{}c@{}}4.11686
\\ 0.00017
\\ 0.06596
\\ 7.83895
\\ 0.00296
\\ 0.00185
\\ 0.00545
\\ 0.10575
\end{tabular}     & \begin{tabular}[c]{@{}c@{}}3.3236
\\ 0.00020
\\ 0.07595
\\ 7.63509
\\ 0.00284
\\ 0.00204
\\ 0.00411
\\ 0.08476
\end{tabular}     & \begin{tabular}[c]{@{}c@{}}3.42535
\\ 0.00015
\\ 0.06152
\\ 6.46595
\\ 0.00218
\\ 0.00167
\\ 0.00320
\\ 0.08708
\end{tabular}     \\ \hline
\end{tabular}\par
\bigskip
\parbox{17.9cm}{\captionof{table}{The possible marginalized 1-$\sigma$ constraints of \textbf{CMB-S4} experiment on bump feature, for CMB-only data and its combination with DESI, EUCLID(CS), DESI+EUCLID(CS), EUCLID(GC) and EUCLID(CS+GC).}} \label{table:CMB-S4-Bump}
\end{minipage}

\setlength{\tabcolsep}{4.8pt} 
\renewcommand{\arraystretch}{1.5} 
\newcolumntype{C}[1]{>{\Centering}m{#1}}
\renewcommand\tabularxcolumn[1]{C{#1}}
\begin{minipage}{\linewidth}
\centering
\captionsetup{font=footnotesize}
\begin{tabular}{|c|c|c|c|c|c|c|c|}
\hline
\textbf{Models}            & \textbf{Parameters}                                                       & \textbf{CMB}                                                              & \textbf{+ DESI}                                                           & \textbf{\begin{tabular}[c]{@{}c@{}}+ Euclid\\ (CS)\end{tabular}}            & \textbf{\begin{tabular}[c]{@{}c@{}}+ DESI +\\  Euclid \\ (CS)\end{tabular}} & \textbf{\begin{tabular}[c]{@{}c@{}}+ Euclid \\ (GC)\end{tabular}}           & \textbf{\begin{tabular}[c]{@{}c@{}}+ Euclid\\ (GC+CS)\end{tabular}}       \\ \hline
\textbf{\begin{tabular}[c]{@{}c@{}}Bump Feature\\ $k_b=0.05$\end{tabular}}           & \begin{tabular}[c]{@{}c@{}}$10^{5}\times\sigma \left(\omega_{\mathrm{b}}\right)$\\ $\sigma \left(\omega_{\mathrm{cdm}}\right)$\\ $\sigma \left( H_0 \right)$\\ $10^{12}\times\sigma \left( A_s \right)$\\ $\sigma \left( n_s \right)$\\ $\sigma \left( \tau_{\mathrm{reio}} \right)$\\ $\sigma \left(B\right)$\\ $\sigma \left(k_b\right)$\end{tabular}     & \begin{tabular}[c]{@{}c@{}}4.01954
\\ 0.00018
\\ 0.07739
\\ 5.97352
\\ 0.00171
\\ 0.00134
\\ 0.00309
\\ 0.02538
\end{tabular}     & \begin{tabular}[c]{@{}c@{}}4.01811
\\ 0.00017
\\ 0.07297
\\ 5.93836
\\ 0.00171
\\ 0.00132
\\ 0.00309
\\ 0.02538
\end{tabular}     & \begin{tabular}[c]{@{}c@{}}4.00549
\\ 0.00013
\\ 0.05919
\\ 6.05598
\\ 0.00170
\\ 0.00131
\\ 0.00306
\\ 0.02533
\end{tabular}     & \begin{tabular}[c]{@{}c@{}}4.00445\\ 0.00013
\\ 0.05739
\\ 6.00275
\\ 0.00170
\\ 0.00129
\\ 0.00306
\\ 0.02533
\end{tabular}     & \begin{tabular}[c]{@{}c@{}}3.60611
\\ 0.00017\\ 0.06914

\\ 5.82747
\\ 0.00160
\\ 0.00132
\\ 0.00247
\\ 0.01976
\end{tabular}     & \begin{tabular}[c]{@{}c@{}}3.69421
\\ 0.00013
\\ 0.05463
\\ 5.75924\\ 0.00158
\\ 0.00125
\\ 0.00244
\\ 0.01977
\end{tabular}     \\ \hline
\textbf{\begin{tabular}[c]{@{}c@{}}Bump Feature\\ $k_b=0.1$\end{tabular}}           & \begin{tabular}[c]{@{}c@{}}$10^{5}\times\sigma \left(\omega_{\mathrm{b}}\right)$\\ $\sigma \left(\omega_{\mathrm{cdm}}\right)$\\ $\sigma \left( H_0 \right)$\\ $10^{12}\times\sigma \left( A_s \right)$\\ $\sigma \left( n_s \right)$\\ $\sigma \left( \tau_{\mathrm{reio}} \right)$\\ $\sigma \left(B\right)$\\ $\sigma \left(k_b\right)$\end{tabular}     & \begin{tabular}[c]{@{}c@{}}4.26359
\\ 0.00018
\\ 0.07729
\\ 5.58565
\\ 0.00159
\\ 0.00135
\\ 0.00193
\\ 0.03776
\end{tabular}     & \begin{tabular}[c]{@{}c@{}}4.26267
\\ 0.00017
\\ 0.07298
\\ 5.55322
\\ 0.00158
\\ 0.00134
\\ 0.00193
\\ 0.03776
\end{tabular}     & \begin{tabular}[c]{@{}c@{}}4.29925
\\ 0.00013
\\ 0.0597
\\ 5.60272
\\ 0.00158
\\ 0.00134
\\ 0.00192
\\ 0.03731
\end{tabular}     & \begin{tabular}[c]{@{}c@{}}4.25468
\\ 0.00013
\\ 0.05762
\\ 5.50792
\\ 0.00157
\\ 0.00131\\ 0.00192
\\ 0.03731
\end{tabular}     & \begin{tabular}[c]{@{}c@{}}3.76261\\ 0.00016
\\ 0.06753
\\ 5.4035\\ 0.00147
\\ 0.00133
\\ 0.00155
\\ 0.03126
\end{tabular}     & \begin{tabular}[c]{@{}c@{}}3.81808\\ 0.00013
\\ 0.05427
\\ 5.22607
\\ 0.00143
\\ 0.00126
\\ 0.00155
\\ 0.03035
\end{tabular}     \\ \hline
\textbf{\begin{tabular}[c]{@{}c@{}}Bump Feature\\ $k_b=0.2$\end{tabular}}           & \begin{tabular}[c]{@{}c@{}}$10^{5}\times\sigma \left(\omega_{\mathrm{b}}\right)$\\ $\sigma \left(\omega_{\mathrm{cdm}}\right)$\\ $\sigma \left( H_0 \right)$\\ $10^{12}\times\sigma \left( A_s \right)$\\ $\sigma \left( n_s \right)$\\ $\sigma \left( \tau_{\mathrm{reio}} \right)$\\ $\sigma \left(B\right)$\\ $\sigma \left(k_b\right)$\end{tabular}     & \begin{tabular}[c]{@{}c@{}}4.55908\\ 0.00019
\\ 0.07669
\\ 6.66455\\ 0.00284
\\ 0.00141
\\ 0.00547
\\ 0.10151
\end{tabular}     & \begin{tabular}[c]{@{}c@{}}4.55581\\ 0.00018
\\ 0.07230
\\ 6.66158
\\ 0.00283
\\ 0.00141
\\ 0.00545
\\ 0.10131
\end{tabular}     & \begin{tabular}[c]{@{}c@{}}4.40484
\\ 0.00015
\\ 0.05931
\\ 6.2053
\\ 0.00254
\\ 0.00135
\\ 0.00487
\\ 0.10059
\end{tabular}     & \begin{tabular}[c]{@{}c@{}}4.39320\\ 0.00014
\\ 0.05730
\\ 6.19387\\ 0.00254
\\ 0.00134
\\ 0.00483
\\ 0.10032
\end{tabular}     & \begin{tabular}[c]{@{}c@{}}3.64917
\\ 0.00017\\ 0.06923
\\ 5.60512
\\ 0.00233
\\ 0.00134
\\ 0.00361
\\ 0.08438
\end{tabular}     & \begin{tabular}[c]{@{}c@{}}3.70696
\\ 0.00013
\\ 0.05493\\ 5.1604
\\ 0.00207\\ 0.00125
\\ 0.00321
\\ 0.0855
\end{tabular}     \\ \hline
\end{tabular}\par
\bigskip
\parbox{17.9cm}{\captionof{table}{The possible marginalized 1-$\sigma$ constraints of \textbf{CORE-M5} experiment on bump feature, for CMB-only data and its combination with DESI, EUCLID(CS), DESI+EUCLID(CS), EUCLID(GC) and EUCLID(CS+GC).}} \label{table:CORE-M5-Bump}
\end{minipage}

\setlength{\tabcolsep}{4.8pt} 
\renewcommand{\arraystretch}{1.5} 
\newcolumntype{C}[1]{>{\Centering}m{#1}}
\renewcommand\tabularxcolumn[1]{C{#1}}
\begin{minipage}{\linewidth}
\centering
\captionsetup{font=footnotesize}
\begin{tabular}{|c|c|c|c|c|c|c|c|}
\hline
\textbf{Models}            & \textbf{Parameters}                                                       & \textbf{CMB}                                                              & \textbf{+ DESI}                                                           & \textbf{\begin{tabular}[c]{@{}c@{}}+ Euclid\\ (CS)\end{tabular}}            & \textbf{\begin{tabular}[c]{@{}c@{}}+ DESI +\\  Euclid \\ (CS)\end{tabular}} & \textbf{\begin{tabular}[c]{@{}c@{}}+ Euclid \\ (GC)\end{tabular}}           & \textbf{\begin{tabular}[c]{@{}c@{}}+ Euclid\\ (GC+CS)\end{tabular}}       \\ \hline
\textbf{\begin{tabular}[c]{@{}c@{}}Bump Feature\\ $k_b=0.05$\end{tabular}}           & \begin{tabular}[c]{@{}c@{}}$10^{5}\times\sigma \left(\omega_{\mathrm{b}}\right)$\\ $\sigma \left(\omega_{\mathrm{cdm}}\right)$\\ $\sigma \left( H_0 \right)$\\ $10^{12}\times\sigma \left( A_s \right)$\\ $\sigma \left( n_s \right)$\\ $\sigma \left( \tau_{\mathrm{reio}} \right)$\\ $\sigma \left(B\right)$\\ $\sigma \left(k_b\right)$\end{tabular}     & \begin{tabular}[c]{@{}c@{}}2.86341
\\ 0.00015
\\ 0.05879
\\ 5.33777
\\ 0.00160
\\ 0.00118
\\ 0.00297
\\ 0.02481
\end{tabular}     & \begin{tabular}[c]{@{}c@{}}2.86116
\\ 0.00014
\\ 0.05641
\\ 5.29678
\\ 0.00159
\\ 0.00117
\\ 0.00297
\\ 0.02481
\end{tabular}     & \begin{tabular}[c]{@{}c@{}}3.12571
\\ 0.00013
\\ 0.05148
\\ 5.44474\\ 0.00170
\\ 0.00118
\\ 0.00296
\\ 0.02491
\end{tabular}     & \begin{tabular}[c]{@{}c@{}}3.12327\\ 0.00012
\\ 0.05003
\\ 5.39847
\\ 0.00170
\\ 0.00117
\\ 0.00296
\\ 0.0249
\end{tabular}     & \begin{tabular}[c]{@{}c@{}}2.75163
\\ 0.00015\\ 0.05795
\\ 5.48258
\\ 0.00150
\\ 0.00124

\\ 0.00240
\\ 0.01948
\end{tabular}     & \begin{tabular}[c]{@{}c@{}}2.92572
\\ 0.00012
\\ 0.0468
\\ 5.10524
\\ 0.00163
\\ 0.00111
\\ 0.00243
\\ 0.01958
\end{tabular}     \\ \hline
\textbf{\begin{tabular}[c]{@{}c@{}}Bump Feature\\ $k_b=0.1$\end{tabular}}           & \begin{tabular}[c]{@{}c@{}}$10^{5}\times\sigma \left(\omega_{\mathrm{b}}\right)$\\ $\sigma \left(\omega_{\mathrm{cdm}}\right)$\\ $\sigma \left( H_0 \right)$\\ $10^{12}\times\sigma \left( A_s \right)$\\ $\sigma \left( n_s \right)$\\ $\sigma \left( \tau_{\mathrm{reio}} \right)$\\ $\sigma \left(B\right)$\\ $\sigma \left(k_b\right)$\end{tabular}     & \begin{tabular}[c]{@{}c@{}}3.11541\\ 0.00015
\\ 0.05855
\\ 4.94359
\\ 0.00143
\\ 0.00120
\\ 0.00180
\\ 0.03279
\end{tabular}     & \begin{tabular}[c]{@{}c@{}}3.1109
\\ 0.00014\\ 0.0562
\\ 4.90052
\\ 0.00142\\ 0.00119
\\ 0.00180
\\ 0.03279
\end{tabular}     & \begin{tabular}[c]{@{}c@{}}3.33718
\\ 0.00013
\\ 0.05013
\\ 4.92633
\\ 0.00152
\\ 0.00119
\\ 0.00178
\\ 0.03257
\end{tabular}     & \begin{tabular}[c]{@{}c@{}}3.39426\\ 0.00012
\\ 0.04989
\\ 4.90887
\\ 0.00151
\\ 0.00119
\\ 0.00179
\\ 0.03258
\end{tabular}     & \begin{tabular}[c]{@{}c@{}}2.89013
\\ 0.00014
\\ 0.0551
\\ 4.73548
\\ 0.00134
\\ 0.00117
\\ 0.00148
\\ 0.02786
\end{tabular}     & \begin{tabular}[c]{@{}c@{}}3.05171
\\ 0.00012
\\ 0.04624
\\ 4.60473\\ 0.00143
\\ 0.00112
\\ 0.00147
\\ 0.02771
\end{tabular}     \\ \hline
\textbf{\begin{tabular}[c]{@{}c@{}}Bump Feature\\ $k_b=0.2$\end{tabular}}           & \begin{tabular}[c]{@{}c@{}}$10^{5}\times\sigma \left(\omega_{\mathrm{b}}\right)$\\ $\sigma \left(\omega_{\mathrm{cdm}}\right)$\\ $\sigma \left( H_0 \right)$\\ $10^{12}\times\sigma \left( A_s \right)$\\ $\sigma \left( n_s \right)$\\ $\sigma \left( \tau_{\mathrm{reio}} \right)$\\ $\sigma \left(B\right)$\\ $\sigma \left(k_b\right)$\end{tabular}     & \begin{tabular}[c]{@{}c@{}}3.47765
\\ 0.00016
\\ 0.06020
\\ 5.85058
\\ 0.00269\\ 0.00127
\\ 0.00512
\\ 0.08666
\end{tabular}     & \begin{tabular}[c]{@{}c@{}}3.46527
\\ 0.00015
\\ 0.05757
\\ 5.84367
\\ 0.00269
\\ 0.00127
\\ 0.00509
\\ 0.08640
\end{tabular}     & \begin{tabular}[c]{@{}c@{}}3.37391
\\ 0.00014
\\ 0.05224
\\ 5.52466
\\ 0.00255
\\ 0.00124
\\ 0.00451
\\ 0.08401
\end{tabular}     & \begin{tabular}[c]{@{}c@{}}3.36267
\\ 0.00013
\\ 0.05056
\\ 5.51273
\\ 0.00255
\\ 0.00123
\\ 0.00448
\\ 0.08378
\end{tabular}     & \begin{tabular}[c]{@{}c@{}}2.88798\\ 0.00015
\\ 0.05811
\\ 5.17179
\\ 0.00222
\\ 0.00127
\\ 0.00349
\\ 0.07120
\end{tabular}     & \begin{tabular}[c]{@{}c@{}}3.10661
\\ 0.00013
\\ 0.04777
\\ 4.76707\\ 0.00236
\\ 0.00119
\\ 0.00320
\\ 0.07582
\end{tabular}     \\ \hline
\end{tabular}\par
\bigskip
\parbox{17.9cm}{\captionof{table}{The possible marginalized 1-$\sigma$ constraints of \textbf{PICO} experiment on bump feature, for CMB-only data and its combination with DESI, EUCLID(CS), DESI+EUCLID(CS), EUCLID(GC) and EUCLID(CS+GC).}} \label{table:PICO-Bump}
\end{minipage}

\setlength{\tabcolsep}{4.8pt} 
\renewcommand{\arraystretch}{1.5} 
\newcolumntype{C}[1]{>{\Centering}m{#1}}
\renewcommand\tabularxcolumn[1]{C{#1}}
\begin{minipage}{\linewidth}
\centering
\captionsetup{font=footnotesize}
\begin{tabular}{|c|c|c|c|c|c|c|c|}
\hline
\textbf{Models}            & \textbf{Parameters}                                                       & \textbf{CMB}                                                              & \textbf{+ DESI}                                                           & \textbf{\begin{tabular}[c]{@{}c@{}}+ Euclid\\ (CS)\end{tabular}}            & \textbf{\begin{tabular}[c]{@{}c@{}}+ DESI +\\  Euclid \\ (CS)\end{tabular}} & \textbf{\begin{tabular}[c]{@{}c@{}}+ Euclid \\ (GC)\end{tabular}}           & \textbf{\begin{tabular}[c]{@{}c@{}}+ Euclid\\ (GC+CS)\end{tabular}}       \\ \hline
\textbf{\begin{tabular}[c]{@{}c@{}}Bump Feature\\ $k_b=0.05$\end{tabular}}           & \begin{tabular}[c]{@{}c@{}}$10^{5}\times\sigma \left(\omega_{\mathrm{b}}\right)$\\ $\sigma \left(\omega_{\mathrm{cdm}}\right)$\\ $\sigma \left( H_0 \right)$\\ $10^{12}\times\sigma \left( A_s \right)$\\ $\sigma \left( n_s \right)$\\ $\sigma \left( \tau_{\mathrm{reio}} \right)$\\ $\sigma \left(B\right)$\\ $\sigma \left(k_b\right)$\end{tabular}     & \begin{tabular}[c]{@{}c@{}}3.40421
\\ 0.00021
\\ 0.08266
\\ 7.9965
\\ 0.00196
\\ 0.00190
\\ 0.00361
\\ 0.03046
\end{tabular}     & \begin{tabular}[c]{@{}c@{}}3.40289
\\ 0.00019
\\ 0.07632
\\ 7.81102
\\ 0.00195
\\ 0.00184
\\ 0.00361
\\ 0.03046
\end{tabular}     & \begin{tabular}[c]{@{}c@{}}3.54103
\\ 0.00016
\\ 0.065
\\ 7.87394
\\ 0.00198
\\ 0.00169
\\ 0.00362
\\ 0.0309
\end{tabular}     & \begin{tabular}[c]{@{}c@{}}3.53949
\\ 0.00015
\\ 0.06214
\\ 7.73978
\\ 0.00198
\\ 0.00165
\\ 0.00362
\\ 0.0309
\end{tabular}     & \begin{tabular}[c]{@{}c@{}}3.12687
\\ 0.00019
\\ 0.0726
\\ 7.45455
\\ 0.00179
\\ 0.00178
\\ 0.00274
\\ 0.02191
\end{tabular}     & \begin{tabular}[c]{@{}c@{}}3.26273
\\ 0.00015
\\ 0.05978
\\ 6.93331
\\ 0.00186
\\ 0.00161
\\ 0.00275
\\ 0.02191
\end{tabular}     \\ \hline
\textbf{\begin{tabular}[c]{@{}c@{}}Bump Feature\\ $k_b=0.1$\end{tabular}}           & \begin{tabular}[c]{@{}c@{}}$10^{5}\times\sigma \left(\omega_{\mathrm{b}}\right)$\\ $\sigma \left(\omega_{\mathrm{cdm}}\right)$\\ $\sigma \left( H_0 \right)$\\ $10^{12}\times\sigma \left( A_s \right)$\\ $\sigma \left( n_s \right)$\\ $\sigma \left( \tau_{\mathrm{reio}} \right)$\\ $\sigma \left(B\right)$\\ $\sigma \left(k_b\right)$\end{tabular}     & \begin{tabular}[c]{@{}c@{}}3.73989
\\ 0.00023
\\ 0.09618
\\ 7.86519
\\ 0.00172
\\ 0.00209
\\ 0.00225
\\ 0.04057\end{tabular}     & \begin{tabular}[c]{@{}c@{}}3.73396
\\ 0.00021
\\ 0.08720
\\ 7.55079\\ 0.00171
\\ 0.00199
\\ 0.00225
\\ 0.04056
\end{tabular}     & \begin{tabular}[c]{@{}c@{}}3.90746
\\ 0.00017
\\ 0.07554
\\ 7.84339
\\ 0.00165
\\ 0.00191
\\ 0.00221
\\ 0.03958
\end{tabular}     & \begin{tabular}[c]{@{}c@{}}3.92105
\\ 0.00015
\\ 0.06214
\\ 6.91017
\\ 0.00174
\\ 0.00168
\\ 0.00224
\\ 0.04017
\end{tabular}     & \begin{tabular}[c]{@{}c@{}}3.29917
\\ 0.00017
\\ 0.06804
\\ 6.34706
\\ 0.00160\\ 0.00164
\\ 0.00171
\\ 0.03251\end{tabular}     & \begin{tabular}[c]{@{}c@{}}3.43907\\ 0.00014
\\ 0.05753
\\ 6.29950
\\ 0.00155\\ 0.00153
\\ 0.00170
\\ 0.03188\end{tabular}     \\ \hline
\textbf{\begin{tabular}[c]{@{}c@{}}Bump Feature\\ $k_b=0.2$\end{tabular}}           & \begin{tabular}[c]{@{}c@{}}$10^{5}\times\sigma \left(\omega_{\mathrm{b}}\right)$\\ $\sigma \left(\omega_{\mathrm{cdm}}\right)$\\ $\sigma \left( H_0 \right)$\\ $10^{12}\times\sigma \left( A_s \right)$\\ $\sigma \left( n_s \right)$\\ $\sigma \left( \tau_{\mathrm{reio}} \right)$\\ $\sigma \left(B\right)$\\ $\sigma \left(k_b\right)$\end{tabular}     & \begin{tabular}[c]{@{}c@{}}4.47385
\\ 0.00021\\ 0.08212
\\ 9.22626
\\ 0.00364
\\ 0.00211
\\ 0.00710
\\ 0.11088
\end{tabular}     & \begin{tabular}[c]{@{}c@{}}4.18544
\\ 0.00021
\\ 0.08631\\ 9.16224
\\ 0.00349
\\ 0.00214
\\ 0.00659
\\ 0.10949
\end{tabular}     & \begin{tabular}[c]{@{}c@{}}3.90123
\\ 0.00019
\\ 0.07859\\ 7.93558\\ 0.00229
\\ 0.00194
\\ 0.00429
\\ 0.10226
\end{tabular}     & \begin{tabular}[c]{@{}c@{}}4.06954
\\ 0.00016\\ 0.06238
\\ 7.32926
\\ 0.00278
\\ 0.00169
\\ 0.00524
\\ 0.10388
\end{tabular}     & \begin{tabular}[c]{@{}c@{}}3.25119
\\ 0.00018
\\ 0.06891
\\ 6.4520
\\ 0.00261
\\ 0.00166
\\ 0.00392
\\ 0.08517
\end{tabular}     & \begin{tabular}[c]{@{}c@{}}3.36010
\\ 0.00015
\\ 0.05886
\\ 6.05604
\\ 0.00208
\\ 0.00154
\\ 0.00312
\\ 0.08581
\end{tabular}     \\ \hline
\end{tabular}\par
\bigskip
\parbox{17.9cm}{\captionof{table}{The possible marginalized 1-$\sigma$ constraints of \textbf{Planck+CMB-S4} experiment, for CMB-only data and its combination with DESI, EUCLID(CS), DESI+EUCLID(CS), EUCLID(GC) and EUCLID(CS+GC) on bump feature.}} \label{table:Planck+CMB-S4-Bump}
\end{minipage}

\setlength{\tabcolsep}{4.8pt} 
\renewcommand{\arraystretch}{1.5} 
\newcolumntype{C}[1]{>{\Centering}m{#1}}
\renewcommand\tabularxcolumn[1]{C{#1}}
\begin{minipage}{\linewidth}
\centering
\captionsetup{font=footnotesize}
\begin{tabular}{|c|c|c|c|c|c|c|c|}
\hline
\textbf{Models}            & \textbf{Parameters}                                                       & \textbf{CMB}                                                              & \textbf{+ DESI}                                                           & \textbf{\begin{tabular}[c]{@{}c@{}}+ Euclid\\ (CS)\end{tabular}}            & \textbf{\begin{tabular}[c]{@{}c@{}}+ DESI +\\  Euclid \\ (CS)\end{tabular}} & \textbf{\begin{tabular}[c]{@{}c@{}}+ Euclid \\ (GC)\end{tabular}}           & \textbf{\begin{tabular}[c]{@{}c@{}}+ Euclid\\ (GC+CS)\end{tabular}}       \\ \hline
\textbf{\begin{tabular}[c]{@{}c@{}}Bump Feature\\ $k_b=0.05$\end{tabular}}           & \begin{tabular}[c]{@{}c@{}}$10^{5}\times\sigma \left(\omega_{\mathrm{b}}\right)$\\ $\sigma \left(\omega_{\mathrm{cdm}}\right)$\\ $\sigma \left( H_0 \right)$\\ $10^{12}\times\sigma \left( A_s \right)$\\ $\sigma \left( n_s \right)$\\ $\sigma \left( \tau_{\mathrm{reio}} \right)$\\ $\sigma \left(B\right)$\\ $\sigma \left(k_b\right)$\end{tabular}     & \begin{tabular}[c]{@{}c@{}}3.19517
\\ 0.00020
\\ 0.08176
\\ 6.44885
\\ 0.00176\\ 0.00147
\\ 0.00344
\\ 0.03005\end{tabular}     & \begin{tabular}[c]{@{}c@{}}3.19455
\\ 0.00018
\\ 0.0758
\\ 6.33291
\\ 0.00176
\\ 0.00143
\\ 0.00344
\\ 0.03004
\end{tabular}     & \begin{tabular}[c]{@{}c@{}}3.41329
\\ 0.00014
\\ 0.05847
\\ 6.40325
\\ 0.00179
\\ 0.00135
\\ 0.00341
\\ 0.03
\end{tabular}     & \begin{tabular}[c]{@{}c@{}}3.41013
\\ 0.00014
\\ 0.05635
\\ 6.33552
\\ 0.00179
\\ 0.00133
\\ 0.0034
\\ 0.03
\end{tabular}     & \begin{tabular}[c]{@{}c@{}}3.05791\\ 0.00017
\\ 0.06747
\\ 6.08073
\\ 0.00165\\ 0.00139
\\ 0.00263
\\ 0.02165
\end{tabular}     & \begin{tabular}[c]{@{}c@{}}3.10899
\\ 0.00013
\\ 0.05268
\\ 5.9251\\ 0.00163
\\ 0.00125
\\ 0.0026
\\ 0.02164
\end{tabular}     \\ \hline
\textbf{\begin{tabular}[c]{@{}c@{}}Bump Feature\\ $k_b=0.1$\end{tabular}}           & \begin{tabular}[c]{@{}c@{}}$10^{5}\times\sigma \left(\omega_{\mathrm{b}}\right)$\\ $\sigma \left(\omega_{\mathrm{cdm}}\right)$\\ $\sigma \left( H_0 \right)$\\ $10^{12}\times\sigma \left( A_s \right)$\\ $\sigma \left( n_s \right)$\\ $\sigma \left( \tau_{\mathrm{reio}} \right)$\\ $\sigma \left(B\right)$\\ $\sigma \left(k_b\right)$\end{tabular}     & \begin{tabular}[c]{@{}c@{}}3.47797
\\ 0.00020\\ 0.08133
\\ 5.96099
\\ 0.00159
\\ 0.00148\\ 0.00218
\\ 0.04073
\end{tabular}     & \begin{tabular}[c]{@{}c@{}}3.47754
\\ 0.00018
\\ 0.07556
\\ 5.85590
\\ 0.00158
\\ 0.00145
\\ 0.00218
\\ 0.04073
\end{tabular}     & \begin{tabular}[c]{@{}c@{}}3.67380
\\ 0.00014
\\ 0.05702
\\ 5.73866
\\ 0.00162
\\ 0.00135
\\ 0.00218
\\ 0.04031
\end{tabular}     & \begin{tabular}[c]{@{}c@{}}3.74636
\\ 0.00014
\\ 0.05634
\\ 5.71246
\\ 0.00161
\\ 0.00135
\\ 0.00219
\\ 0.04032
\end{tabular}     & \begin{tabular}[c]{@{}c@{}}3.21756
\\ 0.00017
\\ 0.06456
\\ 5.57940
\\ 0.00151
\\ 0.00139
\\ 0.00168
\\ 0.03261
\end{tabular}     & \begin{tabular}[c]{@{}c@{}}3.26255
\\ 0.00013
\\ 0.05218
\\ 5.29780
\\ 0.00146
\\ 0.00125
\\ 0.00167
\\ 0.03165
\end{tabular}     \\ \hline
\textbf{\begin{tabular}[c]{@{}c@{}}Bump Feature\\ $k_b=0.2$\end{tabular}}           & \begin{tabular}[c]{@{}c@{}}$10^{5}\times\sigma \left(\omega_{\mathrm{b}}\right)$\\ $\sigma \left(\omega_{\mathrm{cdm}}\right)$\\ $\sigma \left( H_0 \right)$\\ $10^{12}\times\sigma \left( A_s \right)$\\ $\sigma \left( n_s \right)$\\ $\sigma \left( \tau_{\mathrm{reio}} \right)$\\ $\sigma \left(B\right)$\\ $\sigma \left(k_b\right)$\end{tabular}     & \begin{tabular}[c]{@{}c@{}}3.92103
\\ 0.00021
\\ 0.08330
\\ 7.20755
\\ 0.00303
\\ 0.00159
\\ 0.00607
\\ 0.10292
\end{tabular}     & \begin{tabular}[c]{@{}c@{}}3.91816
\\ 0.00019
\\ 0.07703
\\ 7.15688
\\ 0.00303
\\ 0.00156
\\ 0.00605
\\ 0.10271
\end{tabular}     & \begin{tabular}[c]{@{}c@{}}3.92871
\\ 0.00016\\ 0.05945
\\ 6.22062\\ 0.00255
\\ 0.00137
\\ 0.00511
\\ 0.10101
\end{tabular}     & \begin{tabular}[c]{@{}c@{}}3.90512
\\ 0.00015
\\ 0.05694
\\ 6.20559
\\ 0.00255
\\ 0.00135
\\ 0.00507
\\ 0.10054
\end{tabular}     & \begin{tabular}[c]{@{}c@{}}3.208
\\ 0.00018\\ 0.06756
\\ 5.69472
\\ 0.00238
\\ 0.00141
\\ 0.00376
\\ 0.08054
\end{tabular}     & \begin{tabular}[c]{@{}c@{}}3.27366
\\ 0.00014
\\ 0.05457
\\ 5.15839
\\ 0.00195
\\ 0.00127
\\ 0.00305
\\ 0.08453
\end{tabular}     \\ \hline
\end{tabular}\par
\bigskip
\parbox{17.9cm}{\captionof{table}{The possible marginalized 1-$\sigma$ constraints of \textbf{LiteBIRD+CMB-S4} experiment on bump feature, for CMB-only data and its combination with DESI, EUCLID(CS), DESI+EUCLID(CS), EUCLID(GC) and EUCLID(CS+GC).}} \label{table:LiteBIRD+CMB-S4-Bump}
\end{minipage}

\setlength{\tabcolsep}{4.8pt} 
\renewcommand{\arraystretch}{1.5} 
\newcolumntype{C}[1]{>{\Centering}m{#1}}
\renewcommand\tabularxcolumn[1]{C{#1}}
\begin{minipage}{\linewidth}
\centering
\captionsetup{font=footnotesize}
\begin{tabular}{|c|c|c|c|c|c|c|c|}
\hline
\textbf{Models}            & \textbf{Parameters}                                                       & \textbf{CMB}                                                              & \textbf{+ DESI}                                                           & \textbf{\begin{tabular}[c]{@{}c@{}}+ Euclid\\ (CS)\end{tabular}}            & \textbf{\begin{tabular}[c]{@{}c@{}}+ DESI +\\  Euclid \\ (CS)\end{tabular}} & \textbf{\begin{tabular}[c]{@{}c@{}}+ Euclid \\ (GC)\end{tabular}}           & \textbf{\begin{tabular}[c]{@{}c@{}}+ Euclid\\ (GC+CS)\end{tabular}}       \\ \hline
\textbf{\begin{tabular}[c]{@{}c@{}}Bump Feature\\ $k_b=0.05$\end{tabular}}           & \begin{tabular}[c]{@{}c@{}}$10^{5}\times\sigma \left(\omega_{\mathrm{b}}\right)$\\ $\sigma \left(\omega_{\mathrm{cdm}}\right)$\\ $\sigma \left( H_0 \right)$\\ $10^{12}\times\sigma \left( A_s \right)$\\ $\sigma \left( n_s \right)$\\ $\sigma \left( \tau_{\mathrm{reio}} \right)$\\ $\sigma \left(B\right)$\\ $\sigma \left(k_b\right)$\end{tabular}     & \begin{tabular}[c]{@{}c@{}}2.86526
\\ 0.00015
\\ 0.06117
\\ 5.46391\\ 0.00162
\\ 0.00121
\\ 0.003
\\ 0.02495
\end{tabular}     & \begin{tabular}[c]{@{}c@{}}2.86355
\\ 0.00015
\\ 0.0585
\\ 5.42623
\\ 0.00162
\\ 0.0012
\\ 0.003\\ 0.02495
\end{tabular}     & \begin{tabular}[c]{@{}c@{}}3.10029
\\ 0.00013
\\ 0.05235
\\ 5.54595
\\ 0.00172
\\ 0.00121
\\ 0.00298
\\ 0.02503
\end{tabular}     & \begin{tabular}[c]{@{}c@{}}3.09829
\\ 0.00012
\\ 0.05083
\\ 5.49934
\\ 0.00172
\\ 0.00119
\\ 0.00298
\\ 0.02503
\end{tabular}     & \begin{tabular}[c]{@{}c@{}}2.75004
\\ 0.00015
\\ 0.05943
\\ 5.57265
\\ 0.00153
\\ 0.00125
\\ 0.00242
\\ 0.01955
\end{tabular}     & \begin{tabular}[c]{@{}c@{}}2.89042
\\ 0.00012
\\ 0.04777
\\ 5.211
\\ 0.00164
\\ 0.00114
\\ 0.00244
\\ 0.01962
\end{tabular}     \\ \hline
\textbf{\begin{tabular}[c]{@{}c@{}}Bump Feature\\ $k_b=0.1$\end{tabular}}           & \begin{tabular}[c]{@{}c@{}}$10^{5}\times\sigma \left(\omega_{\mathrm{b}}\right)$\\ $\sigma \left(\omega_{\mathrm{cdm}}\right)$\\ $\sigma \left( H_0 \right)$\\ $10^{12}\times\sigma \left( A_s \right)$\\ $\sigma \left( n_s \right)$\\ $\sigma \left( \tau_{\mathrm{reio}} \right)$\\ $\sigma \left(B\right)$\\ $\sigma \left(k_b\right)$\end{tabular}     & \begin{tabular}[c]{@{}c@{}}3.13917
\\ 0.00015
\\ 0.06091
\\ 5.05623\\ 0.00146
\\ 0.00123
\\ 0.00184
\\ 0.03365
\end{tabular}     & \begin{tabular}[c]{@{}c@{}}3.13520
\\ 0.00015
\\ 0.05828
\\ 5.01728\\ 0.00146
\\ 0.00122
\\ 0.00183\\ 0.03365
\end{tabular}     & \begin{tabular}[c]{@{}c@{}}3.34342
\\ 0.00013
\\ 0.05122\\ 5.02794
\\ 0.00155
\\ 0.00122
\\ 0.00182
\\ 0.03352
\end{tabular}     & \begin{tabular}[c]{@{}c@{}}3.40337
\\ 0.00012
\\ 0.05075
\\ 5.00390
\\ 0.00154
\\ 0.00122
\\ 0.00183
\\ 0.03352
\end{tabular}     & \begin{tabular}[c]{@{}c@{}}2.90874
\\ 0.00015
\\ 0.05721
\\ 5.08283
\\ 0.00139
\\ 0.00126
\\ 0.00150
\\ 0.02873
\end{tabular}     & \begin{tabular}[c]{@{}c@{}}3.04324
\\ 0.00012
\\ 0.04725
\\ 4.70964\\ 0.00145
\\ 0.00115
\\ 0.00149
\\ 0.02841
\end{tabular}     \\ \hline
\textbf{\begin{tabular}[c]{@{}c@{}}Bump Feature\\ $k_b=0.2$\end{tabular}}           & \begin{tabular}[c]{@{}c@{}}$10^{5}\times\sigma \left(\omega_{\mathrm{b}}\right)$\\ $\sigma \left(\omega_{\mathrm{cdm}}\right)$\\ $\sigma \left( H_0 \right)$\\ $10^{12}\times\sigma \left( A_s \right)$\\ $\sigma \left( n_s \right)$\\ $\sigma \left( \tau_{\mathrm{reio}} \right)$\\ $\sigma \left(B\right)$\\ $\sigma \left(k_b\right)$\end{tabular}     & \begin{tabular}[c]{@{}c@{}}3.62406
\\ 0.00016
\\ 0.06244
\\ 6.12857
\\ 0.00282
\\ 0.00132
\\ 0.00545
\\ 0.09118
\end{tabular}     & \begin{tabular}[c]{@{}c@{}}3.60876
\\ 0.00016
\\ 0.05949
\\ 6.12596\\ 0.00282
\\ 0.00132\\ 0.00541
\\ 0.09080
\end{tabular}     & \begin{tabular}[c]{@{}c@{}}3.46661
\\ 0.00014
\\ 0.05322
\\ 5.75584
\\ 0.00265
\\ 0.00128
\\ 0.00477
\\ 0.08737
\end{tabular}     & \begin{tabular}[c]{@{}c@{}}3.45318
\\ 0.00013\\ 0.05143
\\ 5.74774\\ 0.00265
\\ 0.00127
\\ 0.00474
\\ 0.08706
\end{tabular}     & \begin{tabular}[c]{@{}c@{}}2.90238
\\ 0.00016
\\ 0.05964
\\ 5.28267
\\ 0.00225
\\ 0.00128
\\ 0.00352
\\ 0.07406
\end{tabular}     & \begin{tabular}[c]{@{}c@{}}3.08646
\\ 0.00013
\\ 0.04859\\ 4.88127
\\ 0.00235
\\ 0.00121
\\ 0.00325
\\ 0.07909
\end{tabular}     \\ \hline
\end{tabular}\par
\bigskip
\parbox{17.9cm}{\captionof{table}{The possible marginalized 1-$\sigma$ constraints of \textbf{CORE-M5+CMB-S4} experiment on bump feature, for CMB-only data and its combination with DESI, EUCLID(CS), DESI+EUCLID(CS), EUCLID(GC) and EUCLID(CS+GC).}} \label{table:CORE-M5+CMB-S4-Bump}
\end{minipage}

\setlength{\tabcolsep}{4.85pt} 
\renewcommand{\arraystretch}{1.5} 
\newcolumntype{C}[1]{>{\Centering}m{#1}}
\renewcommand\tabularxcolumn[1]{C{#1}}
\begin{minipage}{\linewidth}
\centering
\captionsetup{font=footnotesize}
\begin{tabular}{|c|c|c|c|c|c|c|c|}
\hline
\textbf{Models}            & \textbf{Parameters}                                                       & \textbf{CMB}                                                              & \textbf{+ DESI}                                                           & \textbf{\begin{tabular}[c]{@{}c@{}}+ Euclid\\ (CS)\end{tabular}}            & \textbf{\begin{tabular}[c]{@{}c@{}}+ DESI +\\  Euclid \\ (CS)\end{tabular}} & \textbf{\begin{tabular}[c]{@{}c@{}}+ Euclid \\ (GC)\end{tabular}}           & \textbf{\begin{tabular}[c]{@{}c@{}}+ Euclid\\ (GC+CS)\end{tabular}}       \\ \hline
\textbf{\begin{tabular}[c]{@{}c@{}}Sharp Feature\\ $k_s=0.004$\end{tabular}}      & \begin{tabular}[c]{@{}c@{}}$10^{5}\times\sigma \left(\omega_{\mathrm{b}}\right)$\\ $\sigma \left(\omega_{\mathrm{cdm}}\right)$\\ $\sigma \left( H_0 \right)$\\ $10^{12}\times\sigma \left( A_s \right)$\\ $\sigma \left( n_s \right)$\\ $\sigma \left( \tau_{\mathrm{reio}} \right)$\\ $\sigma \left(S\right)$\\ $10^{5}\times\sigma \left(k_s\right)$\\ $\sigma \left( \phi_s \right)$\end{tabular} & \begin{tabular}[c]{@{}c@{}}3.25844
\\ 0.00025\\ 0.09981
\\ 8.34456\\ 0.00165\\ 0.00215\\ 0.00338\\ 1.01404
\\ 0.16058
\end{tabular} & \begin{tabular}[c]{@{}c@{}}3.25607
\\ 0.00022\\ 0.08914\\ 8.04908\\ 0.00163\\ 0.00205\\ 0.00338\\ 1.01\\ 0.16056
\end{tabular} & \begin{tabular}[c]{@{}c@{}}3.31820
\\ 0.00021\\ 0.08107

\\ 7.40434
\\ 0.00165\\ 0.00186\\ 0.00335\\ 0.66992
\\ 0.14301
\end{tabular} & \begin{tabular}[c]{@{}c@{}}3.31510
\\ 0.00019\\ 0.07490\\ 7.30292
\\ 0.00163\\ 0.00182\\ 0.00335\\ 0.65711
\\ 0.14277
\end{tabular} & \begin{tabular}[c]{@{}c@{}}3.09727
\\ 0.00018\\ 0.06736
\\ 6.96812
\\ 0.00150\\ 0.00184\\ 0.00113\\ 0.28042
\\ 0.06215
\end{tabular} & \begin{tabular}[c]{@{}c@{}}2.99878
\\ 0.00015\\ 0.05579
\\ 6.01931
\\ 0.00145\\ 0.00158\\ 0.00112\\ 0.29031
\\ 0.06426
\end{tabular} \\ \hline
\textbf{\begin{tabular}[c]{@{}c@{}}Sharp Feature\\ $k_s=0.03$\end{tabular}}          & \begin{tabular}[c]{@{}c@{}}$10^{5}\times\sigma \left(\omega_{\mathrm{b}}\right)$\\ $\sigma \left(\omega_{\mathrm{cdm}}\right)$\\ $\sigma \left( H_0 \right)$\\ $10^{12}\times\sigma \left( A_s \right)$\\ $\sigma \left( n_s \right)$\\ $\sigma \left( \tau_{\mathrm{reio}} \right)$\\ $\sigma \left(S\right)$\\ $10^{4}\times\sigma \left(k_s\right)$\\ $\sigma \left( \phi_s \right)$\end{tabular} & \begin{tabular}[c]{@{}c@{}}3.37069\\ 0.00026\\ 0.10791
\\ 8.91265\\ 0.00163\\ 0.00232\\ 0.00122\\ 3.60642\\ 0.10979\end{tabular} & \begin{tabular}[c]{@{}c@{}}3.35953\\ 0.00023\\ 0.09506\\ 8.47673\\ 0.00162\\ 0.00219\\ 0.00122\\ 3.57746\\ 0.10875\end{tabular} & \begin{tabular}[c]{@{}c@{}}3.24471\\ 0.00018\\ 0.06958\\ 7.24253\\ 0.00161\\ 0.00192\\ 0.00121\\ 3.26033
\\ 0.10016
\end{tabular} & \begin{tabular}[c]{@{}c@{}}3.2379\\ 0.00017\\ 0.06555\\ 7.07350\\ 0.00160
\\ 0.00186
\\ 0.00121\\ 3.25291
\\ 0.09989\end{tabular} & \begin{tabular}[c]{@{}c@{}}3.16765\\ 0.00019\\ 0.07534
\\ 7.31358\\ 0.00152\\ 0.00195
\\ 0.00087\\ 1.08325\\ 0.04327\end{tabular} & \begin{tabular}[c]{@{}c@{}}3.00661\\ 0.00016\\ 0.05995\\ 6.32688\\ 0.00141
\\ 0.00174
\\ 0.00086\\ 0.94123\\ 0.0401\end{tabular} \\ \hline
\textbf{\begin{tabular}[c]{@{}c@{}}Sharp Feature\\ $k_s=0.1$\end{tabular}}      & \begin{tabular}[c]{@{}c@{}}$10^{5}\times\sigma \left(\omega_{\mathrm{b}}\right)$\\ $\sigma \left(\omega_{\mathrm{cdm}}\right)$\\ $\sigma \left( H_0 \right)$\\ $10^{11}\times\sigma \left( A_s \right)$\\ $\sigma \left( n_s \right)$\\ $\sigma \left( \tau_{\mathrm{reio}} \right)$\\ $\sigma \left(S\right)$\\ $\sigma \left(k_s\right)$\\ $\sigma \left( \phi_s \right)$\end{tabular} & \begin{tabular}[c]{@{}c@{}}4.04315
\\ 0.00034
\\ 0.13053
\\ 1.25487
\\ 0.00762
\\ 0.00292
\\ 0.00540
\\ 0.01342
\\ 0.45201\end{tabular} & \begin{tabular}[c]{@{}c@{}}4.04194
\\ 0.00028
\\ 0.10805
\\ 1.10275
\\ 0.00719
\\ 0.00292
\\ 0.00505
\\ 0.01242
\\ 0.41866
\end{tabular} & \begin{tabular}[c]{@{}c@{}}4.36755
\\ 0.00023
\\ 0.08253
\\ 1.06909
\\ 0.00785
\\ 0.00288
\\ 0.00574\\ 0.01394
\\ 0.46657
\end{tabular} & \begin{tabular}[c]{@{}c@{}}4.31181
\\ 0.00021
\\ 0.07537
\\ 0.99819\\ 0.00757
\\ 0.00287\\ 0.00550
\\ 0.01333
\\ 0.44609
\end{tabular} & \begin{tabular}[c]{@{}c@{}}3.39542
\\ 0.00021\\ 0.08073\\ 0.85856
\\ 0.00287
\\ 0.00220
\\ 0.00154
\\ 0.00164
\\ 0.08323
\end{tabular} & \begin{tabular}[c]{@{}c@{}}3.36552
\\ 0.00015
\\ 0.06004
\\ 0.68853
\\ 0.00282
\\ 0.00163
\\ 0.00148
\\ 0.00154
\\ 0.08083
\end{tabular} \\ \hline
\end{tabular}\par
\bigskip
\parbox{17.9cm}{\captionof{table}{The possible marginalized 1-$\sigma$ constraints of \textbf{CMB-S4} experiment on sharp feature signal, for CMB-only data and its combination with DESI, EUCLID(CS), DESI+EUCLID(CS), EUCLID(GC) and EUCLID(CS+GC).}} \label{table:CMB-S4-Sharp} 
\end{minipage}

\setlength{\tabcolsep}{4.85pt} 
\renewcommand{\arraystretch}{1.5} 
\newcolumntype{C}[1]{>{\Centering}m{#1}}
\renewcommand\tabularxcolumn[1]{C{#1}}
\begin{minipage}{\linewidth}
\centering
\captionsetup{font=footnotesize}
\begin{tabular}{|c|c|c|c|c|c|c|c|}
\hline
\textbf{Models}            & \textbf{Parameters}                                                       & \textbf{CMB}                                                              & \textbf{+ DESI}                                                           & \textbf{\begin{tabular}[c]{@{}c@{}}+ Euclid\\ (CS)\end{tabular}}            & \textbf{\begin{tabular}[c]{@{}c@{}}+ DESI +\\  Euclid \\ (CS)\end{tabular}} & \textbf{\begin{tabular}[c]{@{}c@{}}+ Euclid \\ (GC)\end{tabular}}           & \textbf{\begin{tabular}[c]{@{}c@{}}+ Euclid\\ (GC+CS)\end{tabular}}       \\ \hline
\textbf{\begin{tabular}[c]{@{}c@{}}Sharp Feature\\ $k_s=0.004$\end{tabular}}      & \begin{tabular}[c]{@{}c@{}}$10^{5}\times\sigma \left(\omega_{\mathrm{b}}\right)$\\ $\sigma \left(\omega_{\mathrm{cdm}}\right)$\\ $\sigma \left( H_0 \right)$\\ $10^{12}\times\sigma \left( A_s \right)$\\ $\sigma \left( n_s \right)$\\ $\sigma \left( \tau_{\mathrm{reio}} \right)$\\ $\sigma \left(S\right)$\\ $10^{5}\times\sigma \left(k_s\right)$\\ $\sigma \left( \phi_s \right)$\end{tabular} & \begin{tabular}[c]{@{}c@{}}3.81801\\ 0.00021\\ 0.08505\\ 5.84765\\ 0.00143\\ 0.00136\\ 0.00273\\ 1.28020\\ 0.15184\end{tabular} & \begin{tabular}[c]{@{}c@{}}3.81700\\ 0.00019
\\ 0.07876\\ 5.84563\\ 0.00142\\ 0.00136\\ 0.00273\\ 1.26148\\ 0.15098\end{tabular} & \begin{tabular}[c]{@{}c@{}}3.73573\\ 0.00020\\ 0.08264\\ 5.61134\\ 0.00139\\ 0.00133\\ 0.00272\\ 0.75412\\ 0.12040\end{tabular} & \begin{tabular}[c]{@{}c@{}}3.73459\\ 0.00019\\ 0.07688\\ 5.60742\\ 0.00138\\ 0.00132\\ 0.00272\\ 0.72809\\ 0.11952\end{tabular} & \begin{tabular}[c]{@{}c@{}}3.39047\\ 0.00016\\ 0.06142\\ 5.27304\\ 0.00127\\ 0.00129\\ 0.00110\\ 0.28227\\ 0.06036\end{tabular} & \begin{tabular}[c]{@{}c@{}}3.45673\\ 0.00013\\ 0.05315\\ 4.93090\\ 0.00125\\ 0.00123\\ 0.00109
\\ 0.28139\\ 0.06204\end{tabular} \\ \hline
\textbf{\begin{tabular}[c]{@{}c@{}}Sharp Feature\\ $k_s=0.03$\end{tabular}}          & \begin{tabular}[c]{@{}c@{}}$10^{5}\times\sigma \left(\omega_{\mathrm{b}}\right)$\\ $\sigma \left(\omega_{\mathrm{cdm}}\right)$\\ $\sigma \left( H_0 \right)$\\ $10^{12}\times\sigma \left( A_s \right)$\\ $\sigma \left( n_s \right)$\\ $\sigma \left( \tau_{\mathrm{reio}} \right)$\\ $\sigma \left(S\right)$\\ $10^{4}\times\sigma \left(k_s\right)$\\ $\sigma \left( \phi_s \right)$\end{tabular} & \begin{tabular}[c]{@{}c@{}}3.81428
\\ 0.00019
\\ 0.08113
\\ 5.50894
\\ 0.00140
\\ 0.00133
\\ 0.00107\\ 3.56345
\\ 0.09527
\end{tabular} & \begin{tabular}[c]{@{}c@{}}3.81235\\ 0.00018
\\ 0.07584
\\ 5.49205
\\ 0.00139\\ 0.00132
\\ 0.00107\\ 3.53929
\\ 0.09458\end{tabular} & \begin{tabular}[c]{@{}c@{}}3.61162
\\ 0.00014
\\ 0.06187\\ 5.31421\\ 0.00135
\\ 0.00131
\\ 0.00107
\\ 3.23437
\\ 0.08763\end{tabular} & \begin{tabular}[c]{@{}c@{}}3.61105
\\ 0.00014
\\ 0.05958
\\ 5.27323\\ 0.00135
\\ 0.0013
\\ 0.00107
\\ 3.22066
\\ 0.08727\end{tabular} & \begin{tabular}[c]{@{}c@{}}3.48791
\\ 0.00017\\ 0.06817
\\ 5.3361\\ 0.00128\\ 0.00131\\ 0.00081\\ 1.04361
\\ 0.03936
\end{tabular} & \begin{tabular}[c]{@{}c@{}}3.32847
\\ 0.00013\\ 0.05449
\\ 4.88688\\ 0.00121
\\ 0.00124
\\ 0.00081
\\ 0.97734\\ 0.03796\end{tabular} \\ \hline
\textbf{\begin{tabular}[c]{@{}c@{}}Sharp Feature\\ $k_s=0.1$\end{tabular}}      & \begin{tabular}[c]{@{}c@{}}$10^{5}\times\sigma \left(\omega_{\mathrm{b}}\right)$\\ $\sigma \left(\omega_{\mathrm{cdm}}\right)$\\ $\sigma \left( H_0 \right)$\\ $10^{11}\times\sigma \left( A_s \right)$\\ $\sigma \left( n_s \right)$\\ $\sigma \left( \tau_{\mathrm{reio}} \right)$\\ $\sigma \left(S\right)$\\ $\sigma \left(k_s\right)$\\ $\sigma \left( \phi_s \right)$\end{tabular} & \begin{tabular}[c]{@{}c@{}}4.56014
\\ 0.00028
\\ 0.10424
\\ 1.01585
\\ 0.00598
\\ 0.00173
\\ 0.00467\\ 0.01271
\\ 0.39791
\end{tabular} & \begin{tabular}[c]{@{}c@{}}4.53441
\\ 0.00025
\\ 0.09154
\\ 0.94412
\\ 0.00563
\\ 0.00171
\\ 0.00434
\\ 0.01177
\\ 0.36892
\end{tabular} & \begin{tabular}[c]{@{}c@{}}4.52611
\\ 0.00022
\\ 0.079
\\ 0.97074
\\ 0.00557
\\ 0.00166
\\ 0.00433
\\ 0.01182\\ 0.36805
\end{tabular} & \begin{tabular}[c]{@{}c@{}}4.49704
\\ 0.0002
\\ 0.07326
\\ 0.91681
\\ 0.00537
\\ 0.00165
\\ 0.00413
\\ 0.01125
\\ 0.35062\end{tabular} & \begin{tabular}[c]{@{}c@{}}3.69775
\\ 0.00017
\\ 0.07091
\\ 0.59412\\ 0.00254
\\ 0.00136
\\ 0.00134
\\ 0.00156
\\ 0.08079
\end{tabular} & \begin{tabular}[c]{@{}c@{}}3.68555
\\ 0.00013
\\ 0.05441
\\ 0.56169
\\ 0.00250
\\ 0.00126
\\ 0.00131
\\ 0.00151
\\ 0.07861
\end{tabular} \\ \hline
\end{tabular}\par
\bigskip
\parbox{17.9cm}{\captionof{table}{The possible marginalized 1-$\sigma$ constraints of \textbf{CORE-M5} experiment on sharp feature signal, for CMB-only data and its combination with DESI, EUCLID(CS), DESI+EUCLID(CS), EUCLID(GC) and EUCLID(CS+GC).}} \label{table:CORE-M5-Sharp} 
\end{minipage}

\setlength{\tabcolsep}{4.85pt} 
\renewcommand{\arraystretch}{1.5} 
\newcolumntype{C}[1]{>{\Centering}m{#1}}
\renewcommand\tabularxcolumn[1]{C{#1}}
\begin{minipage}{\linewidth}
\centering
\captionsetup{font=footnotesize}
\begin{tabular}{|c|c|c|c|c|c|c|c|}
\hline
\textbf{Models}            & \textbf{Parameters}                                                       & \textbf{CMB}                                                              & \textbf{+ DESI}                                                           & \textbf{\begin{tabular}[c]{@{}c@{}}+ Euclid\\ (CS)\end{tabular}}            & \textbf{\begin{tabular}[c]{@{}c@{}}+ DESI +\\  Euclid \\ (CS)\end{tabular}} & \textbf{\begin{tabular}[c]{@{}c@{}}+ Euclid \\ (GC)\end{tabular}}           & \textbf{\begin{tabular}[c]{@{}c@{}}+ Euclid\\ (GC+CS)\end{tabular}}       \\ \hline
\textbf{\begin{tabular}[c]{@{}c@{}}Sharp Feature\\ $k_s=0.004$\end{tabular}}      & \begin{tabular}[c]{@{}c@{}}$10^{5}\times\sigma \left(\omega_{\mathrm{b}}\right)$\\ $\sigma \left(\omega_{\mathrm{cdm}}\right)$\\ $\sigma \left( H_0 \right)$\\ $10^{12}\times\sigma \left( A_s \right)$\\ $\sigma \left( n_s \right)$\\ $\sigma \left( \tau_{\mathrm{reio}} \right)$\\ $\sigma \left(S\right)$\\ $10^{5}\times\sigma \left(k_s\right)$\\ $\sigma \left( \phi_s \right)$\end{tabular} & \begin{tabular}[c]{@{}c@{}}2.77413\\ 0.00016\\ 0.06173\\ 4.92407\\ 0.00130\\ 0.00118\\ 0.00259\\ 0.83541\\ 0.12443\end{tabular} & \begin{tabular}[c]{@{}c@{}}2.77089\\ 0.00015\\ 0.05886\\ 4.90458\\ 0.00129\\ 0.00117\\ 0.00259\\ 0.83189\\ 0.12433\end{tabular} & \begin{tabular}[c]{@{}c@{}}2.78245\\ 0.00016\\ 0.06248\\ 4.89717\\ 0.00128\\ 0.00118\\ 0.00257\\ 0.58506\\ 0.11153\end{tabular} & \begin{tabular}[c]{@{}c@{}}2.78011\\ 0.00015\\ 0.05955\\ 4.87677\\ 0.00127\\ 0.00117\\ 0.00257\\ 0.57660\\ 0.11129\end{tabular} & \begin{tabular}[c]{@{}c@{}}2.66150\\ 0.00014\\ 0.05372\\ 4.66605\\ 0.00119\\ 0.00118\\ 0.00109\\ 0.26597\\ 0.05713\end{tabular} & \begin{tabular}[c]{@{}c@{}}2.57923\\ 0.00012\\ 0.04618\\ 4.30070\\ 0.00115\\ 0.00108\\ 0.00108\\ 0.28106\\ 0.05993\end{tabular} \\ \hline
\textbf{\begin{tabular}[c]{@{}c@{}}Sharp Feature\\ $k_s=0.03$\end{tabular}}          & \begin{tabular}[c]{@{}c@{}}$10^{5}\times\sigma \left(\omega_{\mathrm{b}}\right)$\\ $\sigma \left(\omega_{\mathrm{cdm}}\right)$\\ $\sigma \left( H_0 \right)$\\ $10^{12}\times\sigma \left( A_s \right)$\\ $\sigma \left( n_s \right)$\\ $\sigma \left( \tau_{\mathrm{reio}} \right)$\\ $\sigma \left(S\right)$\\ $10^{4}\times\sigma \left(k_s\right)$\\ $\sigma \left( \phi_s \right)$\end{tabular} & \begin{tabular}[c]{@{}c@{}}2.79358
\\ 0.00016
\\ 0.06534\\ 4.95241\\ 0.00127\\ 0.00121\\ 0.00095\\ 2.86997\\ 0.08514\end{tabular} & \begin{tabular}[c]{@{}c@{}}2.79267\\ 0.00016\\ 0.06207
\\ 4.92948\\ 0.00126
\\ 0.0012\\ 0.00095
\\ 2.84824\\ 0.08449
\end{tabular} & \begin{tabular}[c]{@{}c@{}}2.73951
\\ 0.00014\\ 0.05434\\ 4.87358\\ 0.00121
\\ 0.00114
\\ 0.00095\\ 2.53318
\\ 0.07636\end{tabular} & \begin{tabular}[c]{@{}c@{}}2.73774
\\ 0.00013
\\ 0.05248
\\ 4.84057
\\ 0.00121
\\ 0.00113
\\ 0.00095
\\ 2.52127
\\ 0.07601
\end{tabular} & \begin{tabular}[c]{@{}c@{}}2.70770
\\ 0.00015
\\ 0.05855
\\ 4.78981\\ 0.0012
\\ 0.00122
\\ 0.00075
\\ 1.04045
\\ 0.03917
\end{tabular} & \begin{tabular}[c]{@{}c@{}}2.54532
\\ 0.00012
\\ 0.04771\\ 4.29986\\ 0.0011\\ 0.00114\\ 0.00075
\\ 0.87472\\ 0.03553
\end{tabular} \\ \hline
\textbf{\begin{tabular}[c]{@{}c@{}}Sharp Feature\\ $k_s=0.1$\end{tabular}}      & \begin{tabular}[c]{@{}c@{}}$10^{5}\times\sigma \left(\omega_{\mathrm{b}}\right)$\\ $\sigma \left(\omega_{\mathrm{cdm}}\right)$\\ $\sigma \left( H_0 \right)$\\ $10^{11}\times\sigma \left( A_s \right)$\\ $\sigma \left( n_s \right)$\\ $\sigma \left( \tau_{\mathrm{reio}} \right)$\\ $\sigma \left(S\right)$\\ $\sigma \left(k_s\right)$\\ $\sigma \left( \phi_s \right)$\end{tabular} & \begin{tabular}[c]{@{}c@{}}3.61135
\\ 0.00025
\\ 0.08811
\\ 0.90160\\ 0.00579\\ 0.00165
\\ 0.00434
\\ 0.01066
\\ 0.35759
\end{tabular} & \begin{tabular}[c]{@{}c@{}}3.54495
\\ 0.00023\\ 0.07884
\\ 0.83735
\\ 0.00547
\\ 0.00163\\ 0.00404
\\ 0.00992\\ 0.33322
\end{tabular} & \begin{tabular}[c]{@{}c@{}}3.54488
\\ 0.00020
\\ 0.06983
\\ 0.85608
\\ 0.00543
\\ 0.00160
\\ 0.00401
\\ 0.00994
\\ 0.3323
\end{tabular} & \begin{tabular}[c]{@{}c@{}}3.50534\\ 0.00019\\ 0.06515
\\ 0.81327
\\ 0.00524
\\ 0.00159
\\ 0.00384
\\ 0.00952\\ 0.31837\end{tabular} & \begin{tabular}[c]{@{}c@{}}2.93521\\ 0.00016
\\ 0.06079
\\ 0.55952
\\ 0.00233
\\ 0.00132
\\ 0.00135\\ 0.00165
\\ 0.07562
\end{tabular} & \begin{tabular}[c]{@{}c@{}}2.8282
\\ 0.00012
\\ 0.04774
\\ 0.50415\\ 0.00226
\\ 0.00112
\\ 0.00130
\\ 0.00153
\\ 0.07237
\end{tabular} \\ \hline
\end{tabular}\par
\bigskip
\parbox{17.9cm}{\captionof{table}{The possible marginalized 1-$\sigma$ constraints of \textbf{PICO} experiment on sharp feature signal, for CMB-only data and its combination with DESI, EUCLID(CS), DESI+EUCLID(CS), EUCLID(GC) and EUCLID(CS+GC).}} \label{table:PICO-Sharp}
\end{minipage}

\setlength{\tabcolsep}{4.85pt} 
\renewcommand{\arraystretch}{1.5} 
\newcolumntype{C}[1]{>{\Centering}m{#1}}
\renewcommand\tabularxcolumn[1]{C{#1}}
\begin{minipage}{\linewidth}
\centering
\captionsetup{font=footnotesize}
\begin{tabular}{|c|c|c|c|c|c|c|c|}
\hline
\textbf{Models}            & \textbf{Parameters}                                                       & \textbf{CMB}                                                              & \textbf{+ DESI}                                                           & \textbf{\begin{tabular}[c]{@{}c@{}}+ Euclid\\ (CS)\end{tabular}}            & \textbf{\begin{tabular}[c]{@{}c@{}}+ DESI +\\  Euclid \\ (CS)\end{tabular}} & \textbf{\begin{tabular}[c]{@{}c@{}}+ Euclid \\ (GC)\end{tabular}}           & \textbf{\begin{tabular}[c]{@{}c@{}}+ Euclid\\ (GC+CS)\end{tabular}}       \\ \hline
\textbf{\begin{tabular}[c]{@{}c@{}}Sharp Feature\\ $k_s=0.004$\end{tabular}}      & \begin{tabular}[c]{@{}c@{}}$10^{5}\times\sigma \left(\omega_{\mathrm{b}}\right)$\\ $\sigma \left(\omega_{\mathrm{cdm}}\right)$\\ $\sigma \left( H_0 \right)$\\ $10^{12}\times\sigma \left( A_s \right)$\\ $\sigma \left( n_s \right)$\\ $\sigma \left( \tau_{\mathrm{reio}} \right)$\\ $\sigma \left(S\right)$\\ $10^{5}\times\sigma \left(k_s\right)$\\ $\sigma \left( \phi_s \right)$\end{tabular} & \begin{tabular}[c]{@{}c@{}}3.28069\\ 0.00021\\ 0.08380\\ 7.44015\\ 0.00158\\ 0.00186\\ 0.00325\\ 0.77148\\ 0.13750\end{tabular} & \begin{tabular}[c]{@{}c@{}}3.27853\\ 0.00020\\ 0.07703\\ 7.31262\\ 0.00156\\ 0.00181\\ 0.00325\\ 0.76961\\ 0.13750\end{tabular} & \begin{tabular}[c]{@{}c@{}}3.22226\\ 0.00020\\ 0.07808\\ 6.79185\\ 0.00155\\ 0.00169\\ 0.00322\\ 0.66129\\ 0.13626\end{tabular} & \begin{tabular}[c]{@{}c@{}}3.21954\\ 0.00018\\ 0.07255\\ 6.72621\\ 0.00154\\ 0.00166\\ 0.00322\\ 0.64794\\ 0.13598\end{tabular} & \begin{tabular}[c]{@{}c@{}}3.02314\\ 0.00017
\\ 0.06494
\\ 6.43\\ 0.00142\\ 0.00167\\ 0.00112\\ 0.27812\\ 0.06135\end{tabular} & \begin{tabular}[c]{@{}c@{}}2.99176\\ 0.00015\\ 0.05641\\ 5.89417\\ 0.00137\\ 0.00153
\\ 0.00112\\ 0.27742\\ 0.06265\end{tabular} \\ \hline
\textbf{\begin{tabular}[c]{@{}c@{}}Sharp Feature\\ $k_s=0.03$\end{tabular}}          & \begin{tabular}[c]{@{}c@{}}$10^{5}\times\sigma \left(\omega_{\mathrm{b}}\right)$\\ $\sigma \left(\omega_{\mathrm{cdm}}\right)$\\ $\sigma \left( H_0 \right)$\\ $10^{12}\times\sigma \left( A_s \right)$\\ $\sigma \left( n_s \right)$\\ $\sigma \left( \tau_{\mathrm{reio}} \right)$\\ $\sigma \left(S\right)$\\ $10^{4}\times\sigma \left(k_s\right)$\\ $\sigma \left( \phi_s \right)$\end{tabular} & \begin{tabular}[c]{@{}c@{}}3.2785\\ 0.00023\\ 0.08902\\ 7.75451
\\ 0.00156
\\ 0.00196\\ 0.00119\\ 4.33292\\ 0.12592\end{tabular} & \begin{tabular}[c]{@{}c@{}}3.27777\\ 0.00021\\ 0.08107
\\ 7.67021\\ 0.00155
\\ 0.00193
\\ 0.00119\\ 4.27788\\ 0.12424\end{tabular} & \begin{tabular}[c]{@{}c@{}}3.15314\\ 0.00017\\ 0.0663
\\ 6.60613\\ 0.00151\\ 0.00172
\\ 0.00117\\ 3.15005\\ 0.09489\end{tabular} & \begin{tabular}[c]{@{}c@{}}3.14828\\ 0.00016\\ 0.06286
\\ 6.48739\\ 0.00150
\\ 0.00168
\\ 0.00117\\ 3.14091\\ 0.09459
\end{tabular} & \begin{tabular}[c]{@{}c@{}}3.08974\\ 0.00018\\ 0.07215
\\ 6.67738\\ 0.00143\\ 0.00174
\\ 0.00086\\ 1.07205\\ 0.04243\end{tabular} & \begin{tabular}[c]{@{}c@{}}3.00343
\\ 0.00015\\ 0.05943\\ 6.05082
\\ 0.00138
\\ 0.00160
\\ 0.00085
\\ 1.02138
\\ 0.04127
\end{tabular} \\ \hline
\textbf{\begin{tabular}[c]{@{}c@{}}Sharp Feature\\ $k_s=0.1$\end{tabular}}      & \begin{tabular}[c]{@{}c@{}}$10^{5}\times\sigma \left(\omega_{\mathrm{b}}\right)$\\ $\sigma \left(\omega_{\mathrm{cdm}}\right)$\\ $\sigma \left( H_0 \right)$\\ $10^{11}\times\sigma \left( A_s \right)$\\ $\sigma \left( n_s \right)$\\ $\sigma \left( \tau_{\mathrm{reio}} \right)$\\ $\sigma \left(S\right)$\\ $\sigma \left(k_s\right)$\\ $\sigma \left( \phi_s \right)$\end{tabular} & \begin{tabular}[c]{@{}c@{}}3.94233
\\0.00032
\\ 0.12424\\ 1.15904\\ 0.00674\\ 0.00245
\\ 0.00484
\\ 0.01234\\ 0.41138
\end{tabular} & \begin{tabular}[c]{@{}c@{}}3.94056
\\ 0.00027
\\ 0.10437
\\ 1.03344
\\ 0.00636
\\ 0.00245
\\ 0.00451
\\ 0.01140
\\ 0.38048
\end{tabular} & \begin{tabular}[c]{@{}c@{}}4.23910
\\ 0.00023
\\ 0.08011
\\ 1.00477
\\ 0.00686
\\ 0.00244
\\ 0.00508
\\ 0.01259
\\ 0.41765\end{tabular} & \begin{tabular}[c]{@{}c@{}}4.18989
\\ 0.00021\\ 0.07355
\\ 0.94332
\\ 0.00662
\\ 0.00243
\\ 0.00486
\\ 0.01203\\ 0.39938
\end{tabular} & \begin{tabular}[c]{@{}c@{}}3.33833
\\ 0.00020
\\ 0.07603
\\ 0.77239
\\ 0.00271
\\ 0.00192\\ 0.00148\\ 0.00163
\\ 0.08173
\end{tabular} & \begin{tabular}[c]{@{}c@{}}3.27191
\\ 0.00015
\\ 0.06026
\\ 0.68979
\\ 0.00263
\\ 0.00165
\\ 0.00143\\ 0.00152
\\ 0.07824\end{tabular} \\ \hline
\end{tabular}\par
\bigskip
\parbox{17.9cm}{\captionof{table}{The possible marginalized 1-$\sigma$ constraints of \textbf{Planck+CMB-S4} experiment on sharp feature signal, for CMB-only data and its combination with DESI, EUCLID(CS), DESI+EUCLID(CS), EUCLID(GC) and EUCLID(CS+GC).}} \label{table:Planck+CMB-S4-Sharp}
\end{minipage}

\setlength{\tabcolsep}{4.85pt} 
\renewcommand{\arraystretch}{1.5} 
\newcolumntype{C}[1]{>{\Centering}m{#1}}
\renewcommand\tabularxcolumn[1]{C{#1}}
\begin{minipage}{\linewidth}
\centering
\captionsetup{font=footnotesize}
\begin{tabular}{|c|c|c|c|c|c|c|c|}
\hline
\textbf{Models}            & \textbf{Parameters}                                                       & \textbf{CMB}                                                              & \textbf{+ DESI}                                                           & \textbf{\begin{tabular}[c]{@{}c@{}}+ Euclid\\ (CS)\end{tabular}}            & \textbf{\begin{tabular}[c]{@{}c@{}}+ DESI +\\  Euclid \\ (CS)\end{tabular}} & \textbf{\begin{tabular}[c]{@{}c@{}}+ Euclid \\ (GC)\end{tabular}}           & \textbf{\begin{tabular}[c]{@{}c@{}}+ Euclid\\ (GC+CS)\end{tabular}}       \\ \hline
\textbf{\begin{tabular}[c]{@{}c@{}}Sharp Feature\\ $k_s=0.004$\end{tabular}}      & \begin{tabular}[c]{@{}c@{}}$10^{5}\times\sigma \left(\omega_{\mathrm{b}}\right)$\\ $\sigma \left(\omega_{\mathrm{cdm}}\right)$\\ $\sigma \left( H_0 \right)$\\ $10^{12}\times\sigma \left( A_s \right)$\\ $\sigma \left( n_s \right)$\\ $\sigma \left( \tau_{\mathrm{reio}} \right)$\\ $\sigma \left(S\right)$\\ $10^{5}\times\sigma \left(k_s\right)$\\ $\sigma \left( \phi_s \right)$\end{tabular} & \begin{tabular}[c]{@{}c@{}}3.07108\\ 0.00021\\ 0.08619\\ 5.81210\\ 0.00144\\ 0.00145\\ 0.00318\\ 0.97524\\ 0.14548\end{tabular} & \begin{tabular}[c]{@{}c@{}}3.07060\\ 0.00020\\ 0.07908\\ 5.73707\\ 0.00142\\ 0.00142\\ 0.00318\\ 0.96865\\ 0.14532\end{tabular} & \begin{tabular}[c]{@{}c@{}}3.14015\\ 0.00019\\ 0.07409\\ 5.52198\\ 0.00145\\ 0.00135\\ 0.00315\\ 0.64484
\\ 0.13086\end{tabular} & \begin{tabular}[c]{@{}c@{}}3.13708\\ 0.00018
\\ 0.06932\\ 5.49370\\ 0.00143
\\ 0.00133\\ 0.00315\\ 0.63032\\ 0.13052\end{tabular} & \begin{tabular}[c]{@{}c@{}}2.97458\\ 0.00016\\ 0.06172\\ 5.29199\\ 0.00133\\ 0.00134\\ 0.00112\\ 0.27537\\ 0.06058\end{tabular} & \begin{tabular}[c]{@{}c@{}}2.88853\\ 0.00013\\ 0.05151\\ 4.85299\\ 0.00128\\ 0.00123\\ 0.00112\\ 0.28328\\ 0.06264\end{tabular} \\ \hline
\textbf{\begin{tabular}[c]{@{}c@{}}Sharp Feature\\ $k_s=0.03$\end{tabular}}          & \begin{tabular}[c]{@{}c@{}}$10^{5}\times\sigma \left(\omega_{\mathrm{b}}\right)$\\ $\sigma \left(\omega_{\mathrm{cdm}}\right)$\\ $\sigma \left( H_0 \right)$\\ $10^{12}\times\sigma \left( A_s \right)$\\ $\sigma \left( n_s \right)$\\ $\sigma \left( \tau_{\mathrm{reio}} \right)$\\ $\sigma \left(S\right)$\\ $10^{4}\times\sigma \left(k_s\right)$\\ $\sigma \left( \phi_s \right)$\end{tabular} & \begin{tabular}[c]{@{}c@{}}3.16069
\\ 0.00022
\\ 0.0905\\ 5.98247
\\ 0.00142\\ 0.00151
\\ 0.00119
\\ 3.40002\\ 0.10196
\end{tabular} & \begin{tabular}[c]{@{}c@{}}3.15843\\ 0.00020
\\ 0.08251\\ 5.87643
\\ 0.00141
\\ 0.00147
\\ 0.00119\\ 3.35916\\ 0.10068
\end{tabular} & \begin{tabular}[c]{@{}c@{}}3.10586
\\ 0.00016
\\ 0.06153
\\ 5.38206
\\ 0.0014\\ 0.00137
\\ 0.00118\\ 3.08091
\\ 0.09346
\end{tabular} & \begin{tabular}[c]{@{}c@{}}3.10103
\\ 0.00015
\\ 0.0588
\\ 5.3255
\\ 0.00139
\\ 0.00135
\\ 0.00118
\\ 3.06607
\\ 0.09301
\end{tabular} & \begin{tabular}[c]{@{}c@{}}3.03842
\\ 0.00017
\\ 0.0678
\\ 5.41519
\\ 0.00133
\\ 0.00138
\\ 0.00086
\\ 1.06155
\\ 0.04231
\end{tabular} & \begin{tabular}[c]{@{}c@{}}2.88352
\\ 0.00014\\ 0.05373
\\ 4.91919
\\ 0.00123
\\ 0.00129
\\ 0.00085
\\ 0.92556\\ 0.03936\end{tabular} \\ \hline
\textbf{\begin{tabular}[c]{@{}c@{}}Sharp Feature\\ $k_s=0.1$\end{tabular}}      & \begin{tabular}[c]{@{}c@{}}$10^{5}\times\sigma \left(\omega_{\mathrm{b}}\right)$\\ $\sigma \left(\omega_{\mathrm{cdm}}\right)$\\ $\sigma \left( H_0 \right)$\\ $10^{11}\times\sigma \left( A_s \right)$\\ $\sigma \left( n_s \right)$\\ $\sigma \left( \tau_{\mathrm{reio}} \right)$\\ $\sigma \left(S\right)$\\ $\sigma \left(k_s\right)$\\ $\sigma \left( \phi_s \right)$\end{tabular} & \begin{tabular}[c]{@{}c@{}}3.73818
\\ 0.00030
\\ 0.11713
\\ 1.03602
\\ 0.00553
\\ 0.00167
\\ 0.00402
\\ 0.01082
\\ 0.35661
\end{tabular} & \begin{tabular}[c]{@{}c@{}}3.73644
\\ 0.00026
\\ 0.10002
\\ 0.93834
\\ 0.00522
\\ 0.00167
\\ 0.00373
\\ 0.00996
\\ 0.32889
\end{tabular} & \begin{tabular}[c]{@{}c@{}}3.93552\\ 0.00021
\\ 0.07691
\\ 0.91375
\\ 0.00538\\ 0.00166\\ 0.00402
\\ 0.01042\\ 0.34290
\end{tabular} & \begin{tabular}[c]{@{}c@{}}3.90177\\ 0.00020
\\ 0.07115\\ 0.86491
\\ 0.00521
\\ 0.00166
\\ 0.00385
\\ 0.00997
\\ 0.32838
\end{tabular} & \begin{tabular}[c]{@{}c@{}}3.26113
\\ 0.00018
\\ 0.07022
\\ 0.62692
\\ 0.00248
\\ 0.00146
\\ 0.00140
\\ 0.00161\\ 0.07996
\end{tabular} & \begin{tabular}[c]{@{}c@{}}3.24452
\\ 0.00013
\\ 0.05448
\\ 0.56013
\\ 0.00246
\\ 0.00125\\ 0.00136
\\ 0.00153\\ 0.07772
\end{tabular} \\ \hline
\end{tabular}\par
\bigskip
\parbox{17.9cm}{\captionof{table}{The possible marginalized 1-$\sigma$ constraints of \textbf{LiteBIRD+CMB-S4} experiment on sharp feature signal, for CMB-only data and its combination with DESI, EUCLID(CS), DESI+EUCLID(CS), EUCLID(GC) and EUCLID(CS+GC).}} \label{table:LiteBIRD+CMB-S4-Sharp}
\end{minipage}

\setlength{\tabcolsep}{4.85pt} 
\renewcommand{\arraystretch}{1.5} 
\newcolumntype{C}[1]{>{\Centering}m{#1}}
\renewcommand\tabularxcolumn[1]{C{#1}}
\begin{minipage}{\linewidth}
\centering
\captionsetup{font=footnotesize}
\begin{tabular}{|c|c|c|c|c|c|c|c|}
\hline
\textbf{Models}            & \textbf{Parameters}                                                       & \textbf{CMB}                                                              & \textbf{+ DESI}                                                           & \textbf{\begin{tabular}[c]{@{}c@{}}+ Euclid\\ (CS)\end{tabular}}            & \textbf{\begin{tabular}[c]{@{}c@{}}+ DESI +\\  Euclid \\ (CS)\end{tabular}} & \textbf{\begin{tabular}[c]{@{}c@{}}+ Euclid \\ (GC)\end{tabular}}           & \textbf{\begin{tabular}[c]{@{}c@{}}+ Euclid\\ (GC+CS)\end{tabular}}       \\ \hline
\textbf{\begin{tabular}[c]{@{}c@{}}Sharp Feature\\ $k_s=0.004$\end{tabular}}      & \begin{tabular}[c]{@{}c@{}}$10^{5}\times\sigma \left(\omega_{\mathrm{b}}\right)$\\ $\sigma \left(\omega_{\mathrm{cdm}}\right)$\\ $\sigma \left( H_0 \right)$\\ $10^{12}\times\sigma \left( A_s \right)$\\ $\sigma \left( n_s \right)$\\ $\sigma \left( \tau_{\mathrm{reio}} \right)$\\ $\sigma \left(S\right)$\\ $10^{5}\times\sigma \left(k_s\right)$\\ $\sigma \left( \phi_s \right)$\end{tabular} & \begin{tabular}[c]{@{}c@{}}2.77754
\\ 0.00017\\ 0.06443\\ 5.08174\\ 0.00132\\ 0.00121\\ 0.00262
\\ 0.891\\ 0.12818\end{tabular} & \begin{tabular}[c]{@{}c@{}}2.77494
\\ 0.00016\\ 0.06119\\ 5.06655\\ 0.00131\\ 0.0012\\ 0.00262\\ 0.88668\\ 0.12805
\end{tabular} & \begin{tabular}[c]{@{}c@{}}2.78587\\ 0.00017\\ 0.06477\\ 5.04634\\ 0.0013\\ 0.00121
\\ 0.00261\\ 0.60776\\ 0.11308\end{tabular} & \begin{tabular}[c]{@{}c@{}}2.78393\\ 0.00016\\ 0.06152
\\ 5.03051\\ 0.00129
\\ 0.0012\\ 0.00261\\ 0.59756\\ 0.11279\end{tabular} & \begin{tabular}[c]{@{}c@{}}2.66072\\ 0.00014\\ 0.05494\\ 4.7822\\ 0.0012\\ 0.00121\\ 0.00109\\ 0.26922\\ 0.05764\end{tabular} & \begin{tabular}[c]{@{}c@{}}2.58251
\\ 0.00012\\ 0.04706
\\ 4.42437\\ 0.00117\\ 0.00111\\ 0.00108
\\ 0.28039\\ 0.06003\end{tabular} \\ \hline
\textbf{\begin{tabular}[c]{@{}c@{}}Sharp Feature\\ $k_s=0.03$\end{tabular}}          & \begin{tabular}[c]{@{}c@{}}$10^{5}\times\sigma \left(\omega_{\mathrm{b}}\right)$\\ $\sigma \left(\omega_{\mathrm{cdm}}\right)$\\ $\sigma \left( H_0 \right)$\\ $10^{12}\times\sigma \left( A_s \right)$\\ $\sigma \left( n_s \right)$\\ $\sigma \left( \tau_{\mathrm{reio}} \right)$\\ $\sigma \left(S\right)$\\ $10^{4}\times\sigma \left(k_s\right)$\\ $\sigma \left( \phi_s \right)$\end{tabular} & \begin{tabular}[c]{@{}c@{}}2.80682
\\ 0.00017\\ 0.06756\\ 5.08285
\\ 0.00129
\\ 0.00124
\\ 0.00097
\\ 2.96063\\ 0.08659
\end{tabular} & \begin{tabular}[c]{@{}c@{}}2.80616
\\ 0.00016
\\ 0.06397
\\ 5.06396
\\ 0.00128
\\ 0.00123
\\ 0.00097
\\ 2.93746
\\ 0.08590
\end{tabular} & \begin{tabular}[c]{@{}c@{}}2.72569
\\ 0.00014
\\ 0.05554
\\ 4.79204
\\ 0.00125
\\ 0.00121\\ 0.00097
\\ 2.62308
\\ 0.07789
\end{tabular} & \begin{tabular}[c]{@{}c@{}}2.72308
\\ 0.00014
\\ 0.05351
\\ 4.75203
\\ 0.00125
\\ 0.00119
\\ 0.00097\\ 2.61122
\\ 0.07755
\end{tabular} & \begin{tabular}[c]{@{}c@{}}2.71119
\\ 0.00015
\\ 0.06003
\\ 4.88803
\\ 0.00121
\\ 0.00124
\\ 0.00077
\\ 1.0404
\\ 0.03926
\end{tabular} & \begin{tabular}[c]{@{}c@{}}2.55649
\\ 0.00012
\\ 0.04853
\\ 4.41301\\ 0.00112
\\ 0.00116
\\ 0.00076
\\ 0.89367\\ 0.03606
\end{tabular} \\ \hline
\textbf{\begin{tabular}[c]{@{}c@{}}Sharp Feature\\ $k_s=0.1$\end{tabular}}      & \begin{tabular}[c]{@{}c@{}}$10^{5}\times\sigma \left(\omega_{\mathrm{b}}\right)$\\ $\sigma \left(\omega_{\mathrm{cdm}}\right)$\\ $\sigma \left( H_0 \right)$\\ $10^{11}\times\sigma \left( A_s \right)$\\ $\sigma \left( n_s \right)$\\ $\sigma \left( \tau_{\mathrm{reio}} \right)$\\ $\sigma \left(S\right)$\\ $\sigma \left(k_s\right)$\\ $\sigma \left( \phi_s \right)$\end{tabular} & \begin{tabular}[c]{@{}c@{}}3.6584
\\ 0.00026
\\ 0.09162
\\ 0.92295
\\ 0.00588
\\ 0.00170
\\ 0.00448
\\ 0.01117
\\ 0.36893
\end{tabular} & \begin{tabular}[c]{@{}c@{}}3.58802
\\ 0.00023
\\ 0.08133\\ 0.85559
\\ 0.00553
\\ 0.00168\\ 0.00416
\\ 0.01035
\\ 0.34216
\end{tabular} & \begin{tabular}[c]{@{}c@{}}3.58562
\\ 0.00020
\\ 0.07115\\ 0.87569\\ 0.00546
\\ 0.00162
\\ 0.00410\\ 0.01031
\\ 0.33920
\end{tabular} & \begin{tabular}[c]{@{}c@{}}3.54552
\\ 0.00019
\\ 0.06624\\ 0.83128
\\ 0.00527
\\ 0.00162\\ 0.00392
\\ 0.00986
\\ 0.32454
\end{tabular} & \begin{tabular}[c]{@{}c@{}}2.9582
\\ 0.00016
\\ 0.06210
\\ 0.56572
\\ 0.00238
\\ 0.00133
\\ 0.00135
\\ 0.00163
\\ 0.07641
\end{tabular} & \begin{tabular}[c]{@{}c@{}}2.85517
\\ 0.00012
\\ 0.04860
\\ 0.51588\\ 0.00231
\\ 0.00115
\\ 0.00131
\\ 0.00153
\\ 0.07354
\end{tabular} \\ \hline
\end{tabular}\par
\bigskip
\parbox{17.9cm}{\captionof{table}{The possible marginalized 1-$\sigma$ constraints of \textbf{CORE-M5+CMB-S4} experiment on sharp feature signal, for CMB-only data and its combination with DESI, EUCLID(CS), DESI+EUCLID(CS), EUCLID(GC) and EUCLID(CS+GC).}} \label{table:CORE-M5+CMB-S4-Sharp}
\end{minipage}

\setlength{\tabcolsep}{3.25pt} 
\renewcommand{\arraystretch}{1.5} 
\newcolumntype{C}[1]{>{\Centering}m{#1}}
\renewcommand\tabularxcolumn[1]{C{#1}}
\begin{minipage}{\linewidth}
\centering
\captionsetup{font=footnotesize}
\begin{tabular}{|c|c|c|c|c|c|c|c|}
\hline
\textbf{Models}            & \textbf{Parameters}                                                       & \textbf{CMB}                                                              & \textbf{+ DESI}                                                           & \textbf{\begin{tabular}[c]{@{}c@{}}+ Euclid\\ (CS)\end{tabular}}            & \textbf{\begin{tabular}[c]{@{}c@{}}+ DESI +\\  Euclid \\ (CS)\end{tabular}} & \textbf{\begin{tabular}[c]{@{}c@{}}+ Euclid \\ (GC)\end{tabular}}           & \textbf{\begin{tabular}[c]{@{}c@{}}+ Euclid\\ (GC+CS)\end{tabular}}       \\ \hline
\textbf{\begin{tabular}[c]{@{}c@{}}Resonance Feature\\ $k_r=5$\end{tabular}}      & \begin{tabular}[c]{@{}c@{}}$10^{5}\times\sigma \left(\omega_{\mathrm{b}}\right)$\\ $\sigma \left(\omega_{\mathrm{cdm}}\right)$\\ $\sigma \left( H_0 \right)$\\ $10^{12}\times\sigma \left( A_s \right)$\\ $\sigma \left( n_s \right)$\\ $\sigma \left( \tau_{\mathrm{reio}} \right)$\\ $\sigma \left(R\right)$\\ $\sigma \left(k_r\right)$\\ $\sigma \left( \phi_r \right)$\end{tabular} & \begin{tabular}[c]{@{}c@{}}3.62435
\\ 0.00025
\\ 0.10383
\\ 8.95729
\\ 0.00167
\\ 0.00236
\\ 0.00120
\\ 0.08124
\\ 0.11839
\end{tabular} & \begin{tabular}[c]{@{}c@{}}3.61510
\\ 0.00022
\\ 0.09251
\\ 8.46351
\\ 0.00166\\ 0.00221
\\ 0.00120
\\ 0.08123
\\ 0.11837\end{tabular} & \begin{tabular}[c]{@{}c@{}}3.79995\\ 0.00017\\ 0.06859
\\ 7.36231
\\ 0.00175
\\ 0.00189
\\ 0.00119
\\ 0.079
\\ 0.11513
\end{tabular} & \begin{tabular}[c]{@{}c@{}}3.79736\\ 0.00016
\\ 0.06522
\\ 7.17905
\\ 0.00174
\\ 0.00183
\\ 0.00119
\\ 0.079
\\ 0.11513
\end{tabular} & \begin{tabular}[c]{@{}c@{}}3.26987
\\ 0.00019
\\ 0.07548
\\ 7.45964
\\ 0.00154
\\ 0.00200\\ 0.00084
\\ 0.04616
\\ 0.07002\end{tabular} & \begin{tabular}[c]{@{}c@{}}3.43654
\\ 0.00015
\\ 0.05945
\\ 6.36824\\ 0.00157
\\ 0.00167
\\ 0.00083
\\ 0.04554\\ 0.06907
\end{tabular} \\ \hline
\textbf{\begin{tabular}[c]{@{}c@{}}Resonance Feature\\ $k_r=30$\end{tabular}}          & \begin{tabular}[c]{@{}c@{}}$10^{5}\times\sigma \left(\omega_{\mathrm{b}}\right)$\\ $\sigma \left(\omega_{\mathrm{cdm}}\right)$\\ $\sigma \left( H_0 \right)$\\ $10^{12}\times\sigma \left( A_s \right)$\\ $\sigma \left( n_s \right)$\\ $\sigma \left( \tau_{\mathrm{reio}} \right)$\\ $\sigma \left(R\right)$\\ $\sigma \left(k_r\right)$\\ $\sigma \left( \phi_r \right)$\end{tabular} & \begin{tabular}[c]{@{}c@{}}3.50837
\\ 0.00023\\ 0.08986
\\ 8.65570
\\ 0.00169\\ 0.00223\\ 0.00228
\\ 0.18065
\\ 0.27502
\end{tabular} & \begin{tabular}[c]{@{}c@{}}3.50730
\\ 0.00021
\\ 0.08180
\\ 8.35256
\\ 0.00168
\\ 0.00213
\\ 0.00228
\\ 0.18061
\\ 0.27485
\end{tabular} & \begin{tabular}[c]{@{}c@{}}3.44019
\\ 0.00017
\\ 0.07104
\\ 7.39265
\\ 0.00163
\\ 0.00188
\\ 0.00227
\\ 0.15935
\\ 0.24286
\end{tabular} & \begin{tabular}[c]{@{}c@{}}3.43912
\\ 0.00016
\\ 0.06716
\\ 7.23061
\\ 0.00162
\\ 0.00183
\\ 0.00227\\ 0.15851
\\ 0.24130
\end{tabular} & \begin{tabular}[c]{@{}c@{}}3.22370
\\ 0.00019
\\ 0.07249
\\ 7.42659
\\ 0.00153\\ 0.00197
\\ 0.00110\\ 0.05608\\ 0.08678\end{tabular} & \begin{tabular}[c]{@{}c@{}}3.17239
\\ 0.00015
\\ 0.06032
\\ 6.52831
\\ 0.00146
\\ 0.00171
\\ 0.00109
\\ 0.05556
\\ 0.08589
\end{tabular} \\ \hline
\textbf{\begin{tabular}[c]{@{}c@{}}Resonance Feature\\ $k_r=100$\end{tabular}}      & \begin{tabular}[c]{@{}c@{}}$10^{5}\times\sigma \left(\omega_{\mathrm{b}}\right)$\\ $\sigma \left(\omega_{\mathrm{cdm}}\right)$\\ $\sigma \left( H_0 \right)$\\ $10^{12}\times\sigma \left( A_s \right)$\\ $\sigma \left( n_s \right)$\\ $\sigma \left( \tau_{\mathrm{reio}} \right)$\\ $\sigma \left(R\right)$\\ $\sigma \left(k_r\right)$\\ $\sigma \left( \phi_r \right)$\end{tabular} & \begin{tabular}[c]{@{}c@{}}3.25103
\\ 0.00024
\\ 0.09594
\\ 8.27515
\\ 0.00162
\\ 0.00215
\\ 0.00583
\\ 0.37981
\\ 0.62952
\end{tabular} & \begin{tabular}[c]{@{}c@{}}3.24798
\\ 0.00021
\\ 0.0865
\\ 7.91445
\\ 0.00161
\\ 0.00204
\\ 0.00583
\\ 0.37979
\\ 0.62905
\end{tabular} & \begin{tabular}[c]{@{}c@{}}3.53867
\\ 0.00018
\\ 0.07427
\\ 7.29354
\\ 0.00173
\\ 0.00186
\\ 0.00582
\\ 0.26119
\\ 0.39417
\end{tabular} & \begin{tabular}[c]{@{}c@{}}3.53579
\\ 0.00017
\\ 0.06970
\\ 7.11502
\\ 0.00172
\\ 0.00180
\\ 0.00582
\\ 0.25908
\\ 0.39152
\end{tabular} & \begin{tabular}[c]{@{}c@{}}3.13846
\\ 0.00009\\ 0.02907\\ 6.2265
\\ 0.00140
\\ 0.00158
\\ 0.00134
\\ 0.07501
\\ 0.11341
\end{tabular} & \begin{tabular}[c]{@{}c@{}}3.26755
\\ 0.00009\\ 0.02857
\\ 5.07505
\\ 0.00148
\\ 0.00126
\\ 0.00134
\\ 0.07338
\\ 0.11027
\end{tabular} \\ \hline
\end{tabular}\par
\bigskip
\parbox{17.9cm}{\captionof{table}{The possible marginalized 1-$\sigma$ constraints of \textbf{CMB-S4} experiment on resonance feature signal, for CMB-only data and its combination with DESI, EUCLID(CS), DESI+EUCLID(CS), EUCLID(GC) and EUCLID(CS+GC).}} \label{table:CMB-S4-Resonance}
\end{minipage}

\setlength{\tabcolsep}{3.25pt} 
\renewcommand{\arraystretch}{1.5} 
\newcolumntype{C}[1]{>{\Centering}m{#1}}
\renewcommand\tabularxcolumn[1]{C{#1}}
\begin{minipage}{\linewidth}
\centering
\captionsetup{font=footnotesize}
\begin{tabular}{|c|c|c|c|c|c|c|c|}
\hline
\textbf{Models}            & \textbf{Parameters}                                                       & \textbf{CMB}                                                              & \textbf{+ DESI}                                                           & \textbf{\begin{tabular}[c]{@{}c@{}}+ Euclid\\ (CS)\end{tabular}}            & \textbf{\begin{tabular}[c]{@{}c@{}}+ DESI +\\  Euclid \\ (CS)\end{tabular}} & \textbf{\begin{tabular}[c]{@{}c@{}}+ Euclid \\ (GC)\end{tabular}}           & \textbf{\begin{tabular}[c]{@{}c@{}}+ Euclid\\ (GC+CS)\end{tabular}}       \\ \hline
\textbf{\begin{tabular}[c]{@{}c@{}}Resonance Feature\\ $k_r=5$\end{tabular}}      & \begin{tabular}[c]{@{}c@{}}$10^{5}\times\sigma \left(\omega_{\mathrm{b}}\right)$\\ $\sigma \left(\omega_{\mathrm{cdm}}\right)$\\ $\sigma \left( H_0 \right)$\\ $10^{12}\times\sigma \left( A_s \right)$\\ $\sigma \left( n_s \right)$\\ $\sigma \left( \tau_{\mathrm{reio}} \right)$\\ $\sigma \left(R\right)$\\ $\sigma \left(k_r\right)$\\ $\sigma \left( \phi_r \right)$\end{tabular} & \begin{tabular}[c]{@{}c@{}}3.96603\\ 0.00018\\ 0.07713
\\ 5.44393
\\ 0.00142
\\ 0.00133
\\ 0.00096\\ 0.07071
\\ 0.11253\end{tabular} & \begin{tabular}[c]{@{}c@{}}3.96454
\\ 0.00017
\\ 0.07274
\\ 5.41688\\ 0.00141
\\ 0.00131
\\ 0.00096\\ 0.07070
\\ 0.11252
\end{tabular} & \begin{tabular}[c]{@{}c@{}}4.01849
\\ 0.00013
\\ 0.05955
\\ 5.37220
\\ 0.00144
\\ 0.00131
\\ 0.00096
\\ 0.06919
\\ 0.11009
\end{tabular} & \begin{tabular}[c]{@{}c@{}}4.01769
\\ 0.00013
\\ 0.05775
\\ 5.32054
\\ 0.00144
\\ 0.00130\\ 0.00096\\ 0.06918
\\ 0.11008
\end{tabular} & \begin{tabular}[c]{@{}c@{}}3.61522\\ 0.00017\\ 0.06891\\ 5.29332
\\ 0.00129\\ 0.00132\\ 0.00074
\\ 0.04345\\ 0.06910\end{tabular} & \begin{tabular}[c]{@{}c@{}}3.69586\\ 0.00013\\ 0.05405
\\ 4.98985
\\ 0.00131
\\ 0.00125
\\ 0.00074
\\ 0.04301
\\ 0.06841
\end{tabular} \\ \hline
\textbf{\begin{tabular}[c]{@{}c@{}}Resonance Feature\\ $k_r=30$\end{tabular}}          & \begin{tabular}[c]{@{}c@{}}$10^{5}\times\sigma \left(\omega_{\mathrm{b}}\right)$\\ $\sigma \left(\omega_{\mathrm{cdm}}\right)$\\ $\sigma \left( H_0 \right)$\\ $10^{12}\times\sigma \left( A_s \right)$\\ $\sigma \left( n_s \right)$\\ $\sigma \left( \tau_{\mathrm{reio}} \right)$\\ $\sigma \left(R\right)$\\ $\sigma \left(k_r\right)$\\ $\sigma \left( \phi_r \right)$\end{tabular} & \begin{tabular}[c]{@{}c@{}}4.04171\\ 0.00018
\\ 0.07875
\\ 5.50034
\\ 0.00142
\\ 0.00132
\\ 0.00211
\\ 0.16264
\\ 0.26441
\end{tabular} & \begin{tabular}[c]{@{}c@{}}4.03829
\\ 0.00017\\ 0.07425
\\ 5.47246
\\ 0.00141
\\ 0.00131
\\ 0.00211
\\ 0.16250
\\ 0.26412
\end{tabular} & \begin{tabular}[c]{@{}c@{}}3.97704
\\ 0.00014
\\ 0.06423\\ 5.44361
\\ 0.00141
\\ 0.00132
\\ 0.00211
\\ 0.14701\\ 0.23892
\end{tabular} & \begin{tabular}[c]{@{}c@{}}3.97645\\ 0.00014
\\ 0.06194
\\ 5.40322
\\ 0.00140\\ 0.00130
\\ 0.00211
\\ 0.14611
\\ 0.23732
\end{tabular} & \begin{tabular}[c]{@{}c@{}}3.63979
\\ 0.00016
\\ 0.06693
\\ 5.29969
\\ 0.00128
\\ 0.00131
\\ 0.00108
\\ 0.05498
\\ 0.08634
\end{tabular} & \begin{tabular}[c]{@{}c@{}}3.62466
\\ 0.00012
\\ 0.0538
\\ 4.94317
\\ 0.00127
\\ 0.00122
\\ 0.00108
\\ 0.05460
\\ 0.08569
\end{tabular} \\ \hline
\textbf{\begin{tabular}[c]{@{}c@{}}Resonance Feature\\ $k_r=100$\end{tabular}}      & \begin{tabular}[c]{@{}c@{}}$10^{5}\times\sigma \left(\omega_{\mathrm{b}}\right)$\\ $\sigma \left(\omega_{\mathrm{cdm}}\right)$\\ $\sigma \left( H_0 \right)$\\ $10^{12}\times\sigma \left( A_s \right)$\\ $\sigma \left( n_s \right)$\\ $\sigma \left( \tau_{\mathrm{reio}} \right)$\\ $\sigma \left(R\right)$\\ $\sigma \left(k_r\right)$\\ $\sigma \left( \phi_r \right)$\end{tabular} & \begin{tabular}[c]{@{}c@{}}3.82620
\\ 0.00018
\\ 0.07684
\\ 5.46762
\\ 0.00139
\\ 0.00132
\\ 0.00502
\\ 0.35570
\\ 0.64410\end{tabular} & \begin{tabular}[c]{@{}c@{}}3.82445
\\ 0.00017
\\ 0.07245
\\ 5.43881
\\ 0.00139
\\ 0.00131\\ 0.00502
\\ 0.35568
\\ 0.64410
\end{tabular} & \begin{tabular}[c]{@{}c@{}}3.85677
\\ 0.00016
\\ 0.06759
\\ 5.37103
\\ 0.00142
\\ 0.00130
\\ 0.00501
\\ 0.25320
\\ 0.42912
\end{tabular} & \begin{tabular}[c]{@{}c@{}}3.85630
\\ 0.00015\\ 0.06467
\\ 5.33290
\\ 0.00141
\\ 0.00129
\\ 0.00501
\\ 0.25063
\\ 0.42503
\end{tabular} & \begin{tabular}[c]{@{}c@{}}3.51651
\\ 0.00008\\ 0.03029
\\ 5.07480
\\ 0.00123
\\ 0.00122
\\ 0.00133
\\ 0.07415
\\ 0.11324
\end{tabular} & \begin{tabular}[c]{@{}c@{}}3.64414
\\ 0.00008\\ 0.03004
\\ 4.47549
\\ 0.00126
\\ 0.00107\\ 0.00133
\\ 0.07272
\\ 0.11081
\end{tabular} \\ \hline
\end{tabular}\par
\bigskip
\parbox{17.9cm}{\captionof{table}{The possible marginalized 1-$\sigma$ constraints of \textbf{CORE-M5} experiment on resonance feature signal, for CMB-only data and its combination with DESI, EUCLID(CS), DESI+EUCLID(CS), EUCLID(GC) and EUCLID(CS+GC).}} \label{table:CORE-M5-Resonance}
\end{minipage}

\setlength{\tabcolsep}{3.25pt} 
\renewcommand{\arraystretch}{1.5} 
\newcolumntype{C}[1]{>{\Centering}m{#1}}
\renewcommand\tabularxcolumn[1]{C{#1}}
\begin{minipage}{\linewidth}
\centering
\captionsetup{font=footnotesize}
\begin{tabular}{|c|c|c|c|c|c|c|c|}
\hline
\textbf{Models}            & \textbf{Parameters}                                                       & \textbf{CMB}                                                              & \textbf{+ DESI}                                                           & \textbf{\begin{tabular}[c]{@{}c@{}}+ Euclid\\ (CS)\end{tabular}}            & \textbf{\begin{tabular}[c]{@{}c@{}}+ DESI +\\  Euclid \\ (CS)\end{tabular}} & \textbf{\begin{tabular}[c]{@{}c@{}}+ Euclid \\ (GC)\end{tabular}}           & \textbf{\begin{tabular}[c]{@{}c@{}}+ Euclid\\ (GC+CS)\end{tabular}}       \\ \hline
\textbf{\begin{tabular}[c]{@{}c@{}}Resonance Feature\\ $k_r=5$\end{tabular}}      & \begin{tabular}[c]{@{}c@{}}$10^{5}\times\sigma \left(\omega_{\mathrm{b}}\right)$\\ $\sigma \left(\omega_{\mathrm{cdm}}\right)$\\ $\sigma \left( H_0 \right)$\\ $10^{12}\times\sigma \left( A_s \right)$\\ $\sigma \left( n_s \right)$\\ $\sigma \left( \tau_{\mathrm{reio}} \right)$\\ $\sigma \left(R\right)$\\ $\sigma \left(k_r\right)$\\ $\sigma \left( \phi_r \right)$\end{tabular} & \begin{tabular}[c]{@{}c@{}}2.99134
\\ 0.00015
\\ 0.06009
\\ 4.85559
\\ 0.00129
\\ 0.00120
\\ 0.00092
\\ 0.06238
\\ 0.09211
\end{tabular} & \begin{tabular}[c]{@{}c@{}}2.98960
\\ 0.00014
\\ 0.05762
\\ 4.81740
\\ 0.00128
\\ 0.00118
\\ 0.00092
\\ 0.06238
\\ 0.09210
\end{tabular} & \begin{tabular}[c]{@{}c@{}}2.98103
\\ 0.00012
\\ 0.05150
\\ 4.82063
\\ 0.00128
\\ 0.00119
\\ 0.00091
\\ 0.06137
\\ 0.09063
\end{tabular} & \begin{tabular}[c]{@{}c@{}}2.97952
\\ 0.00012
\\ 0.05004
\\ 4.77436
\\ 0.00127
\\ 0.00118
\\ 0.00091
\\ 0.06136
\\ 0.09062
\end{tabular} & \begin{tabular}[c]{@{}c@{}}2.80456\\ 0.00015
\\ 0.05745
\\ 4.82153
\\ 0.00121
\\ 0.00124
\\ 0.00072
\\ 0.04141
\\ 0.06259
\end{tabular} & \begin{tabular}[c]{@{}c@{}}2.95407
\\ 0.00012
\\ 0.04718
\\ 4.45841
\\ 0.00126\\ 0.00111\\ 0.00071
\\ 0.04104
\\ 0.06207
\end{tabular} \\ \hline
\textbf{\begin{tabular}[c]{@{}c@{}}Resonance Feature\\ $k_r=30$\end{tabular}}          & \begin{tabular}[c]{@{}c@{}}$10^{5}\times\sigma \left(\omega_{\mathrm{b}}\right)$\\ $\sigma \left(\omega_{\mathrm{cdm}}\right)$\\ $\sigma \left( H_0 \right)$\\ $10^{12}\times\sigma \left( A_s \right)$\\ $\sigma \left( n_s \right)$\\ $\sigma \left( \tau_{\mathrm{reio}} \right)$\\ $\sigma \left(R\right)$\\ $\sigma \left(k_r\right)$\\ $\sigma \left( \phi_r \right)$\end{tabular} & \begin{tabular}[c]{@{}c@{}}2.90927
\\ 0.00015
\\ 0.06067
\\ 4.89429
\\ 0.00128
\\ 0.00119
\\ 0.00179
\\ 0.14434
\\ 0.22411\end{tabular} & \begin{tabular}[c]{@{}c@{}}2.90851
\\ 0.00014
\\ 0.05816
\\ 4.85629
\\ 0.00127
\\ 0.00118
\\ 0.00179
\\ 0.14426
\\ 0.22394
\end{tabular} & \begin{tabular}[c]{@{}c@{}}2.89512
\\ 0.00013
\\ 0.054
\\ 4.85731\\ 0.00127
\\ 0.00119
\\ 0.00179
\\ 0.13046
\\ 0.20269\end{tabular} & \begin{tabular}[c]{@{}c@{}}2.89435
\\ 0.00013
\\ 0.05231
\\ 4.81757
\\ 0.00126
\\ 0.00117\\ 0.00179
\\ 0.13001
\\ 0.20187
\end{tabular} & \begin{tabular}[c]{@{}c@{}}2.77256
\\ 0.00014
\\ 0.05618
\\ 4.82806
\\ 0.00121\\ 0.00123
\\ 0.00103
\\ 0.05383
\\ 0.08387
\end{tabular} & \begin{tabular}[c]{@{}c@{}}2.70048
\\ 0.00011\\ 0.04677
\\ 4.37976
\\ 0.00116
\\ 0.00109
\\ 0.00102
\\ 0.05340
\\ 0.08308\end{tabular} \\ \hline
\textbf{\begin{tabular}[c]{@{}c@{}}Resonance Feature\\ $k_r=100$\end{tabular}}      & \begin{tabular}[c]{@{}c@{}}$10^{5}\times\sigma \left(\omega_{\mathrm{b}}\right)$\\ $\sigma \left(\omega_{\mathrm{cdm}}\right)$\\ $\sigma \left( H_0 \right)$\\ $10^{12}\times\sigma \left( A_s \right)$\\ $\sigma \left( n_s \right)$\\ $\sigma \left( \tau_{\mathrm{reio}} \right)$\\ $\sigma \left(R\right)$\\ $\sigma \left(k_r\right)$\\ $\sigma \left( \phi_r \right)$\end{tabular} & \begin{tabular}[c]{@{}c@{}}2.75877\\ 0.00015
\\ 0.05841
\\ 4.77954
\\ 0.00127
\\ 0.00118
\\ 0.00455
\\ 0.30053\\ 0.51053
\end{tabular} & \begin{tabular}[c]{@{}c@{}}2.75695\\ 0.00014
\\ 0.05606
\\ 4.74294
\\ 0.00126
\\ 0.00116
\\ 0.00455
\\ 0.30052
\\ 0.51041\end{tabular} & \begin{tabular}[c]{@{}c@{}}2.78369\\ 0.00014
\\ 0.05501\\ 4.80023
\\ 0.00126
\\ 0.00118
\\ 0.00454
\\ 0.21860
\\ 0.34613
\end{tabular} & \begin{tabular}[c]{@{}c@{}}2.78234
\\ 0.00013
\\ 0.05312
\\ 4.75558\\ 0.00125
\\ 0.00116
\\ 0.00454
\\ 0.21771
\\ 0.34497
\end{tabular} & \begin{tabular}[c]{@{}c@{}}2.69631
\\ 0.00008\\ 0.02704
\\ 4.34767\\ 0.00114
\\ 0.00106
\\ 0.00132
\\ 0.07382
\\ 0.11155
\end{tabular} & \begin{tabular}[c]{@{}c@{}}2.81392
\\ 0.00007\\ 0.02634
\\ 3.91888
\\ 0.00121
\\ 0.00094
\\ 0.00132
\\ 0.07211
\\ 0.10840
\end{tabular} \\ \hline
\end{tabular}\par
\bigskip
\parbox{17.9cm}{\captionof{table}{The possible marginalized 1-$\sigma$ constraints of \textbf{PICO} experiment on resonance feature signal, for CMB-only data and its combination with DESI, EUCLID(CS), DESI+EUCLID(CS), EUCLID(GC) and EUCLID(CS+GC).}} \label{table:PICO-Resonance}
\end{minipage}

\setlength{\tabcolsep}{3.25pt} 
\renewcommand{\arraystretch}{1.5} 
\newcolumntype{C}[1]{>{\Centering}m{#1}}
\renewcommand\tabularxcolumn[1]{C{#1}}
\begin{minipage}{\linewidth}
\centering
\captionsetup{font=footnotesize}
\begin{tabular}{|c|c|c|c|c|c|c|c|}
\hline
\textbf{Models}            & \textbf{Parameters}                                                       & \textbf{CMB}                                                              & \textbf{+ DESI}                                                           & \textbf{\begin{tabular}[c]{@{}c@{}}+ Euclid\\ (CS)\end{tabular}}            & \textbf{\begin{tabular}[c]{@{}c@{}}+ DESI +\\  Euclid \\ (CS)\end{tabular}} & \textbf{\begin{tabular}[c]{@{}c@{}}+ Euclid \\ (GC)\end{tabular}}           & \textbf{\begin{tabular}[c]{@{}c@{}}+ Euclid\\ (GC+CS)\end{tabular}}       \\ \hline
\textbf{\begin{tabular}[c]{@{}c@{}}Resonance Feature\\ $k_r=5$\end{tabular}}      & \begin{tabular}[c]{@{}c@{}}$10^{5}\times\sigma \left(\omega_{\mathrm{b}}\right)$\\ $\sigma \left(\omega_{\mathrm{cdm}}\right)$\\ $\sigma \left( H_0 \right)$\\ $10^{12}\times\sigma \left( A_s \right)$\\ $\sigma \left( n_s \right)$\\ $\sigma \left( \tau_{\mathrm{reio}} \right)$\\ $\sigma \left(R\right)$\\ $\sigma \left(k_r\right)$\\ $\sigma \left( \phi_r \right)$\end{tabular} & \begin{tabular}[c]{@{}c@{}}3.56132
\\ 0.00021
\\ 0.08223
\\ 7.47855
\\ 0.00159
\\ 0.00191
\\ 0.00115
\\ 0.07602
\\ 0.11288
\end{tabular} & \begin{tabular}[c]{@{}c@{}}3.55942
\\ 0.00019
\\ 0.07603\\ 7.31251
\\ 0.00158
\\ 0.00185
\\ 0.00115
\\ 0.07602
\\ 0.11288
\end{tabular} & \begin{tabular}[c]{@{}c@{}}3.69189
\\ 0.00016\\ 0.06488
\\ 6.74134
\\ 0.00163
\\ 0.00170
\\ 0.00113
\\ 0.07422\\ 0.11028
\end{tabular} & \begin{tabular}[c]{@{}c@{}}3.68933
\\ 0.00015
\\ 0.06204
\\ 6.60829
\\ 0.00162
\\ 0.00166
\\ 0.00113
\\ 0.07422
\\ 0.11027\end{tabular} & \begin{tabular}[c]{@{}c@{}}3.20043
\\ 0.00018
\\ 0.07185\\ 6.78526
\\ 0.00144
\\ 0.00178\\ 0.00082
\\ 0.04504
\\ 0.06892
\end{tabular} & \begin{tabular}[c]{@{}c@{}}3.28888
\\ 0.00015
\\ 0.059\\ 6.17094
\\ 0.00145
\\ 0.00161
\\ 0.00081
\\ 0.04452
\\ 0.06813
\end{tabular} \\ \hline
\textbf{\begin{tabular}[c]{@{}c@{}}Resonance Feature\\ $k_r=30$\end{tabular}}          & \begin{tabular}[c]{@{}c@{}}$10^{5}\times\sigma \left(\omega_{\mathrm{b}}\right)$\\ $\sigma \left(\omega_{\mathrm{cdm}}\right)$\\ $\sigma \left( H_0 \right)$\\ $10^{12}\times\sigma \left( A_s \right)$\\ $\sigma \left( n_s \right)$\\ $\sigma \left( \tau_{\mathrm{reio}} \right)$\\ $\sigma \left(R\right)$\\ $\sigma \left(k_r\right)$\\ $\sigma \left( \phi_r \right)$\end{tabular} & \begin{tabular}[c]{@{}c@{}}3.38907
\\ 0.00021
\\ 0.08371\\ 7.65578
\\ 0.00158
\\ 0.00194\\ 0.00223\\ 0.17423
\\ 0.26845
\end{tabular} & \begin{tabular}[c]{@{}c@{}}3.38810
\\ 0.00019
\\ 0.07716
\\ 7.47025\\ 0.00156
\\ 0.00188
\\ 0.00222
\\ 0.17417
\\ 0.26825\end{tabular} & \begin{tabular}[c]{@{}c@{}}3.32894
\\ 0.00016
\\ 0.06747
\\ 6.76076\\ 0.00153
\\ 0.00170
\\ 0.00222
\\ 0.15457
\\ 0.23848
\end{tabular} & \begin{tabular}[c]{@{}c@{}}3.32783
\\ 0.00016
\\ 0.06415
\\ 6.64478
\\ 0.00152
\\ 0.00166
\\ 0.00222
\\ 0.15375
\\ 0.23696
\end{tabular} & \begin{tabular}[c]{@{}c@{}}3.14051
\\ 0.00018
\\ 0.06915
\\ 6.76639
\\ 0.00144
\\ 0.00176
\\ 0.00109
\\ 0.05569
\\ 0.08648
\end{tabular} & \begin{tabular}[c]{@{}c@{}}3.09443
\\ 0.00014
\\ 0.05786
\\ 6.0735
\\ 0.00138
\\ 0.00157
\\ 0.00109
\\ 0.05520
\\ 0.08564
\end{tabular} \\ \hline
\textbf{\begin{tabular}[c]{@{}c@{}}Resonance Feature\\ $k_r=100$\end{tabular}}      & \begin{tabular}[c]{@{}c@{}}$10^{5}\times\sigma \left(\omega_{\mathrm{b}}\right)$\\ $\sigma \left(\omega_{\mathrm{cdm}}\right)$\\ $\sigma \left( H_0 \right)$\\ $10^{12}\times\sigma \left( A_s \right)$\\ $\sigma \left( n_s \right)$\\ $\sigma \left( \tau_{\mathrm{reio}} \right)$\\ $\sigma \left(R\right)$\\ $\sigma \left(k_r\right)$\\ $\sigma \left( \phi_r \right)$\end{tabular} & \begin{tabular}[c]{@{}c@{}}3.26562\\ 0.00020
\\ 0.08151
\\ 7.38862
\\ 0.00156
\\ 0.00187\\ 0.00568
\\ 0.36008
\\ 0.60931
\end{tabular} & \begin{tabular}[c]{@{}c@{}}3.26439
\\ 0.00019
\\ 0.07536
\\ 7.22193
\\ 0.00155
\\ 0.00181
\\ 0.00568\\ 0.36007
\\ 0.609
\end{tabular} & \begin{tabular}[c]{@{}c@{}}3.41568
\\ 0.00017
\\ 0.07031
\\ 6.67632
\\ 0.00162
\\ 0.00168
\\ 0.00567
\\ 0.25166\\ 0.38877
\end{tabular} & \begin{tabular}[c]{@{}c@{}}3.41324
\\ 0.00016
\\ 0.06644
\\ 6.55015
\\ 0.00161
\\ 0.00164
\\ 0.00567
\\ 0.24977
\\ 0.38626
\end{tabular} & \begin{tabular}[c]{@{}c@{}}3.06453
\\ 0.00009\\ 0.02885
\\ 5.89517
\\ 0.00133
\\ 0.00146
\\ 0.00134
\\ 0.07473
\\ 0.11312
\end{tabular} & \begin{tabular}[c]{@{}c@{}}3.16891
\\ 0.00009\\ 0.02869
\\ 4.97945
\\ 0.00137
\\ 0.00123
\\ 0.00134
\\ 0.07313
\\ 0.11010
\end{tabular} \\ \hline
\end{tabular}\par
\bigskip
\parbox{17.9cm}{\captionof{table}{The possible marginalized 1-$\sigma$ constraints of \textbf{Planck+CMB-S4} experiment on resonance feature signal, for CMB-only data and its combination with DESI, EUCLID(CS), DESI+EUCLID(CS), EUCLID(GC) and EUCLID(CS+GC).}} \label{table:Planck+CMB-S4-Resonance}
\end{minipage}

\setlength{\tabcolsep}{3.25pt} 
\renewcommand{\arraystretch}{1.5} 
\newcolumntype{C}[1]{>{\Centering}m{#1}}
\renewcommand\tabularxcolumn[1]{C{#1}}
\begin{minipage}{\linewidth}
\centering
\captionsetup{font=footnotesize}
\begin{tabular}{|c|c|c|c|c|c|c|c|}
\hline
\textbf{Models}            & \textbf{Parameters}                                                       & \textbf{CMB}                                                              & \textbf{+ DESI}                                                           & \textbf{\begin{tabular}[c]{@{}c@{}}+ Euclid\\ (CS)\end{tabular}}            & \textbf{\begin{tabular}[c]{@{}c@{}}+ DESI +\\  Euclid \\ (CS)\end{tabular}} & \textbf{\begin{tabular}[c]{@{}c@{}}+ Euclid \\ (GC)\end{tabular}}           & \textbf{\begin{tabular}[c]{@{}c@{}}+ Euclid\\ (GC+CS)\end{tabular}}       \\ \hline
\textbf{\begin{tabular}[c]{@{}c@{}}Resonance Feature\\ $k_r=5$\end{tabular}}      & \begin{tabular}[c]{@{}c@{}}$10^{5}\times\sigma \left(\omega_{\mathrm{b}}\right)$\\ $\sigma \left(\omega_{\mathrm{cdm}}\right)$\\ $\sigma \left( H_0 \right)$\\ $10^{12}\times\sigma \left( A_s \right)$\\ $\sigma \left( n_s \right)$\\ $\sigma \left( \tau_{\mathrm{reio}} \right)$\\ $\sigma \left(R\right)$\\ $\sigma \left(k_r\right)$\\ $\sigma \left( \phi_r \right)$\end{tabular} & \begin{tabular}[c]{@{}c@{}}3.37691
\\ 0.00020
\\ 0.08368
\\ 5.97760
\\ 0.00144
\\ 0.00152
\\ 0.00116
\\ 0.07045
\\ 0.10649
\end{tabular} & \begin{tabular}[c]{@{}c@{}}3.37584
\\ 0.00019
\\ 0.07745
\\ 5.85464
\\ 0.00144
\\ 0.00148
\\ 0.00116
\\ 0.07045
\\ 0.10649
\end{tabular} & \begin{tabular}[c]{@{}c@{}}3.56437
\\ 0.00014
\\ 0.05858\\ 5.48258
\\ 0.00151
\\ 0.00136
\\ 0.00115
\\ 0.06903
\\ 0.10438\end{tabular} & \begin{tabular}[c]{@{}c@{}}3.56089\\ 0.00014
\\ 0.05651
\\ 5.41676
\\ 0.00150
\\ 0.00134\\ 0.00115
\\ 0.06902
\\ 0.10437
\end{tabular} & \begin{tabular}[c]{@{}c@{}}3.13827
\\ 0.00017
\\ 0.06678
\\ 5.45758
\\ 0.00135
\\ 0.00140
\\ 0.00082
\\ 0.04359
\\ 0.06729
\end{tabular} & \begin{tabular}[c]{@{}c@{}}3.27675
\\ 0.00013
\\ 0.05319
\\ 5.03323
\\ 0.00139
\\ 0.00127
\\ 0.00082\\ 0.04319\\ 0.06672
\end{tabular} \\ \hline
\textbf{\begin{tabular}[c]{@{}c@{}}Resonance Feature\\ $k_r=30$\end{tabular}}          & \begin{tabular}[c]{@{}c@{}}$10^{5}\times\sigma \left(\omega_{\mathrm{b}}\right)$\\ $\sigma \left(\omega_{\mathrm{cdm}}\right)$\\ $\sigma \left( H_0 \right)$\\ $10^{12}\times\sigma \left( A_s \right)$\\ $\sigma \left( n_s \right)$\\ $\sigma \left( \tau_{\mathrm{reio}} \right)$\\ $\sigma \left(R\right)$\\ $\sigma \left(k_r\right)$\\ $\sigma \left( \phi_r \right)$\end{tabular} & \begin{tabular}[c]{@{}c@{}}3.24215
\\ 0.00020
\\ 0.08453
\\ 5.99539
\\ 0.00143
\\ 0.00151
\\ 0.00222
\\ 0.16722
\\ 0.25989
\end{tabular} & \begin{tabular}[c]{@{}c@{}}3.24018
\\ 0.00019
\\ 0.07811
\\ 5.87104
\\ 0.00142
\\ 0.00147
\\ 0.00222
\\ 0.16707\\ 0.25952
\end{tabular} & \begin{tabular}[c]{@{}c@{}}3.25230
\\ 0.00015\\ 0.06191
\\ 5.49172
\\ 0.00142\\ 0.00136
\\ 0.00222
\\ 0.14862
\\ 0.23101
\end{tabular} & \begin{tabular}[c]{@{}c@{}}3.25086\\ 0.00014
\\ 0.05936
\\ 5.43554
\\ 0.00142\\ 0.00134
\\ 0.00222
\\ 0.14777
\\ 0.22948
\end{tabular} & \begin{tabular}[c]{@{}c@{}}3.08972
\\ 0.00017
\\ 0.06469\\ 5.45924
\\ 0.00134
\\ 0.00139
\\ 0.00109
\\ 0.05517
\\ 0.08596
\end{tabular} & \begin{tabular}[c]{@{}c@{}}3.11454
\\ 0.00013
\\ 0.05330
\\ 4.92330
\\ 0.00130
\\ 0.00124
\\ 0.00109\\ 0.05490\\ 0.08527
\end{tabular} \\ \hline
\textbf{\begin{tabular}[c]{@{}c@{}}Resonance Feature\\ $k_r=100$\end{tabular}}      & \begin{tabular}[c]{@{}c@{}}$10^{5}\times\sigma \left(\omega_{\mathrm{b}}\right)$\\ $\sigma \left(\omega_{\mathrm{cdm}}\right)$\\ $\sigma \left( H_0 \right)$\\ $10^{12}\times\sigma \left( A_s \right)$\\ $\sigma \left( n_s \right)$\\ $\sigma \left( \tau_{\mathrm{reio}} \right)$\\ $\sigma \left(R\right)$\\ $\sigma \left(k_r\right)$\\ $\sigma \left( \phi_r \right)$\end{tabular} & \begin{tabular}[c]{@{}c@{}}3.07105
\\ 0.00020
\\ 0.08080
\\ 5.77668
\\ 0.00141
\\ 0.00146
\\ 0.00566
\\ 0.33776
\\ 0.58244
\end{tabular} & \begin{tabular}[c]{@{}c@{}}3.07041
\\ 0.00018
\\ 0.075
\\ 5.67520
\\ 0.00140
\\ 0.00142
\\ 0.00566
\\ 0.33772
\\ 0.58202\end{tabular} & \begin{tabular}[c]{@{}c@{}}3.31431
\\ 0.00016\\ 0.06428
\\ 5.45424
\\ 0.00150
\\ 0.00135
\\ 0.00565
\\ 0.24281\\ 0.38194
\end{tabular} & \begin{tabular}[c]{@{}c@{}}3.31116\\ 0.00015
\\ 0.06132
\\ 5.39346\\ 0.00149
\\ 0.00133
\\ 0.00565
\\ 0.24110
\\ 0.37949
\end{tabular} & \begin{tabular}[c]{@{}c@{}}3.01722
\\ 0.00009\\ 0.02860
\\ 5.02426\\ 0.00125
\\ 0.00122
\\ 0.00134\\ 0.07445
\\ 0.11286
\end{tabular} & \begin{tabular}[c]{@{}c@{}}3.11794
\\ 0.00008\\ 0.02805
\\ 4.37941
\\ 0.00132\\ 0.00106
\\ 0.00134
\\ 0.07283
\\ 0.10980
\end{tabular} \\ \hline
\end{tabular}\par
\bigskip
\parbox{17.9cm}{\captionof{table}{The possible marginalized 1-$\sigma$ constraints of \textbf{LiteBIRD+CMB-S4} experiment on resonance feature signal, for CMB-only data and its combination with DESI, EUCLID(CS), DESI+EUCLID(CS), EUCLID(GC) and EUCLID(CS+GC).}} \label{table:LiteBIRD+CMB-S4-Resonance}
\end{minipage}

\setlength{\tabcolsep}{3.25pt} 
\renewcommand{\arraystretch}{1.5} 
\newcolumntype{C}[1]{>{\Centering}m{#1}}
\renewcommand\tabularxcolumn[1]{C{#1}}
\begin{minipage}{\linewidth}
\centering
\captionsetup{font=footnotesize}
\begin{tabular}{|c|c|c|c|c|c|c|c|}
\hline
\textbf{Models}            & \textbf{Parameters}                                                       & \textbf{CMB}                                                              & \textbf{+ DESI}                                                           & \textbf{\begin{tabular}[c]{@{}c@{}}+ Euclid\\ (CS)\end{tabular}}            & \textbf{\begin{tabular}[c]{@{}c@{}}+ DESI +\\  Euclid \\ (CS)\end{tabular}} & \textbf{\begin{tabular}[c]{@{}c@{}}+ Euclid \\ (GC)\end{tabular}}           & \textbf{\begin{tabular}[c]{@{}c@{}}+ Euclid\\ (GC+CS)\end{tabular}}       \\ \hline
\textbf{\begin{tabular}[c]{@{}c@{}}Resonance Feature\\ $k_r=5$\end{tabular}}      & \begin{tabular}[c]{@{}c@{}}$10^{5}\times\sigma \left(\omega_{\mathrm{b}}\right)$\\ $\sigma \left(\omega_{\mathrm{cdm}}\right)$\\ $\sigma \left( H_0 \right)$\\ $10^{12}\times\sigma \left( A_s \right)$\\ $\sigma \left( n_s \right)$\\ $\sigma \left( \tau_{\mathrm{reio}} \right)$\\ $\sigma \left(R\right)$\\ $\sigma \left(k_r\right)$\\ $\sigma \left( \phi_r \right)$\end{tabular} & \begin{tabular}[c]{@{}c@{}}3.01259
\\ 0.00015
\\ 0.06224
\\ 4.98610
\\ 0.00131\\ 0.00122
\\ 0.00093
\\ 0.06403
\\ 0.09604
\end{tabular} & \begin{tabular}[c]{@{}c@{}}3.01110
\\ 0.00015
\\ 0.05949
\\ 4.95293
\\ 0.00131
\\ 0.00121
\\ 0.00093
\\ 0.06403
\\ 0.09603
\end{tabular} & \begin{tabular}[c]{@{}c@{}}3.15035
\\ 0.00013\\ 0.05233
\\ 4.96385
\\ 0.00136\\ 0.00121
\\ 0.00092
\\ 0.06293
\\ 0.09443
\end{tabular} & \begin{tabular}[c]{@{}c@{}}3.14832
\\ 0.00012\\ 0.05083
\\ 4.92007
\\ 0.00136
\\ 0.00120\\ 0.00092
\\ 0.06292
\\ 0.09442\end{tabular} & \begin{tabular}[c]{@{}c@{}}2.81581
\\ 0.00015
\\ 0.05893
\\ 4.92440
\\ 0.00123
\\ 0.00126
\\ 0.00073
\\ 0.04190
\\ 0.06407
\end{tabular} & \begin{tabular}[c]{@{}c@{}}2.94256
\\ 0.00012
\\ 0.04803\\ 4.57095
\\ 0.00127
\\ 0.00114
\\ 0.00072
\\ 0.04152
\\ 0.06355\end{tabular} \\ \hline
\textbf{\begin{tabular}[c]{@{}c@{}}Resonance Feature\\ $k_r=30$\end{tabular}}          & \begin{tabular}[c]{@{}c@{}}$10^{5}\times\sigma \left(\omega_{\mathrm{b}}\right)$\\ $\sigma \left(\omega_{\mathrm{cdm}}\right)$\\ $\sigma \left( H_0 \right)$\\ $10^{12}\times\sigma \left( A_s \right)$\\ $\sigma \left( n_s \right)$\\ $\sigma \left( \tau_{\mathrm{reio}} \right)$\\ $\sigma \left(R\right)$\\ $\sigma \left(k_r\right)$\\ $\sigma \left( \phi_r \right)$\end{tabular} & \begin{tabular}[c]{@{}c@{}}2.92309
\\ 0.00015
\\ 0.06280\\ 5.02275
\\ 0.00130
\\ 0.00122
\\ 0.00185
\\ 0.14501
\\ 0.22661
\end{tabular} & \begin{tabular}[c]{@{}c@{}}2.92253
\\ 0.00015
\\ 0.06003
\\ 4.98925
\\ 0.00129
\\ 0.00120
\\ 0.00185
\\ 0.14494
\\ 0.22645
\end{tabular} & \begin{tabular}[c]{@{}c@{}}2.90828\\ 0.00013
\\ 0.05489\\ 4.98140
\\ 0.00129
\\ 0.00121
\\ 0.00185
\\ 0.13139
\\ 0.20549
\end{tabular} & \begin{tabular}[c]{@{}c@{}}2.90756
\\ 0.00013
\\ 0.05311
\\ 4.94304
\\ 0.00128
\\ 0.00120
\\ 0.00185
\\ 0.13090\\ 0.20460
\end{tabular} & \begin{tabular}[c]{@{}c@{}}2.77208
\\ 0.00015
\\ 0.05752\\ 4.93060\\ 0.00122
\\ 0.00124
\\ 0.00104
\\ 0.05391\\ 0.08419
\end{tabular} & \begin{tabular}[c]{@{}c@{}}2.70176
\\ 0.00012
\\ 0.04760
\\ 4.49634
\\ 0.00117
\\ 0.00112
\\ 0.00104
\\ 0.05347
\\ 0.08340
\end{tabular} \\ \hline
\textbf{\begin{tabular}[c]{@{}c@{}}Resonance Feature\\ $k_r=100$\end{tabular}}      & \begin{tabular}[c]{@{}c@{}}$10^{5}\times\sigma \left(\omega_{\mathrm{b}}\right)$\\ $\sigma \left(\omega_{\mathrm{cdm}}\right)$\\ $\sigma \left( H_0 \right)$\\ $10^{12}\times\sigma \left( A_s \right)$\\ $\sigma \left( n_s \right)$\\ $\sigma \left( \tau_{\mathrm{reio}} \right)$\\ $\sigma \left(R\right)$\\ $\sigma \left(k_r\right)$\\ $\sigma \left( \phi_r \right)$\end{tabular} & \begin{tabular}[c]{@{}c@{}}2.76257
\\ 0.00015\\ 0.06075
\\ 4.91394
\\ 0.00129
\\ 0.00120
\\ 0.00464
\\ 0.30486
\\ 0.52297
\end{tabular} & \begin{tabular}[c]{@{}c@{}}2.76121
\\ 0.00015
\\ 0.05812
\\ 4.88119
\\ 0.00128
\\ 0.00119
\\ 0.00464
\\ 0.30485
\\ 0.52286\end{tabular} & \begin{tabular}[c]{@{}c@{}}2.94451
\\ 0.00014
\\ 0.05644
\\ 4.94225
\\ 0.00135
\\ 0.00120
\\ 0.00463
\\ 0.22294
\\ 0.35597
\end{tabular} & \begin{tabular}[c]{@{}c@{}}2.94277
\\ 0.00013
\\ 0.05442
\\ 4.90184
\\ 0.00134
\\ 0.00119
\\ 0.00463
\\ 0.22186
\\ 0.35452
\end{tabular} & \begin{tabular}[c]{@{}c@{}}2.69495
\\ 0.00008\\ 0.02723
\\ 4.53101
\\ 0.00116
\\ 0.00110
\\ 0.00132
\\ 0.07385
\\ 0.11173
\end{tabular} & \begin{tabular}[c]{@{}c@{}}2.79164\\ 0.00008\\ 0.02661
\\ 4.039
\\ 0.00122
\\ 0.00097\\ 0.00132
\\ 0.07220
\\ 0.10870
\end{tabular} \\ \hline
\end{tabular}\par
\bigskip
\parbox{17.9cm}{\captionof{table}{The possible marginalized 1-$\sigma$ constraints of \textbf{CORE-M5+CMB-S4} experiment on resonance feature signal, for CMB-only data and its combination with DESI, EUCLID(CS), DESI+EUCLID(CS), EUCLID(GC) and EUCLID(CS+GC).}} \label{table:CORE-M5+CMB-S4-Resonance}
\end{minipage}


\section{\textbf{CMB Experimental Specifications}}\label{CMB:expspec}

\setlength{\tabcolsep}{2.5pt} 
\renewcommand{\arraystretch}{1.0} 
\newcolumntype{C}[1]{>{\Centering}m{#1}}
\renewcommand\tabularxcolumn[1]{C{#1}}
\begin{minipage}{\linewidth}
\centering
\captionsetup{font=footnotesize}
\begin{tabular}{|cccccccc|}
\hline
\multicolumn{1}{|c|}{Experiment}                             & \multicolumn{1}{c|}{$\ell_{\text{min}}$}                 & \multicolumn{1}{c|}{$\ell_{\text{max}}$}                   & \multicolumn{1}{c|}{$f_{\text{sky}}$}                   & \multicolumn{1}{c|}{Channel~[GHz]} & \multicolumn{1}{c|}{FWHM~[arcmin]} & \multicolumn{1}{c|}{$\Delta T$~[$\mu$K arcmin]} & $\Delta P$~[$\mu$K arcmin] \\ \hline
\multicolumn{8}{|c|}{\textbf{CMB-Only}}                                                                                                                                                                                                                                                                                                                                               \\ \hline
\multicolumn{1}{|c|}{\multirow{6}{*}{CORE-M5}}        & \multicolumn{1}{c|}{\multirow{6}{*}{2}}  & \multicolumn{1}{c|}{\multirow{6}{*}{3000}} & \multicolumn{1}{c|}{\multirow{6}{*}{0.70}}  & \multicolumn{1}{c|}{130}                   & \multicolumn{1}{c|}{8.51}                  & \multicolumn{1}{c|}{3.9}               & 5.5               \\ 
\multicolumn{1}{|c|}{}                                         & \multicolumn{1}{c|}{}                             & \multicolumn{1}{c|}{}                               & \multicolumn{1}{c|}{}                               & \multicolumn{1}{c|}{145}                   & \multicolumn{1}{c|}{7.68}                  & \multicolumn{1}{c|}{3.6}               & 5.1               \\ 
\multicolumn{1}{|c|}{}                                         & \multicolumn{1}{c|}{}                             & \multicolumn{1}{c|}{}                               & \multicolumn{1}{c|}{}                               & \multicolumn{1}{c|}{160}                   & \multicolumn{1}{c|}{7.01}                  & \multicolumn{1}{c|}{3.7}               & 5.2               \\ 
\multicolumn{1}{|c|}{}                                         & \multicolumn{1}{c|}{}                             & \multicolumn{1}{c|}{}                               & \multicolumn{1}{c|}{}                               & \multicolumn{1}{c|}{175}                   & \multicolumn{1}{c|}{6.45}                  & \multicolumn{1}{c|}{3.6}               & 5.1               \\ 
\multicolumn{1}{|c|}{}                                         & \multicolumn{1}{c|}{}                             & \multicolumn{1}{c|}{}                               & \multicolumn{1}{c|}{}                               & \multicolumn{1}{c|}{195}                   & \multicolumn{1}{c|}{5.84}                  & \multicolumn{1}{c|}{3.5}               & 4.9               \\ 
\multicolumn{1}{|c|}{}                                         & \multicolumn{1}{c|}{}                             & \multicolumn{1}{c|}{}                               & \multicolumn{1}{c|}{}                               & \multicolumn{1}{c|}{220}                   & \multicolumn{1}{c|}{5.23}                  & \multicolumn{1}{c|}{3.8}               & 5.4               \\ \hline
\multicolumn{1}{|c|}{CMB-S4}                          & \multicolumn{1}{c|}{30}                  & \multicolumn{1}{c|}{3000}                  & \multicolumn{1}{c|}{0.40}                   & \multicolumn{1}{c|}{150}                   & \multicolumn{1}{c|}{3.0}                   & \multicolumn{1}{c|}{1.0}               & 1.41              \\ \hline
\multicolumn{1}{|c|}{\multirow{8}{*}{PICO}}           & \multicolumn{1}{c|}{\multirow{8}{*}{2}}  & \multicolumn{1}{c|}{\multirow{8}{*}{3000}} & \multicolumn{1}{c|}{\multirow{8}{*}{0.70}}  & \multicolumn{1}{c|}{62.2}                  & \multicolumn{1}{c|}{12.8}                  & \multicolumn{1}{c|}{2.76}              & 3.9               \\ 
\multicolumn{1}{|c|}{}                                         & \multicolumn{1}{c|}{}                             & \multicolumn{1}{c|}{}                               & \multicolumn{1}{c|}{}                               & \multicolumn{1}{c|}{74.6}                  & \multicolumn{1}{c|}{10.7}                  & \multicolumn{1}{c|}{2.26}              & 3.2               \\ 
\multicolumn{1}{|c|}{}                                         & \multicolumn{1}{c|}{}                             & \multicolumn{1}{c|}{}                               & \multicolumn{1}{c|}{}                               & \multicolumn{1}{c|}{89.6}                  & \multicolumn{1}{c|}{9.5}                   & \multicolumn{1}{c|}{1.41}              & 2.0                 \\ 
\multicolumn{1}{|c|}{}                                         & \multicolumn{1}{c|}{}                             & \multicolumn{1}{c|}{}                               & \multicolumn{1}{c|}{}                               & \multicolumn{1}{c|}{107.5}                 & \multicolumn{1}{c|}{7.9}                   & \multicolumn{1}{c|}{1.20}               & 1.7               \\ 
\multicolumn{1}{|c|}{}                                         & \multicolumn{1}{c|}{}                             & \multicolumn{1}{c|}{}                               & \multicolumn{1}{c|}{}                               & \multicolumn{1}{c|}{129.0}                   & \multicolumn{1}{c|}{7.4}                   & \multicolumn{1}{c|}{1.13}              & 1.6               \\ 
\multicolumn{1}{|c|}{}                                         & \multicolumn{1}{c|}{}                             & \multicolumn{1}{c|}{}                               & \multicolumn{1}{c|}{}                               & \multicolumn{1}{c|}{154.8}                 & \multicolumn{1}{c|}{6.2}                   & \multicolumn{1}{c|}{0.99}              & 1.4               \\ 
\multicolumn{1}{|c|}{}                                         & \multicolumn{1}{c|}{}                             & \multicolumn{1}{c|}{}                               & \multicolumn{1}{c|}{}                               & \multicolumn{1}{c|}{185.8}                 & \multicolumn{1}{c|}{4.3}                   & \multicolumn{1}{c|}{1.84}              & 2.6               \\ 
\multicolumn{1}{|c|}{}                                         & \multicolumn{1}{c|}{}                             & \multicolumn{1}{c|}{}                               & \multicolumn{1}{c|}{}                               & \multicolumn{1}{c|}{222.9}                 & \multicolumn{1}{c|}{3.6}                   & \multicolumn{1}{c|}{2.19}              & 3.1               \\ \hline
\multicolumn{8}{|c|}{\textbf{CMB-CMB Combination}}                                                                                                                                                                                                                                                                                                                                    \\ \hline
\multicolumn{8}{|c|}{PLANCK+CMB-S4}                                                                                                                                                                                                                                                                                                                                          \\ \hline
\multicolumn{1}{|c|}{\multirow{2}{*}{PLANCK:~low-$\ell$}}  & \multicolumn{1}{c|}{\multirow{2}{*}{2}}  & \multicolumn{1}{c|}{\multirow{2}{*}{50}}   & \multicolumn{1}{c|}{\multirow{2}{*}{0.57}} & \multicolumn{1}{c|}{100}                   & \multicolumn{1}{c|}{10.0}                  & \multicolumn{1}{c|}{6.8}               & 10.9              \\ 
\multicolumn{1}{|c|}{}                                         & \multicolumn{1}{c|}{}                             & \multicolumn{1}{c|}{}                               & \multicolumn{1}{c|}{}                               & \multicolumn{1}{c|}{143}                   & \multicolumn{1}{c|}{7.1}                   & \multicolumn{1}{c|}{6.0}               & 11.4              \\ \hline
\multicolumn{1}{|c|}{CMB-S4:~high-$\ell$}                  & \multicolumn{1}{c|}{51}                  & \multicolumn{1}{c|}{3000}                  & \multicolumn{1}{c|}{0.40}                   & \multicolumn{1}{c|}{150}                   & \multicolumn{1}{c|}{3.0}                   & \multicolumn{1}{c|}{1.0}               & 1.41              \\ \hline
\multicolumn{1}{|c|}{\multirow{2}{*}{PLANCK:~high-$\ell$}} & \multicolumn{1}{c|}{\multirow{2}{*}{51}} & \multicolumn{1}{c|}{\multirow{2}{*}{3000}} & \multicolumn{1}{c|}{\multirow{2}{*}{0.17}} & \multicolumn{1}{c|}{100}                   & \multicolumn{1}{c|}{10.0}                  & \multicolumn{1}{c|}{6.8}               & 10.9              \\ 
\multicolumn{1}{|c|}{}                                         & \multicolumn{1}{c|}{}                             & \multicolumn{1}{c|}{}                               & \multicolumn{1}{c|}{}                               & \multicolumn{1}{c|}{143}                   & \multicolumn{1}{c|}{7.1}                   & \multicolumn{1}{c|}{6.0}               & 11.4              \\ \hline
\multicolumn{8}{|c|}{LiteBIRD+CMB-S4}                                                                                                                                                                                                                                                                                                                                        \\ \hline
\multicolumn{1}{|c|}{LiteBIRD:~low-$\ell$}                 & \multicolumn{1}{c|}{2}                   & \multicolumn{1}{c|}{50}                    & \multicolumn{1}{c|}{0.70}                   & \multicolumn{1}{c|}{140}                   & \multicolumn{1}{c|}{31}                    & \multicolumn{1}{c|}{4.1}               & 5.8               \\ \hline
\multicolumn{1}{|c|}{CMB-S4:~high-$\ell$}                  & \multicolumn{1}{c|}{51}                  & \multicolumn{1}{c|}{3000}                  & \multicolumn{1}{c|}{0.40}                   & \multicolumn{1}{c|}{150}                   & \multicolumn{1}{c|}{3.0}                   & \multicolumn{1}{c|}{1.0}               & 1.41              \\ \hline
\multicolumn{1}{|c|}{LiteBIRD:~high-$\ell$}                & \multicolumn{1}{c|}{51}                  & \multicolumn{1}{c|}{1350}                  & \multicolumn{1}{c|}{0.30}                   & \multicolumn{1}{c|}{140}                   & \multicolumn{1}{c|}{31}                    & \multicolumn{1}{c|}{4.1}               & 5.8               \\ \hline
\multicolumn{8}{|c|}{CORE-M5+CMB-S4}                                                                                                                                                                                                                                                                                                                                         \\ \hline
\multicolumn{1}{|c|}{\multirow{6}{*}{CORE-M5:~low-$\ell$}} & \multicolumn{1}{c|}{\multirow{6}{*}{2}}  & \multicolumn{1}{c|}{\multirow{6}{*}{50}}   & \multicolumn{1}{c|}{\multirow{6}{*}{0.70}}  & \multicolumn{1}{c|}{130}                   & \multicolumn{1}{c|}{8.51}                  & \multicolumn{1}{c|}{3.9}               & 5.5               \\ 
\multicolumn{1}{|c|}{}                                         & \multicolumn{1}{c|}{}                             & \multicolumn{1}{c|}{}                               & \multicolumn{1}{c|}{}                               & \multicolumn{1}{c|}{145}                   & \multicolumn{1}{c|}{7.68}                  & \multicolumn{1}{c|}{3.6}               & 5.1               \\ 
\multicolumn{1}{|c|}{}                                         & \multicolumn{1}{c|}{}                             & \multicolumn{1}{c|}{}                               & \multicolumn{1}{c|}{}                               & \multicolumn{1}{c|}{160}                   & \multicolumn{1}{c|}{7.01}                  & \multicolumn{1}{c|}{3.7}               & 5.2               \\ 
\multicolumn{1}{|c|}{}                                         & \multicolumn{1}{c|}{}                             & \multicolumn{1}{c|}{}                               & \multicolumn{1}{c|}{}                               & \multicolumn{1}{c|}{175}                   & \multicolumn{1}{c|}{6.45}                  & \multicolumn{1}{c|}{3.6}               & 5.1               \\ 
\multicolumn{1}{|c|}{}                                         & \multicolumn{1}{c|}{}                             & \multicolumn{1}{c|}{}                               & \multicolumn{1}{c|}{}                               & \multicolumn{1}{c|}{195}                   & \multicolumn{1}{c|}{5.84}                  & \multicolumn{1}{c|}{3.5}               & 4.9               \\ 
\multicolumn{1}{|c|}{}                                         & \multicolumn{1}{c|}{}                             & \multicolumn{1}{c|}{}                               & \multicolumn{1}{c|}{}                               & \multicolumn{1}{c|}{220}                   & \multicolumn{1}{c|}{5.23}                  & \multicolumn{1}{c|}{3.8}               & 5.4               \\ \hline
\multicolumn{1}{|c|}{CMB-S4:~high-$\ell$}                  & \multicolumn{1}{c|}{51}                  & \multicolumn{1}{c|}{3000}                  & \multicolumn{1}{c|}{0.40}                   & \multicolumn{1}{c|}{150}                   & \multicolumn{1}{c|}{3.0}                   & \multicolumn{1}{c|}{1.0}               & 1.41              \\ \hline
\multicolumn{1}{|c|}{\multirow{6}{*}{CORE-M5:~high-$\ell$}} & \multicolumn{1}{c|}{\multirow{6}{*}{51}}  & \multicolumn{1}{c|}{\multirow{6}{*}{3000}}   & \multicolumn{1}{c|}{\multirow{6}{*}{0.30}}  & \multicolumn{1}{c|}{130}                   & \multicolumn{1}{c|}{8.51}                  & \multicolumn{1}{c|}{3.9}               & 5.5               \\ 
\multicolumn{1}{|c|}{}                                         & \multicolumn{1}{c|}{}                             & \multicolumn{1}{c|}{}                               & \multicolumn{1}{c|}{}                               & \multicolumn{1}{c|}{145}                   & \multicolumn{1}{c|}{7.68}                  & \multicolumn{1}{c|}{3.6}               & 5.1               \\ 
\multicolumn{1}{|c|}{}                                         & \multicolumn{1}{c|}{}                             & \multicolumn{1}{c|}{}                               & \multicolumn{1}{c|}{}                               & \multicolumn{1}{c|}{160}                   & \multicolumn{1}{c|}{7.01}                  & \multicolumn{1}{c|}{3.7}               & 5.2               \\ 
\multicolumn{1}{|c|}{}                                         & \multicolumn{1}{c|}{}                             & \multicolumn{1}{c|}{}                               & \multicolumn{1}{c|}{}                               & \multicolumn{1}{c|}{175}                   & \multicolumn{1}{c|}{6.45}                  & \multicolumn{1}{c|}{3.6}               & 5.1               \\ 
\multicolumn{1}{|c|}{}                                         & \multicolumn{1}{c|}{}                             & \multicolumn{1}{c|}{}                               & \multicolumn{1}{c|}{}                               & \multicolumn{1}{c|}{195}                   & \multicolumn{1}{c|}{5.84}                  & \multicolumn{1}{c|}{3.5}               & 4.9               \\ 
\multicolumn{1}{|c|}{}                                         & \multicolumn{1}{c|}{}                             & \multicolumn{1}{c|}{}                               & \multicolumn{1}{c|}{}                               & \multicolumn{1}{c|}{220}                   & \multicolumn{1}{c|}{5.23}                  & \multicolumn{1}{c|}{3.8}               & 5.4               \\ \hline
\end{tabular}\par
\bigskip
\parbox{17.3578cm}{\captionof{table}{From left to right the CMB experiments and their associated specifications, multipole range~($\ell$), sky fraction, frequency channel, beam width, temperature and polarisation sensitivity are listed. These are adopted from Ref.~\cite{Brinckmann:2018owf}.}}\label{tab:CMBExpspec}
\end{minipage}

\section{\textbf{Power Spectra and Mock Likelihoods}}\label{pwrspec:likhd}
Here, we have presented a concise review on the galaxy power spectrum and the weak lensing power spectrum modelling, and the schemes to model their likelihoods, which have been adopted from the Ref.~\cite{Sprenger:2018tdb,Audren:2012vy}. Subsequently, a brief review on CMB likelihood is presented that is taken from the Ref.~\cite{Brinckmann:2018owf,Perotto:2006rj}. 
Here, we have only described the necessary information relevant to the current work; for detailed prescription we refer to the following Ref.~\cite{Sprenger:2018tdb,Audren:2012vy,Brinckmann:2018owf,Perotto:2006rj}.   

\subsection{Galaxy Clustering}\label{GC}
In order to extract the information of the primordial power spectrum from the LSS surveys, we need to convert the primordial power spectrum into the measured observables. In reality, we measure the distribution of galaxies over the sky. There are several non-trivial astrophysical and astronomical factors that we need to take into account to be capable of drawing the connection between the primordial power spectrum and the actual visible galaxy distribution. Basically, the galaxies are tracing the distribution of the underlying invisible cold dark matter. Hence, to derive the desired outcome, firstly, we need to build the relation between the primordial power spectrum and the matter power spectrum 

\be
\label{Matter:Power}
P_{\rm m}(k,z)=\frac{8 \pi^2}{25}k \frac{T(k)^2 D(z)^2}{\Omega_{\rm m}^2 H_0^4}
P({k})~,
\ee    

where, $ T(k) $ and $ D(z) $ are the \textit{Transfer function} and the \textit{Growth factor}, respectively, and $ P_{\rm m}(k,z) $ stands for matter power spectrum. One can find this above eq.~(\ref{Matter:Power}) using Poisson and Euler equations, for an extensive review one can see to~\cite{Dodelson:2003ft}. Now, our next goal is to construct the galaxy power spectrum from the above matter power spectrum, which is explicitly illustrated below~\cite{Sprenger:2018tdb}:

\be
\label{eqn:Pgal}
\begin{split}
P_{\rm g}(k, \mu, z) = \underbrace{
\left[ \frac{D_{\rm A}(z)}{D_{\rm A,t}(z)} \right]^2
\frac{H_{\rm t}(z)}{H(z)}}_{\textsf{Alcock-Paczinsky Effect}} \,
\underbrace{ e^{-k^2\left[\mu^2\cdot\left(\sigma_{\shortparallel}^2(z)-\sigma_{\perp}^2(z)\right)+\sigma_{\perp}^2(z)\right]}}_{\textsf{Resolution Effect}} \,
\underbrace{\overbrace{\left( 1+\beta(k_{\rm t},z) \, \mu_{\rm t}^2 \right)^2}^{\textsf{Kaiser Effect}}\overbrace{ e^{-k_{\rm t}^2\mu_{\rm t}^2\sigma_{\text{nl}}^2}}^{\textsf{FoG Effect}}}_{\textsf{Redshift Space Distortions}} \\
b_{\rm g}^2(z)P_{\rm m}(k_{\rm t},z) \,
\end{split}
\ee

where, $ \vec{k} $ and $ \hat{r} $ are the Fourier mode and the unit vector along the line-of-sight, respectively, $ \mu $ is the cosine of the angle between the mode vector and the line-of-sight.
Now, we have,

\begin{equation}
k = \vert\vec{k}\vert =\sqrt{k_{\perp}^2 + k_{\shortparallel}^2} \ ~\text{and} \ \mu = \frac{\vec{k}\cdot\hat{r}}{k}=\frac{k_{\shortparallel}}{k} \ ,
\end{equation}

where, $k_{\shortparallel}$ stands for the parallel part of a mode vector with respect to the line-of-sight, and the perpendicular part is $k_{\perp}$.

\begin{itemize}[itemsep=-.3em]
\item[\ding{113}] \textbf{\underline{Alcock-Paczinsky~Effect(AP)}:} To probe the galaxy power spectrum, we measure the redshifts and the positions of the galaxies. However, to compensate for the dearth of knowledge about true cosmology, a geometrical correction term is being introduced in modelling the observed galaxy power spectrum that is called the \textit{Alcock-Paczinsky effect}~\cite{Alcock:1979mp,Seo:2003pu}. The term has been shown explicitly in eq.~(\ref{eqn:Pgal}). In eq.~(\ref{eqn:Pgal}), where there is a '\textit{t}' in the suffix, that term corresponds to the true cosmology, which may differ from our assumed or fiducial cosmology. A simple coordinate transformation yields the true geometry from the fiducial one

\begin{equation}
k_\perp = \frac{D_{\rm A,t}(z)}{D_{\rm A}(z)}\,k_{\perp\,,\rm t}~ ,
\qquad k_\shortparallel=\frac{H(z)}{H_{\rm t}(z)}\,k_{\shortparallel\,,{\rm t}}.
\end{equation}

Using the above equations one can easily derive the conversion rule of the Fourier modes for true and fiducial geometry. The relation is shown below: 

\begin{equation}
\label{k^s}
\frac{k_{\rm t}}{k} = \left[\left(\frac{H_{\rm t}}{H}\right)^2\mu^2 + \left(\frac{D_{\rm A}}{D_{\rm A,t}}\right)^2\left(1-\mu^2\right)\right]^{1/2} 
\end{equation}

and

\begin{equation}
\label{mu^s}
\frac{\mu_{\rm t}}{\mu} = \left(\frac{H_{\rm t}}{H}\right)\cdot \frac{k}{k_{\rm t}} \ ~.
\end{equation}

Here, $H=\frac{\dot{a}}{a}$ and $D_\text{A} = \frac{r(z)}{(1+z)}$ are representing the Hubble parameter and the angular diameter distance, respectively, where $r(z)$ corresponds to the comoving distance and $ a $ is the scale factor. 

\item[\ding{113}] \textbf{\underline{Resolution~Effect}:} The term related to this effect is shown in eq.~(\ref{eqn:Pgal}). 
On small scales we encounter an apparent reduction in power because of instrumental resolving capacity. This reduction can be modelled by an exponential scale-dependent function shown in eq.~(\ref{eqn:Pgal}). Here, $\sigma_{\shortparallel}(z)$ and $\sigma_{\perp}(z)$ are redshift dependent Gaussian errors associated to the coordinates, along and perpendicular to the line of sight, respectively. 

\item[\ding{113}] \textbf{\underline{Redshift Space Distortion(RSD)}:} This effect consists of two distinct phenomena namely: (i) \textit{Kaiser effect}~\cite{Kaiser:1987qv}, (ii) \textit{Finger of God effect}(FoG)~\cite{Jackson:1971sky}. The template expressing the above stated composite effect has been shown in eq.~(\ref{eqn:Pgal}) along with a distinct depiction of each effect. The former effect arises due to Doppler effect. Apart from the cosmological redshift, the usual Doppler phenomena also contribute to the redshift-space power spectrum on large scales. Such contributions from Doppler effect give rise to an anisotropy in the picture. This effect is known as the \textit{Kaiser effect}.
The latter one is the result of peculiar velocities of the galaxies, which further add on to the redshift-space power spectrum a new correction; this phenomenon influences the small scales of the power spectrum~\cite{Bull:2014rha}. 
\end{itemize}

Here, the factor $ \beta $ in the redshift space distortion term is defined as

\begin{equation}
\label{eq:beta}
\beta(k_{\rm t},z) = -\frac{1+z}{2b(z)}\cdot\frac{\dd \ln P_{\rm m}(k_{\rm t},z)}{\dd z} \ .
\end{equation}

The fiducial value considered in our Fisher analysis for the parameter $ \sigma_{\text{nl}} $ associated with the FoG effect is 7 Mpc. The $ b(z) $ is the galaxy bias, whose role is to connect the underlying dark matter power spectrum to the galaxy power spectrum. In this analysis we have considered the binning of the entire survey volume with a bin width of $\Delta z = 0.1$ following Ref~\cite{Sprenger:2018tdb}.
To calculate the survey volume of each redshift bin, we need the following equation:

\begin{equation}
\label{eq:Vr}
 V_{\text{rs}}(\bar{z}) = \frac{4\pi}{3} f_{\text{sky}} \cdot\left[r^3\left(\bar{z}+\frac{\Delta z}{2}\right)-r^3\left(\bar{z}-\frac{\Delta z}{2}\right)\right] \ ,
\end{equation} 

where, $ r(z)$ stands for the comoving distance, which is defined as

\be
r(z) = \int_0^z \dfrac{c}{H(z)} dz.
\ee

In eq.~(\ref{eq:Vr}), the $\bar{z}$ and $f_{\text{sky}}$, respectively, symbolize the mean redshift and the fraction of the sky.
In each bin we have only a finite number of galaxies. Thus, to incorporate this effect, we need to add one more term to the galaxy power spectrum, which is called the Shot Noise ($ P_{\text{sn}}(\bar{z}) $); this allows us to get the observed power spectrum from the galaxy power spectrum. After considering all the effects ranging from the transfer function ($T(k)$), growth factor ($D(z)$) to AP, RSD and shot noise, we have finally ended up connecting the primordial perturbations to the observed galaxy power spectrum. The observed power spectrum is given by

\begin{equation}
\label{eq:pobs}
P_\mathrm{obs}(k,\mu,\bar{z}) = P_{\rm g}(k,\mu,\bar{z}) + P_{\rm sn}(\bar{z})~.
\end{equation}

The chi-square ($\chi^2$) function for this observed galaxy power spectrum can be expressed as follows\footnote{The details of the approach is given in Appendix A of the Ref.~\cite{Sprenger:2018tdb}.}:

\begin{equation}
\label{chi2-result}
\chi^2 = \sum_{\bar{z}}\int \dd^3\vec{k} \frac{V_{\rm rs}(\bar{z})}{2(2\pi)^3}\frac{\left(P_\mathrm{obs,t}(\vec{k},\bar{z})-P_\mathrm{obs}(\vec{k},\bar{z})\right)^2}{P_\mathrm{obs}^2(\vec{k},\bar{z})} \ .
\end{equation}

One can interchange chi-square to likelihood or vice-verse using this relation $\chi^2=-2\ln{\mathscr L}$. 
Here, the expression of $ P_\mathrm{obs}(\vec{k},\bar{z}) $ is given in eq.~(\ref{eq:pobs}). The shot noise $P_{\text{sn}}(\bar{z})$ can be obtained from the number density of galaxy per bin, which is given by,

\begin{equation}
P_{\text{sn}}(\bar{z}) = \frac{V_{\rm rs}(\bar{z})}{N(\bar{z})} \ ,
\end{equation}

where, $ V_{\rm rs}(\bar{z}) $ can be obtained from eq.~(\ref{eq:Vr}) and $ N(\bar{z}) $ is the number of galaxies per bin.

\subsection{Cosmic Shear}\label{CS}

Like the previous section, here also, we have given a brief review of the necessary details of the weak lensing angular power spectrum modelling, required to bridge the gap between our analysis and the prerequisites for it. Weak gravitational lensing or cosmic shear is another excellent probe of the underlying dark matter distributions, which play the role of the host of the visible galaxies. Measurement of cosmic shear provides the subtle picture of the distortions in the shapes of the background galaxies caused by the cosmic structures present in-between the lensed galaxies and the observer. In order to achieve the measured convergence angular power spectrum from the matter power spectrum, we need to convolute the 3-D matter power spectrum with the lensing kernel encapsulating all the lensing attributes of the intermediate space, inhabited by lensing objects. The final relation appears as~\cite{Sprenger:2018tdb,Audren:2012vy}

\begin{equation}\label{Cl_Pk}
{\cal C}_\ell^{ij} = \int_0^{\infty}\frac{\dd r}{r^2} {\cal K}_i(r) {\cal K}_j(r) P\left(k=\frac{\ell}{r},z\right) \ .
\end{equation}

The individual factors in eq.~(\ref{Cl_Pk}) and their connecting relations are given below,

\begin{align}
{\cal K}_i(r) &= \frac{3\Omega_mH_0^2}{2a(r)}{\cal F}_i(r)\\
{\cal F}_i(r) &=\int_r^\infty \dd r' H(r'){\cal n}_i(z)r\frac{(r'-r)}{r'}\\
{\cal n}_i(z) &= \frac{{\cal G}_i(z)}{\int_0^\infty {\cal G}_i(z') \dd z'}\\
{\cal G}_i(z) &= \int_{z_i^{\rm min}}^{z_i^{\rm max}}  {\cal E}(z,z') \, \frac{\dd {\cal n}_{\text{gal}}}{\dd z}(z') \, \dd z' \ .
\end{align}

There is a noise factor (${\cal N}_{\ell}$) too, like the galaxy clustering power spectrum. This noise factor  

\begin{equation}\label{Shear:Noise}
{\cal N}_{\ell}^{ij} = \delta_{ij}\frac{\sigma_{\text{lensing}}^2}{{\cal n}_i}  \ 
\end{equation}

arises because galaxies have their own innate ellipticities. Here, $\sigma_{\text{lensing}}$ is a parameter that measures the rms value of the intrinsic ellipticity of galaxies, having a numerical value of $0.3$. The likelihood or the chi-square for weak lensing angular power spectrum takes the following form\footnote{The detailed scheme is given in Appendix B of the Ref.~\cite{Sprenger:2018tdb}.}:

\begin{equation}
\label{LensingLikelihood}
-2\ln{\cal L}\equiv \chi^2 \equiv \sum_\ell(2\ell+1)f_{\text{sky}}\left(\frac{{\cal D}_\ell^{\text{mix}}}{{\cal D}_\ell^{\text{th}}}+\ln\frac{{\cal D}_\ell^{\text{th}}}{{\cal D}_\ell^{\text{obs}}}-{\cal N}\right) \ 
\end{equation}
where,
\begin{align}
{\cal D}_\ell^{\rm th} &=  \mid {\cal C}_\ell^{{\rm{th}}\,ij} +{\cal N}_\ell^{ij} \mid \ , \\
{\cal D}_\ell^{\rm obs} &= \mid {\cal C}_\ell^{{\rm{fid}}\,ij} +{\cal N}_\ell^{ij} \mid \ , \\
{\cal D}_\ell^{\rm mix} &= \sum_k  \Bigg| {\cal N}_\ell^{ij} + \begin{cases}{\cal C}_\ell^{{\rm{th}}\,ij} &, j\neq k\\[10pt]{\cal C}_\ell^{{\rm{fid}}\,ij} &, j=k \end{cases} \Bigg|.
\end{align}

Here, ${\cal D}_\ell$'s are the  determinants of the angular power spectra (${\cal C}_\ell$) along with corresponding noises and $ f_\text{sky} $ is the sky fraction. The dimension of the ${\cal C}_\ell$ matrices is equal to the number of redshift bins, which is designated by ${\cal N}$.

\subsection{Cosmic Microwave Background}\label{PWR:CMB}



In this section, we have introduced the likelihood of the CMB experiments used for our analysis~\cite{Perotto:2006rj}. Taking into consideration that the statistical distribution of both the CMB multipoles and its corresponding noises is Gaussian in nature, the likelihood function of the observed CMB data given the fiducial model can be illustrated as follows:

\begin{equation}\label{likelihood:CMB}
{\mathscr L}\propto \prod_{\ell~m} \exp \left( -\frac{1}{2} {\mathscr {D}^\dagger}_{\ell m}
\mathscr{C}^{-1} {\mathscr {D}}_{\ell m} \right),
\end{equation}

where, ${\mathscr {D}}_{\ell m} = \{a^ \text {T}_{\ell m},a^\text {E}_{\ell m},a^\text {d}_{\ell m} \}$ implies the data
vector and $\mathscr{C}$ is the 
theoretical covariance matrix made out of the theoretical CMB power spectra. The ${\mathscr {D}}_{\ell m}$ constitutes of $ a^ \text {T}_{\ell m} $,  $ a^ \text {E}_{\ell m} $ and $ a^ \text {d}_{\ell m} $, which are the coefficients of the spherical harmonic expansions of temperature, polarization, and CMB lensing potential maps, respectively.
In this study it has been assumed that there is no statistical correlation between CMB temperature and polarisation noises, thus applying $ N^{TE}_\ell =0$ for our analysis, where, $ N^{TE}_\ell$ implies noise spectrum. The quadratic estimator~\cite{Okamoto:2003zw} has been used to compute the error on CMB lensing potential measurement for a given fiducial model and noise spectra. However, in these quadratic estimators, the contributions coming from the auto-correlated multipoles of B-mode maps~($ a^ \text {B}_{\ell m} $) have not been considered because of their non-Gaussian nature. 
The chi-square ($\chi^2 \equiv - 2 \ln {\mathscr L}$) corresponding to the likelihood function described in eq.~(\ref{likelihood:CMB}) takes the following form\footnote{For detailed prescription one can see the Ref.~\cite{Perotto:2006rj}}:

\begin{equation}\label{chieff}
\chi^2 = \sum_{\ell} (2\ell+1) f_{\rm sky} \left(
\frac{\mathfrak{C}}{|\mathscr{C}|} + \ln{\frac{|\mathscr{C}|}{|\cal{C}|}} - 3 \right),
\end{equation}

where $\mathfrak{C}$ explicitly reads

\begin{eqnarray}
\mathfrak{C} &=&
{\cal C}_\ell^{\text {TT}}\mathscr{C}_\ell^{\text {EE}}\mathscr{C}_\ell^{\text {dd}} +
\mathscr{C}_\ell^{\text {TT}}{\cal C}_\ell^{\text {EE}}\mathscr{C}_\ell^{\text {dd}} +
\mathscr{C}_\ell^{\text {TT}}\mathscr{C}_\ell^{\text {EE}}{\cal C}_\ell^{\text {dd}} \nonumber\\
&&- \mathscr{C}_\ell^{\text {TE}}\left(\mathscr{C}_\ell^{\text {TE}}{\cal C}_\ell^{\text {dd}} +
2{\cal C}_\ell^{\text {TE}}\mathscr{C}_\ell^{\text {dd}} \right)
- \mathscr{C}_\ell^{\text {Td}}\left(\mathscr{C}_\ell^{\text {Td}}{\cal C}_\ell^{\text {EE}} +
2{\cal C}_\ell^{\text {Td}}\mathscr{C}_\ell^{\text {EE}} \right).
\end{eqnarray}

In eq.~(\ref{chieff}), the $|\mathscr{C}|$, ${|\cal C|}$ and $ f_\text{sky} $ represent the determinants of
the theoretical, observed covariance matrices and fraction of sky, respectively. Explicitly, $|\mathscr{C}|$ and ${|\cal C|}$ are as follows:

\be
|\mathtt{C}|=
\begin{vmatrix} 
\mathtt{C}_{\ell}^{\text {TT}} & \mathtt{C}_{\ell}^{\text {TE}} & \mathtt{C}_{\ell}^{\text {Td}} \\
\mathtt{C}_{\ell}^{\text {TE}} & \mathtt{C}_{\ell}^{\text {EE}} & 0\\
\mathtt{C}_{\ell}^{\text {Td}} & 0 & \mathtt{C}_{\ell}^{\text {dd}} \\
\end{vmatrix}
\quad
\ee
where, $\mathtt{C}$ denotes $\mathscr{C}$ or ${\cal C}$.


\section*{Acknowledgments}

DC thanks ISI Kolkata for financial support through Senior Research Fellowship. We gratefully acknowledge the computational facilities of
ISI Kolkata. SP thanks Department of Science and Technology, Govt. of India for partial support through Grant No. NMICPS/006/MD/2020-21.

\end{document}